\documentclass[]{pasj02} %図を表示する 新しいテンプレート 
\usepackage[switch,mathlines]{lineno} % add line number to manuscript
\usepackage{textcomp}
\usepackage[T1]{fontenc} % 既にあるかもしれませんが、念のため確認
\Received{$\langle$09--Dec--2025$\rangle$}
\Accepted{$\langle$27--July--2026$\rangle$}
\Published{$\langle$publication date$\rangle$}

\usepackage{natbib} %\citepのため必要
\usepackage{color}
\def\la{\ifmmode{\,\lesssim\,}\else$\,\lesssim$\,\fi}
\def\ga{\ifmmode{\,\gtrsim\,}\else$\,\gtrsim$\,\fi}

\begin{document}

\title{Hot  molecular cores in the W49A molecular cloud complex}

\author{Ryosuke Miyawaki$^1$, Masahiko Hayashi$^{2,3}$, and Tetsuo Hasegawa$^2$}%
% ORCID
%Ryosuke Miyawaki: https://orcid.org/0000-0001-5259-4080
%Masahiko Hayashi: https://orcid.org/0000-0002-4790-7940
%Tetsuo Hasegawa: https://orcid.org/0000-0003-1853-0184 

%
\altaffiltext{1}{College of Arts and Sciences, J.F. Oberlin University, Machida, Tokyo 194-0294, Japan}
\email{miyawaki@obirin.ac.jp}
\altaffiltext{2}{National Astronomical Observatory of Japan, 
2-21-1 Osawa, Mitaka, Tokyo 181-8588, Japan}
\altaffiltext{3}{JSPS Bonn Office, Ahrstr. 58, 53175 Bonn, Germany}

%\footnotetext[$\dag$]{Present address: ....}


\maketitle

%:ABSTRACT
\begin{abstract}\label{ABSTRACT}
We present a comprehensive dataset of hot molecular cores (HMCs) in the W49A 
molecular cloud complex based on high-resolution ALMA observations, including 
the 1.3~mm continuum, 12 molecular lines, and the H30$\alpha$ 
recombination line.
In total, 18 HMCs are identified in the CH$_{3}$CN ($J_K=12_3-11_3$) map, 
together with 20 continuum sources in the 1.3~mm map.
Ten HMCs have peaks coincident within 0\farcs1 of the 1.3~mm continuum 
peaks, indicating that thermal dust emission dominates the 1.3~mm emission 
for these sources.
Correlation analyses of the line luminosities suggest a common structural 
picture for HMCs, in which five distinct regions with different physical and 
chemical properties coexist: hot and dense gas (CH$_3$CN, HC$_3$N, HNCO, OCS, 
H$_2$CO), outflow gas (SiO, SO), envelope gas (SO$_2$, CH$_3$OH), extended gas 
($^{13}$CS, DCN, C$^{18}$O), and ionized gas (H30$\alpha$).
We find an empirical relation $X(\mathrm{CH_3CN})=2.5\times10^{-7}\exp[-490/T_\mathrm{rot}]$ 
between the fractional abundance and rotation temperature of CH$_{3}$CN, 
suggesting that $T\gtrsim200$--300~K is required to achieve high abundances of 
$\sim10^{-7}$.
We find that HMCs without embedded H/UCHII regions are more numerous than, or 
at least comparable in number to, HMCs with such regions, suggesting that the 
former may have longer lifetimes ($\sim$10$^{5}$~yr) than the latter 
($\sim$10$^{4}$~yr).
We discuss the implications of these results for the core accretion and 
competitive accretion models.

\end{abstract}
%\linenumbers

%:INTRODUCTION
\section{Introduction}\label{INTRODUCTION}

Hot molecular cores (HMCs) represent one of the most prominent stages in massive star formation.
They correspond to an evolutionary phase in which protostars are actively accreting circumstellar material \citep[e.g.,][]{Kurtz2000, Beuther2007}.
High-angular-resolution observations indicate that HMCs are internally heated by embedded high-mass protostellar objects (HMPOs), also referred to as massive young stellar objects (MYSOs) \citep[e.g.,][]{Rolffs2011, Serra2012, Sanna2014, Silva2017}.

An HMC is generally defined as a compact ($\la 0.1$~pc), dense ($n(\mathrm{H_2}) \ga 10^6$~cm$^{-3}$), and warm ($T_\mathrm{gas} \ga 100$~K) gas cocoon surrounding a deeply embedded HMPO \citep{Walmsley1995, Kurtz2000}.
Chemically, HMCs are characterized by a rich inventory of complex organic molecules (COMs), such as methyl cyanide (CH$_3$CN) and methanol (CH$_3$OH),
which are sublimated from grain mantles due to intense radiative heating from the central source \citep{Nomura2004}.
The lifetime of an HMC is estimated to be in the range of $10^4$--$10^5$~yr \citep{Herbst2009, Battersby2017}.

According to the standard scenario, massive star formation begins with high-mass starless cores (HMSCs), which correspond to the earliest evolutionary phase \citep{Motte2018}.
The formation and growth of HMPOs within HMSCs heat and chemically enrich the surrounding core, driving the transition from a cold HMSC to an HMC.

HMCs and their embedded HMPOs typically exhibit strong millimeter continuum and mid-infrared emission, but often lack detectable centimeter emission \citep{Sridharan2002}.
Since centimeter emission arises from free-free emission in ionized gas, its absence indicates that these HMPOs have not yet reached a stage at which they produce sufficient Lyman continuum (Lyc) photons to ionize their surroundings.

As evolution proceeds, HMPOs begin to emit sufficient Lyc photons to form hyper-compact HII (HCHII) regions.
This stage is followed by the development of ultracompact HII (UCHII) regions, which eventually evolve into classical HII regions \citep{DePree2004}.

An intermediate stage, called the Hollow Hot Molecular Core (HHMC), or “Hollow Hot Core,” exists between the HMC and the HII region \citep{Stephan2018}.
This stage represents a later phase of the HMC, retaining a density structure similar to that of a typical HMC but featuring an ionized central cavity corresponding to an HCHII or UCHII region \citep[e.g.,][]{Furuya2011, Rolffs2011, Serra2012, Fuente2018}.
Accordingly, the early evolutionary sequence of high-mass star formation can be divided into four stages: the “Dense Core” corresponding to the HMSC stage, the “Hot Core,” the “Hollow Hot Core,” and the “Post-HMC” stage, characterized by an expanding UCHII region.

While HMCs were traditionally considered a short-lived stage immediately preceding the formation of an ultra-compact HII (UCHII) region, recent surveys suggest that the HMC phase can persist into the HII region stage, acting as a long-lived reservoir of warm molecular gas throughout the early evolution of massive stars \citep{Meng2026}.

In this paper, we present a comprehensive study of HMCs in W49A, one of the major star-forming regions in our Galaxy.
\citet{Wilner2001} identified six hot cores as compact 1.3~mm dust continuum sources associated with CH$_3$CN emission toward W49N, the central part of W49A.
More recently, \citet{Nony2024} extracted 129 cores from an ALMA 1.3~mm (mean frequency of 226~GHz) continuum image of W49A.
Among these cores, they identified 40 associated with UCHII and HCHII regions, as well as 19 HMCs, suggesting that HMCs can coexist with UCHII and HCHII regions during a brief phase of the core lifetime---approximately 2$\times$10$^4$~yr.

We use the $\mathrm{CH_3CN}$ ($J_K=12_3-11_3$) line to identify HMCs, supplemented by information from other molecular lines, including $^{13}$CS, SO, SiO, CH$_{3}$OH, HNCO, HC$_{3}$N, OCS, H$_{2}$CO, DCN, and C$^{18}$O, as well as the recombination line H30$\alpha$.
We also utilize the ALMA 1.3~mm continuum map to identify cores and examine their relationship with the HMCs.

In \S2, we describe the radio continuum and molecular line observations, as well as the data reduction process, based on ALMA archival data.
\S3 presents the 1.3~mm continuum map, where we identify 20 continuum sources, including marginal cases.
In \S4, we present line maps and identify 18 HMC candidates based on spatially compact CH$_{3}$CN features, with detailed discussions of individual HMCs in Appendix2.
\S5 provides a statistical analysis of the HMCs, classifies the observed line properties, examines the fractional abundance of $\mathrm{CH_3CN}$, and discusses the implications for massive star formation.
Finally, \S6 summarizes our main findings.

Throughout this paper, we adopt a distance to W49N of 11.11$^{+0.79}_{-0.69}$kpc, following \citet{Zhang2013}.
For the 1.3~mm continuum cores identified by \citet{Nony2024}, we use the notation starting with the capital letter N followed by the core number; for example, N10 refers to core number 10 in their Table~A.1.

\section{ALMA archival data}\label{ALMA archival data}

For our study of HMCs in W49A, we used six archival data sets obtained with the Atacama Large Millimeter/submillimeter Array (ALMA).
These data sets were collected by several principal investigators (PIs), including B\'aez-Rubio, A. (\#2015.1.01535.S), Ginsburg, A. (\#2016.1.00620.S), Fu, X. (\#2017.1.01499.S), Galv\'an-Madrid, R. (\#2017.1.00318.S, \#2018.1.00589.S), and Wilner, D. (\#2018.1.00520.S).

%:%%% Table 1:ALMA archival data %%%
\begin{table*}[htbp]
\caption{ALMA archival data}
\begin{center}
\scalebox{0.7}[0.7]
{
\begin{tabular}{llcccccccccccccccl}

\hline\hline
\hfil Project &\hfil PI & Frequency & Total FoV & \multicolumn{2}{c}{Phase Center} & Number & Number & Continuum  & Position & Continuum & Maximum\\
 &  & (GHz) & Diameter  & $\alpha$(ICRS) & $\delta$(ICRS) & of 12-m & of 7-m  & Beam Size & Angle & Noise Level & Recoverable\\
 &  &  & ($''$)  &  19$^{\rm h}$10${\rm ^m}$ & 9\degree06$'$ & Antennas & Antennas & ($''\,\times\,''$) & (degree) & (mJy/beam) & Scale ($''$)\\
 \hline
\#2015.1.01535.S & B\'aez-Rubio, A. & 352.677--367.571 & 16.2 & 13\,\fs\,096 & 12$\,\farcs$416 & 40 & -- & 0\farcs18 $\times$ 0\farcs16& PA=$-48.4\degree$ & 3.3$^{2)}$ &7.0\\
\#2016.1.00620.S & Ginsburg, A. & 216.904--234.733 & 132.9 & 13\,\fs\,080 & 12$\,\farcs$165 & 43--45 & 9--12 & 0\farcs32 $\times$ 0\farcs28$^{3)}$ & PA=$-66.9\degree$$^{3)}$ & 1.4$^{3)}$ & 25\\
\#2017.1.00318.S & Galvan-Madrid, R. & 342.373--356.982 & 115.8 & 13\,\fs\,000 & 12$\,\farcs$479 & -- & 10--12 & 6\farcs8 $\times$ 3\farcs9$^{1)}$ & PA=$-73.2\degree$$^{1)}$ & 3.3$^{2)}$ &12\\ 
\#2017.1.01499.S & Fu, X. & 440.587--456.460 & 13.0 & 13\,\fs\,300 & 13$\,\farcs$000 & 44 & -- & 0\farcs66 $\times$ 0\farcs50 & PA=$70.1\degree$ & 9.1 &5.6\\
\#2018.1.00520.S & Wilner, D. & 240.757--259.569 & 23.3 & 13\,\fs\,300 & 14$\,\farcs$000 & 43--44 & -- & 0\farcs046$\times$ 0\farcs032$^{4)}$ & PA=$70.3\degree$$^{4)}$ & 0.7$^{4)}$ &3.9\\ 
\#2018.1.00589.S & Galvan-Madrid, R. & 85.386--100.919 & 163.2 & 13\,\fs\,300 & 12$\,\farcs$000 & 43 & -- & 1\farcs9 $\times$ 1\farcs6 & PA=$65.1\degree$ & 5.5 &46\\ 
\#2018.1.00589.S & Galvan-Madrid, R. & 251.489--270.485 & 137.9 & 13\,\fs\,300 & 12$\,\farcs$000 & -- & 9--11 & 6\farcs8 $\times$ 3\farcs9$^{1)}$ & PA=$-73.2\degree$$^{1)}$ & 30$^{1)}$ &17\\
\hline

\end{tabular}
}
\end{center}

$^{1)}$ Reffered from ALMA Science Archive\\
$^{2)}$ These two data sets are combined\\
$^{3)}$ \citet{Miyawaki2022b}. See text.\\
$^{4)}$ \citet{Miyawaki2023}\\

\label{Table: ALMA archival data}
\end{table*}
%%% Table 1 %%%

\subsection{Continuum data}\label{Continuum data}

We obtained continuum maps from all the archival data sets, with mean frequencies of 93, 226, 250, 350, and 450~GHz, corresponding to 3.2~mm, 1.3~mm, 1.2~mm, 0.86~mm, and 0.67~mm, respectively.
A summary of these data sets is given in Table~\ref{Table: ALMA archival data}.
Each data set consists of four spectral windows that together cover a total bandwidth of $\sim$2~GHz.
For example, the data set with an mean frequency of 226~GHz comprises four windows spanning 216.90--218.90~GHz, 218.85--220.85~GHz, 230.86--232.86~GHz, and 232.73--234.73~GHz.

The image analysis was performed using the CASA 5.8.0 \citep{CASA2022} and CARTA~5.0.3 \citep{Comrie2021} software.
The {\tt uvcontsub} task in CASA was used to separate line and continuum emissions for each spectral window.
A linear baseline was then fitted to the line-free channels, and the line-free intensities were averaged to derive the continuum data for each spectral window.
The 1.2~mm continuum map provides the highest angular resolution.
% 250~GHz
To construct the 0.86~mm continuum map, we combined the data sets obtained with the 12 m (\#2015.1.01535.S) and 7 m (\#2017.1.00318.S) arrays. %350~GHz

The {\tt tclean} parameters were set using {\tt weighting = briggs} with a {\tt robust} value of 0.5.
This choice yields a point-spread function (PSF) that smoothly interpolates between natural and uniform weighting, providing a balance between angular resolution and sensitivity in accordance with the signal-to-noise ratio of the maps, and allowing flexible control of the noise level.

The maximum recoverable scales (MRS) of these observations, listed in the last column of Table~\ref{Table: ALMA archival data}, range from 3\farcs9 to 46$''$ and are significantly larger than the typical sizes of the detected continuum sources and HMCs ($\sim$1$''$).
In particular, we used both the 12~m and 7~m array data to produce the 1.3~mm maps from project \#2016.1.00620.S.
The continuum and line datasets of this project have a minimum baseline of 8.9~m, corresponding to an MRS of $\sim$25$''$, which is an order of magnitude larger than the sizes of the detected continuum sources and HMCs ($\lesssim$3$''$).
The missing-flux effect on core-scale measurements is therefore expected to be minimal.
The largest uncertainty therefore arises from systematic errors in the absolute flux calibration using quasars and is at least 5\% \citep{Francis2021} or practically 5\%--10\% \citep{ALMA2023}.
From our flux density measurements on the 1.3~mm map, the values vary at the $\sim$10\% level when different fitting radii are adopted in the 2D Gaussian fits, which is consistent with this level of uncertainty.

Because we identify continuum sources on the 1.3~mm map in \S\ref{Continuum emission}, it is important to carefully assess its noise characteristics.
In Table~\ref{ALMA archival data}, we list a representative 1$\sigma$ noise level of 1.4~mJy\,beam$^{-1}$ for the 1.3~mm map, consistent with the noise level around MCN-a reported in our previous study \citep{Miyawaki2022b}.
However, the effective noise level---which includes residual dirty-beam patterns and artifacts from strong sources---varies across the map.

In the outer regions, at distances larger than $\sim$30$''$ from the UCHII ring, the noise floor is relatively uniform, with a 1$\sigma$ level of $\lesssim$1~mJy\,beam$^{-1}$.
In contrast, the central region (within a diameter of $\sim$30$''$) exhibits a non-uniform, concave noise floor with numerous weak spurious features.
This structure, as well as the elevated noise level, is caused by the presence of strong UCHII ring sources within the primary beam.
In this central region, the typical noise level ranges from 3 to 6~mJy\,beam$^{-1}$, and the effective detection limit increases to 10--20~mJy\,beam$^{-1}$.

\subsection{Line data}\label{Line Data}
We made maps of the emission lines of CH$_{3}$CN ($J_K=12_3-11_3$), $^{13}$CS ($J=5-4$), SO ($N_J=6_5-5_4$), SiO ($J=5-4$), CH$_{3}$OH ($J_K=4_3-3_1$), HNCO ($J_{K_a,K_c}=10_{0,10}-9_{0,9}$), HC$_{3}$N ($J=24-23$), OCS ($J=19-18$), H$_{2}$CO ($J_{K_a,K_c}=3_{2,2}-2_{2,1}$), SO$_{2}$ ($J_{K_a,K_c}=28_{3,25}-28_{2,26}$), DCN ($J=3=2$), C$^{18}$O ($J=2-1$) and H30$\alpha$.
The observing parameters for these lines are listed in Table~\ref{Table: Observing parameters for emission lines}.

%:%%% Table 2:Observing parameters for emission lines %%%
\begin{table*}[htbp]
\caption{Observing parameters for emission lines}
\begin{center}
\scalebox{0.9}[0.9]
{
\begin{tabular}{llllcr}

\hline\hline
\hfil Molecule &\hfil Transition &\hfil Frequency$^{1)}$ &\hfil E$_{\rm u}$$^{1)}$ &Resolution &PA \hfil\hfil\\
 &  &\hfil (GHz) &\hfil (K)& ($''\,\times\,''$) &(deg) \hfil\\
 \hline
%Continuum &  &  & 0.324$\times$0.275 & -66.9144 \\
CH$_3$CN & $J_K=12_3-11_3$ & 220.709017 &133.15712 & 0\farcs46$\times$0\farcs44 & $63.6\degree$ \\
$^{13}$CS & $J=5-4$ & 231.2206852 &33.29137 & 0\farcs42$\times$0\farcs38 & $-69.8\degree$ \\
SO & $N_J=6_5-5_4$ & 219.949442 &34.9847 & 0\farcs46$\times$0\farcs44 & $65.7\degree$ \\
SiO & $J=5-4$ & 217.104919 &31.25889 & 0\farcs44$\times$0\farcs38 & $-67.5\degree$ \\
CH$_3$OH & $J_K=4_3-3_1$ & 218.440063 & 45.45988& 0\farcs43$\times$0\farcs38 & $-66.9\degree$ \\
HNCO &$J_{K_a,K_c}=10_{0,10}-9_{0,9}$ &219.798274 & 58.01933& 0\farcs46$\times$0\farcs44 & $65.9\degree$ \\
HC$_3$N & $J=24-23$ & 218.324723 &130.98209 & 0\farcs43$\times$0\farcs38 & $-70.0\degree$ \\
OCS & $J=19-18$ & 231.0609934 &110.89923 & 0\farcs42$\times$0\farcs38 & $-66.9\degree$ \\
H$_2$CO &$J_K=3_{2,2}-2_{2,1}$ &218.475632 & 68.0937 & 0\farcs43$\times$0\farcs38 & $-67.2\degree$ \\
SO$_2$ & $J_{K_a,K_c}=28_{3,25}-28_{2,26}$ &234.1870526 & 403.03211 & 0\farcs42$\times$0\farcs38 & $-$70.0\degree \\ 
DCN & $J=3-2$ & 217.2386307 &20.85164 & 0\farcs44$\times$0\farcs38 & $-67.5\degree$ \\
C$^{18}$O &$J=2-1$ & 219.5603541 &15.8058  & 0\farcs47$\times$0\farcs44 & $68.4\degree$ \\
\hline
H30$\alpha$ &  &231.90092784 &\hfil ---&0\farcs42$\times$0\farcs38 & $-69.9\degree$ \\
\hline
\end{tabular}
}
\end{center}

{$^{1)}$ Taken from https://splatalogue.online}

\label{Table: Observing parameters for emission lines}
\end{table*}
%%% Table 2 %%%

To create the line maps, we applied the {\tt clean} task to the data from which the continuum emission had been removed.
Data cubes were produced with a spectral resolution of 2~km\,s$^{-1}$ for most of the analyzed spectral lines.
This resolution was adopted because of differences in the frequency-to-channel relationship among data sets obtained on different days.
To account for this variability and to facilitate comparison of the velocity structures observed in different lines, the flux in each original spectral channel was redistributed into the nearest 2~km\,s$^{-1}$ bin.
This was achieved by setting the width parameter of the {\tt tclean} task to 2~km\,s$^{-1}$.

The same weighting scheme was applied to the line observations; that is, the {\tt tclean} parameters were set with {\tt weighting = briggs} and a {\tt robust} value of 0.5.
Both the 12~m and 7~m array datasets were used, as in the continuum analysis, resulting in an MRS of $\sim$25$''$.
The largest uncertainty arises from systematic errors and is estimated to be 5\%--10\%.
Details of the spectral line analysis in each frequency band are provided in Appendix~\ref{Line data reduction}.

%RESULTS
\section{1.3~mm Continuum Sources}\label{Continuum emission}

\subsection{Source selection}\label{Source selection}

\begin{figure}[htbp]
\includegraphics[bb= 20 300 200 830, scale=0.48]{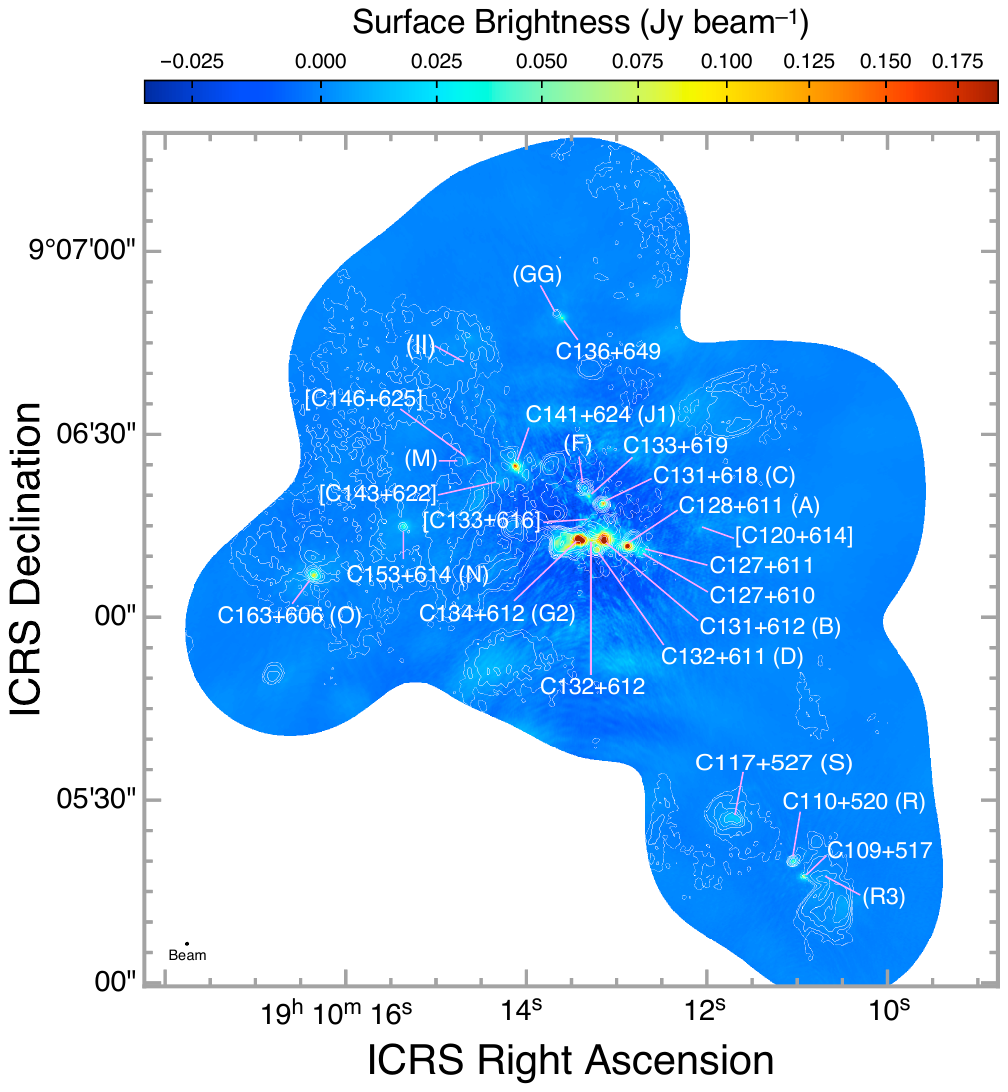}
\caption{1.3~mm continuum map (color).
The 3.6~cm continuum map from \citet{DePree1997} is overlaid with contours drawn at 1, 2, 5, 10, 20, and 50\% of the peak brightness.
The names of the 3.6~cm compact sources are indicated in parentheses.
When a 3.6~cm compact source coincides with a 1.3~mm continuum sourcewithin the 1.3~mm beam size, its name is given in parentheses after the corresponding 1.3~mm source.
The four sources enclosed in square brackets---[C120+614], [C133+616], [C143+622], and [C146+625]---show marginal 1.3~mm emission.
{Alt text: Color map of the 1.3~mm continuum emission over W49A. Contours overlay the 3.6 cm continuum from \citet{DePree1997} at 1, 2, 5, 10, 20, and 50\% of the peak. Compact 3.6 cm sources are labeled in parentheses; when a cm source coincides with a 1.3~mm continuum source(within the 1.3~mm beam), its cm name is appended in parentheses. Marginal 1.3~mm sources are shown in square brackets.}
}
\label{Fig: 226G continuum map}
\end{figure}

The 1.3~mm continuum map of W49A is shown in Figure~\ref{Fig: 226G continuum map}.
We identified 15 compact sources with high confidence by adopting a peak brightness threshold of 20 mJy\,beam$^{-1}$.
Given the 1$\sigma$ noise level of 3--6 mJy\,beam$^{-1}$ in the central region around the UCHII ring, this threshold ensures the robust detection of genuine sources, particularly in areas affected by artifacts from the unsubtracted dirty beam where the noise level reaches 10 mJy\,beam$^{-1}$.
At distances beyond 30$''$ from the map center, the noise decreases to $\sim$1 mJy\,beam$^{-1}$, enabling the detection of one additional faint source, C117+527, with a peak brightness of 15~mJy\,beam$^{-1}$.

%:Figure 1%%% 226GHz continuum map %%%
%%% Fig 1 226GHz continuum %%%

Four additional marginal continuum sources---C120+614, C133+616, C143+622, and C146+625---are marked with square brackets in Figure~\ref{Fig: 226G continuum map}.
We included these sources for reference when discussing their positional relationships with the corresponding CH$_{3}$CN peaks in \S\ref{Individual HMCs}.
[C120+614], [C143+622], and [C146+625] show peak brightnesses of 10--12 mJy\,beam$^{-1}$, corresponding to approximately the 3$\sigma$ level, given the typical noise of 2--3 mJy\,beam$^{-1}$ in the region 15$''$--30$''$ from the map center.
[C133+616], located within the UCHII ring, has a peak brightness of 25 mJy\,beam$^{-1}$, with a surrounding noise level of 10 mJy\,beam$^{-1}$.

Using the 2D Gaussian fitting routine in CASA, we measured the peak positions, peak brightness, major and minor axis lengths, position angles, and flux densities ($F_{226}$) of all 20 continuum sources.
These parameters are listed in Table~\ref{Table: 226 GHz continuum sources}.
The uncertainties in the brightness and flux density measurements are 5\%--10\%, as mentioned in the previous section.
We assigned source names based on their peak positions in right ascension and declination.
The positional errors were smaller than the pixel size, equivalent to 0\farcs05.

If available, the flux densities at 3.2~mm (93~GHz), 860~$\micron$ (350~GHz), and 650~$\micron$ (450~GHz) (referred to as $F_{93}$, $F_{350}$, and $F_{450}$, respectively) are also included in Table\ref{Table: 226 GHz continuum sources}.
A comparison of the listed objects with the 129 cores identified by \citet{Nony2024} is presented in Appendix~\ref{Nony sources}.
Details of the positional relationships with the CH$_{3}$CN peaks are provided in Appendix~\ref{Individual HMCs}.

%:%%% Table 3: 226 GHz continuum sources %%%
\begin{table*}[htbp]
\caption{1.3 mm continuum sources}
\begin{center}
\scalebox{0.8}[0.8]
{
\begin{tabular}{ccccccccccll}

\hline\hline\noalign{\vspace{1pt}}
Name$^{1)}$ & RA & DEC & $I_{\rm p}$ & FWHM Size & PA & $F_{226}$ & $F_{93}$ & $F_{350}$ & $F_{450}$ &\hfil 3.6~cm UCHII$^{2)}$ & Nony Cores$^{3)}$\\
&&&& major $\times$ minor &&&&&&&\\
 & 19$^{\rm h}$ 10$^{\rm m}$ &\hfil +9\degree & (Jy\,beam$^{-1}$) & ($''\,\times\,''$) & (\degree) & (Jy) &  (Jy) &  (Jy) & (Jy)&& \\
\noalign{\vspace{1pt}}\hline\noalign{\vspace{1pt}}
C109+517 & 10\fs943 & 5$'$17\farcs48 & 0.052 & 0.93 $\times$ 0.79 & 72 & 0.49 & 0.11 & & & 0\farcs2 SW of a weak peak &N4\\
C110+520 & 11\fs054 & 5$'$20\farcs07 & 0.041 & 0.92 $\times$ 0.83 & 103 & 0.40 & 0.11 &  &  & R &N9\\ 
C117+527 & 11\fs710 & 5$'$27\farcs04 & 0.015  & 3.21 $\times$ 2.00 &   85 & 1.09 & 0.37 & &&S &N39, N112$^{4)}$\\
{[}C120+614] & 12\fs089 & 6$'$14\farcs58 &0.010& 1.80 $\times$1.00 & 146 & 0.21 &<0.02&&& no emission &N36 \\
C127+611 & 12\fs707 & 6$'$11\farcs13 & 0.037 & 1.09 $\times$ 0.76 & 135 & 0.34 & $<$0.13 & 0.85 & 0.55 & no peak & N44 \\
C127+610 & 12\fs737 & 6$'$10\farcs60 & 0.031 & 1.65 $\times$ 1.02 & 151 & 0.58 & $<$0.13 & 0.24 & $<$0.1 &  no peak & N70 \\
C128+611 & 12\fs893 & 6$'$11\farcs72 & 0.207 & 1.33 $\times$ 1.04 &   71 & 3.40 & 0.87 & 2.02 & 3.84 & A & N3 \\ 
C131+618 & 13\fs150 & 6$'$18\farcs70 & 0.114 & 0.87 $\times$ 0.71 & 143 & 0.91 & 0.53 & 0.60 & 0.37 & C &N7, N72\\
C131+612 & 13\fs153 & 6$'$12\farcs69 & 0.327 & 1.55 $\times$ 1.40 &   62 & 8.33 & 1.40 & 5.35 & 15.6 & B & N2\\ 
C132+611 & 13\fs214 & 6$'$11\farcs25 & 0.078 & 1.30 $\times$ 1.05 &   91 & 1.28 & 0.3 & 0.92 & $<$0.2 & D & N14\\
C132+612 & 13\fs279 & 6$'$12\farcs68 & 0.096 & 0.96 $\times$ 0.62 &   80 & 0.64 & $<$0.3 & 1.57 & 3.30 & 0\farcs5 NE of E& N20\\ 
C133+619 & 13\fs319 & 6$'$19\farcs86 & 0.054 & 0.95 $\times$ 0.75 &   45 & 0.43 & $<$0.06 & 1.55 & 1.12 & no peak & N18 \\ 
{[}C133+616] & 13\fs331 & 6$'$11\farcs85 & 0.025 & 0.63 $\times$ 0.46 & 44 & 0.82 &<0.03&&& no peak &N54 \\
C134+612 & 13\fs426 & 6$'$12\farcs94 & 0.602 & 1.19 $\times$ 0.87 & 109 & 7.64 & 2.57 & 4.87 & 16.7 & G2 & N1 (N10, N16)\\
C136+649 & 13\fs618 & 6$'$49\farcs08 & 0.042 & 1.03 $\times$ 0.97 &   21 & 0.51 & 0.012&&& 1\farcs5 SW of GG & N6 \\
C141+624 & 14\fs131 & 6$'$24\farcs89 & 0.129 & 1.26 $\times$ 0.98 &   36 & 1.93 & 0.055 &&& J1 &N8 \\
{[}C143+622] &14\fs307 & 6$'$22\farcs25 & 0.012 & 1.63 $\times$ 0.89 & 97 & 0.20 &<0.01&&& no peak & N31 (N24)\\
{[}C146+625] & 14\fs655 & 6$'$25\farcs88 & 0.012 & 1.23 $\times$ 0.87 & 149 & 0.14 &<0.03&&& 1\farcs5 W of M& N19 \\
C153+614 & 15\fs367 & 6$'$14\farcs97 & 0.037 & 1.27 $\times$ 1.09 & 119 & 0.61 & 0.07 &&& N &N11 \\
C163+606 & 16\fs368 & 6$'$06\farcs88 & 0.082 & 1.16 $\times$ 1.07 & 102 & 1.21 & 0.51 &&& O & N12 (N5) \\
\noalign{\vspace{1pt}}\hline\noalign{\vspace{1pt}}

\end{tabular}
}
\end{center}

$^{1)}$ Names in square brackets indicate marginal sources.\\[4pt]
$^{2)}$ UCHII names assigned by \citet{DePree1997}.\\[4pt]
$^{3)}$ Core numbers assigned to the 1.3~mm continuum sources by \citet{Nony2024}. Numbers in parentheses denote secondary cores located in the vicinity of the main core.\\[4pt]
$^{4)}$ A shell-shaped UCHII region identified as a double source by \citet{Nony2024}.\\

\label{Table: 226 GHz continuum sources}
\end{table*}
%%% Table 3 %%%

Of the 20 sources, 10 coincide with UCHII regions previously identified by \citet{DePree1997} at 3.6 cm (8.3~GHz).
These matches were found within the 0\farcs3 beam size in the current data.
The UCHII regions corresponding to these sources are indicated in Figure~\ref{Fig: 226G continuum map} by adding their names in parentheses next to the source names.
The second-to-last column of Table~\ref{Table: 226 GHz continuum sources} lists the names of the corresponding UCHII regions.
Additionally, one further source, C109+517, nearly coincides with a weak 3.6~cm peak located at the western edge of UCHII R3, although its peak is offset by 0\farcs2 from the 3.6~cm peak.
Three sources---C132+612, C136+649, and [C146+625]---are located near UCHII regions, at offsets of 0\farcs5--1\farcs5 from the corresponding 3.6~cm peaks.
The remaining six sources have no corresponding 3.6~cm peaks.
% beam size: 0.324$\times$0.275 (PA=-66.9)

\subsection{Spectral energy distribution}\label{Spectral energy distribution}

\begin{figure}[htbp]
\includegraphics[bb= 100 330 500 760, scale=0.54]{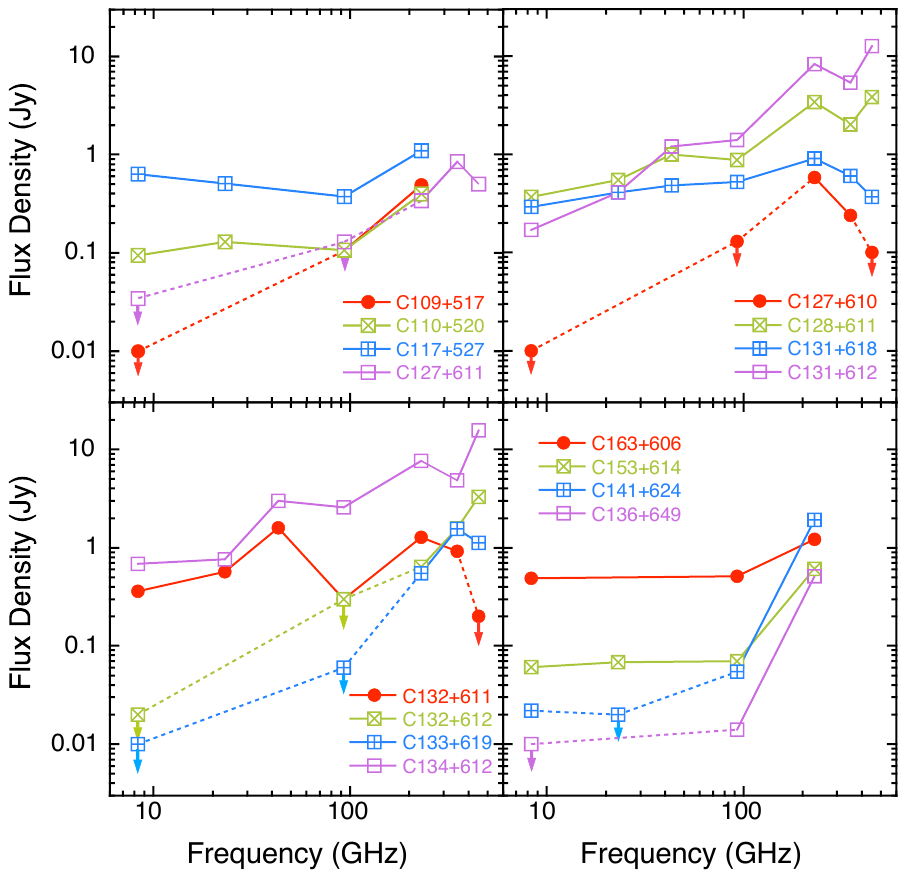}
\caption{SEDs for the 1.3~mm continuum sources.
{Alt text: Set of spectral energy distribution (SED) plots for the 1.3~mm continuum sources, showing flux density versus observing frequency from centimeter to (sub)millimeter wavelengths. 
Each SED traces how the continuum brightness changes with frequency for an individual source, highlighting flat/rising spectra and steeply rising high-frequency components.}
}
\label{Fig: SEDs}
\end{figure}

Figure~\ref{Fig: SEDs} shows the spectral energy distributions (SEDs) for the 16 continuum sources detected with certainty.
The flux densities at 3.6~cm, 1.4~cm, and 7.0~mm were taken from \citet{DePree1997, DePree2000, DePree2004}.
Most of the sources exhibit SEDs that are either flat or rising with frequency.

%:%%% Figure 2:SED %%%
%%% Figure 2  %%%

%:%%% Table 4:226-8.3 Correspondence %%%
\begin{table*}[htbp]
\caption{1.3~mm - 3.6~cm correspondence}
\begin{center}
\scalebox{0.8}[0.8]
{
\begin{tabular}{clccl}

\hline\hline\noalign{\vspace{1pt}}
Name &\hfill 3.6~cm UCHII & 93-226 GHz Spectral Index &H30$\alpha^{1)}$ & \hfill 226 GHz Continuum\\
 &\hfill Coincidence & $\alpha=\log(F_{226}/F_{93})/\log(226/93)$ &(Jy\,beam$^{-1}$~kms$^{-1}$) &\hfill Dust or Free-Free \\
\noalign{\vspace{1pt}}\hline\noalign{\vspace{1pt}}
C109+517 & shifted ($0\farcs2$)&1.7&6.7 & dust? + free-free \\
C110+520 & good &1.5&3.6 & dust? + free-free \\
C117+527 & good &1.2&1.7& dust? + free-free \\ % newly measured
{[}C120+614] & no 3.6~cm detected&$>$2.6&0.6 & dust \\
C127+611 & no 3.6~cm peak&$>$1.1&1.7 &dust\\
C127+610 & no 3.6~cm peak&$>$1.7&0.5&dust\\ % newly measured
C128+611 & good &1.5&22.5& dust + free-free? \\ % newly measured
C131+618 & good &0.6&10.3 & free-free \\ % need re-check
C131+612 & good &2.0&45.1 & free-free\\
C132+611 & shifted ($0\farcs2$) &1.6&22.6& free-free \\ % newly measured
C132+612 & no 3.6~cm peak&$>$0.9& $<$0.2 & dust + free-free? \\
C133+619 & no 3.6~cm detected&$>$2.2&$-$1.1 & dust \\
{[}C133+616] & no 3.6~cm peak&$>$3.6&0.9 & dust \\
C134+612 & good &1.2&87.0 & dust + free-free? \\
C136+649 & no 3.6~cm detected&4.2&1.5 & dust \\
C141+624 & good &4.0&8.6 & dust\\
{[}C143+622] & no 3.6~cm peak&$>$3.4&0.5 & dust\\
{[}C146+625] & no 3.6~cm detected&$>$1.7&0.5 & dust\\
C153+614 & good&2.5&2.4 & dust\\
C163+606 & good&1.0&7.9 & dust? + free-free \\
\noalign{\vspace{1pt}}\hline\noalign{\vspace{1pt}}
% Certain -- Probable -- Marginal -- Impossible
\end{tabular}
}
\end{center}
$^{1)}$ Velocity-integrated surface brightness of the line emission (or absorption, for negative values) at the H30$\alpha$ frequency toward each 1.3~mm continuum source.\\
\label{Table: 226-8.3 correspondence}
\end{table*}
%Table 4

Specifically, the sources C136+649, C141+624, and C153+614 exhibit SEDs that rise steeply between 3.2~mm and 1.3~mm.
Their flux densities at 1.3~mm are approximately an order of magnitude higher than those at 3.2~mm.
This sharp increase in flux indicates a spectral index greater than 2.5, consistent with optically thin thermal dust emission dominating at frequencies above 226~GHz for these sources.

The sources C127+611, C127+610, C132+612, and C133+619 were detected only at 226~GHz and higher frequencies.
The high-frequency emission from these sources may also be dominated by thermal dust emission.

The sources C110+520, C117+527, C131+618, and C163+606 exhibit flat SEDs up to 93~GHz, suggesting the presence of optically thin thermal free-free emission at these frequencies.
Among these, C110+520, C117+527, and C163+606 show signs of a flux increase toward higher frequencies beyond 93~GHz, indicating an additional contribution from thermal dust emission at frequencies above 226~GHz, although their flux densities at 350~GHz and 450~GHz are not available.

The SEDs of C109+517, C128+611, C131+612, and C134+612 show a gradual increase with frequency.
The latter three sources are located within clusters of UCHII and HCHII regions at the center of W49N, and their SEDs likely represent the combined emission from individual UCHII/HCHII regions at different evolutionary stages.
Consequently, these SEDs may comprise a mixture of thermal free-free and dust emission.

Table~\ref{Table: 226-8.3 correspondence} summarizes the positional relationships of the 20 1.3 mm sources with the 3.6~cm UCHII regions \citep{DePree1997}.
The table also lists the spectral indices between 93~GHz and 226~GHz, the integrated surface brightness of the H30$\alpha$ emission, and our assessment of whether the 1.3~mm continuum is dominated by free-free or thermal dust emission.

In some cases, the 1.3~mm continuum peaks are significantly offset from the nearby 3.6~cm peaks but are more closely aligned with the CH$_3$CN peaks. 
This suggests that the 1.3~mm emission toward these CH$_3$CN peaks is dominated by thermal dust emission, although a contribution from free-free emission associated with embedded H/UCHII regions cannot be excluded. 
These cases are discussed in detail in \S\ref{Origin of 226 GHz continuum}.

\section{Hot Molecular Cores}\label{Hot Molecular Cores}

\subsection{Line maps}\label{Line emission}

The emission of CH$_3$CN is a good indicator of HMCs \citep[e.g.][]{Wilner2001, Remijan2004}, and we therefore use CH$_3$CN emission to identify HMCs.  
It is also a valuable tracer of the physical and chemical conditions in HMCs.  
Additionally, it serves as a probe of complex molecular species that are otherwise difficult to observe, as demonstrated by \citet{Remijan2004}.  
Among the various CH$_3$CN transitions with different $K$ components, we adopt the $J_K=12_3-11_3$ transition to identify HMCs because it is the strongest unblended line, as illustrated in Figure~3 of \citet{Miyawaki2022b}.

\begin{figure*}[htbp]
\includegraphics[bb= 150 170 500 400, scale=1.1]{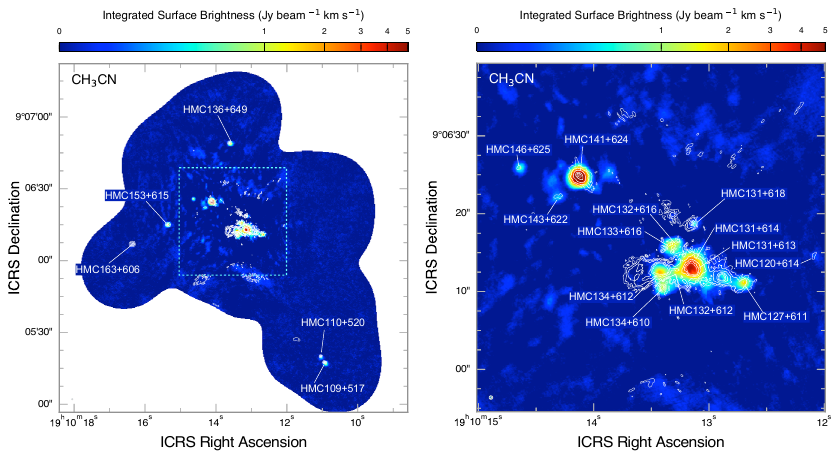}
\caption{
Maps of CH$_{3}$CN ($J_K=12_3-11_3$) emission integrated over the velocity range from $-$4 to 30~km\,s$^{-1}$.
The left panel shows the full observed field, while the right panel presents a zoomed view of the central region corresponding to the area outlined by the dashed blue rectangle in the left panel.
White contours represent the 1.3~mm continuum emission, plotted at 5, 10, 20, 40, 60, and 80\% of the peak brightness.
{Alt text: Two-panel integrated-intensity map of CH$_3$CN (J$_K$=12$_3$-11$_3$) emission summed over $V_{\rm LSR}$ = -4 to 30 km~s$^{-1}$. The left panel shows the full field and the right panel zooms into the central region (as indicated by the dashed blue rectangle). White contours show 1.3~mm continuum emission at 5, 10, 20, 40, 60, and 80\% of the peak.}
}
\label{Fig: CH3CN map}
\end{figure*}

\begin{figure*}[htbp]
\includegraphics[bb= 150 170 500 400, scale=1.1]{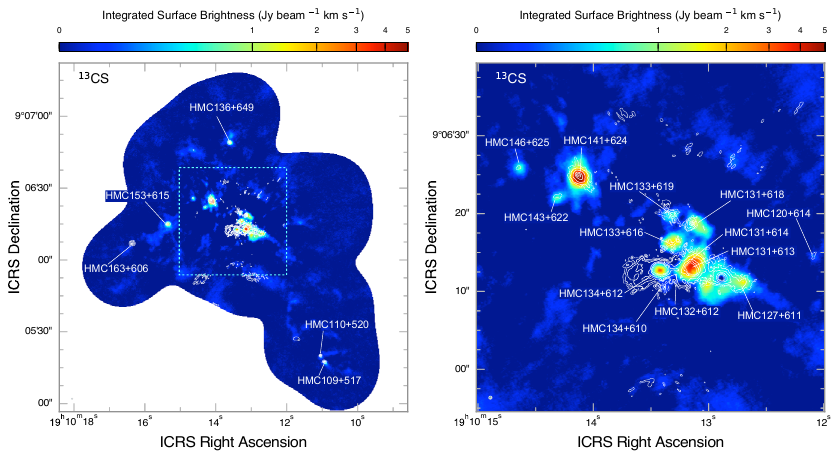}
\caption{
Same as Fig.~\ref{Fig: CH3CN map}, but for the $^{13}$CS  ($J=5-4$) emission. 
{Alt text: Two-panel integrated-intensity map of $^{13}$CS (J=5-4) emission. Layout matches Fig. 5: left panel shows the full field and right panel zooms into the central region. White contours show 1.3~mm continuum emission at 5, 10, 20, 40, 60, and 80\% of the peak.}
}
\label{Fig: 13CS map}
\end{figure*}

\begin{figure*}[htbp]
\includegraphics[bb= 150 170 500 400, scale=1.1]{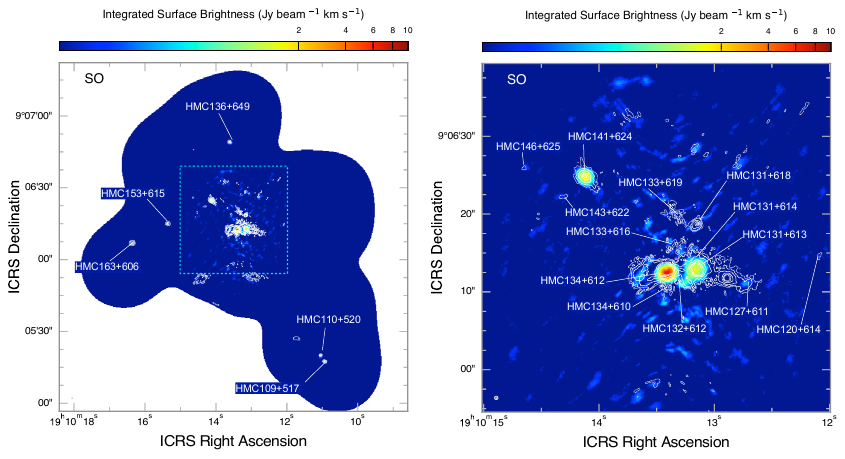}
\caption{
Same as Fig.~\ref{Fig: CH3CN map}, but for the SO ($N_J=6_5-5_4$) emission. 
{Alt text: Two-panel integrated-intensity map of SO (N$_J$=6$_5$-5$_4$) emission. Layout matches Fig. 5: left panel shows the full field and right panel zooms into the central region. White contours show 1.3~mm continuum emission at 5, 10, 20, 40, 60, and 80\% of the peak.}
}
\label{Fig: SO map}
\end{figure*}

\begin{figure*}[htbp]
\includegraphics[bb= 150 170 500 400, scale=1.1]{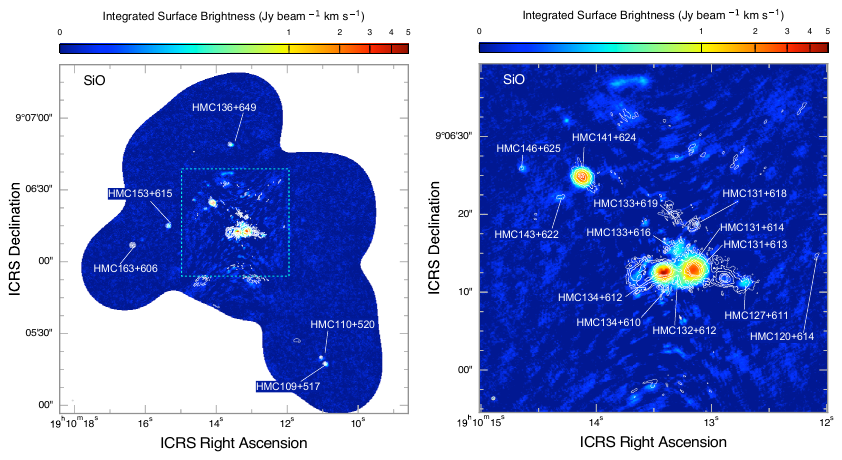}
\caption{
Same as Fig.~\ref{Fig: CH3CN map}, but for the SiO ($J=5-4$) emission.
{Alt text: Two-panel integrated-intensity map of SiO (J=5-4) emission. Layout matches Fig. 5: left panel shows the full field and right panel zooms into the central region. White contours show 1.3~mm continuum emission at 5, 10, 20, 40, 60, and 80\% of the peak.}}
\label{Fig: SiO map}
\end{figure*}

\begin{figure*}[htbp]
\includegraphics[bb= 135 170 500 400, scale=1.1]{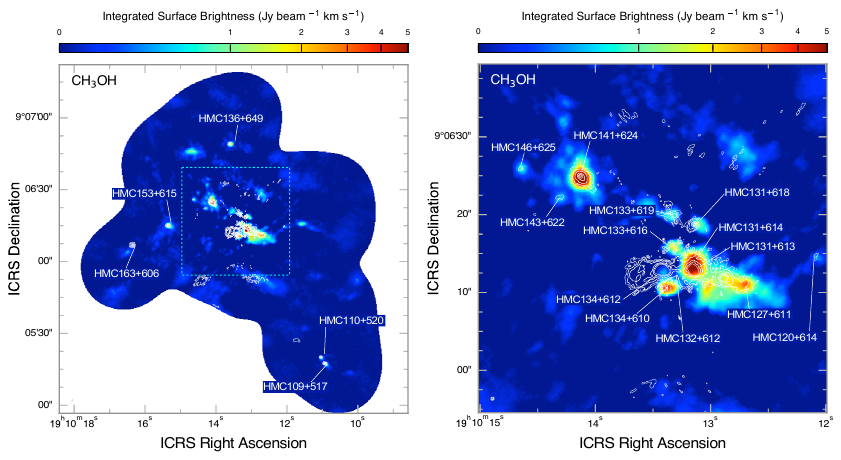}
\caption{
Same as Fig.~\ref{Fig: CH3CN map}, but for the CH$_{3}$OH ($J_K=4_{3}-3_{2}$) emission.
{Alt text: Two-panel integrated-intensity map of CH$_3$OH (J$_K$=4$_3$-3$_2$) emission. Layout matches Fig. 5: left panel shows the full field and right panel zooms into the central region. White contours show 1.3~mm continuum emission at 5, 10, 20, 40, 60, and 80\% of the peak.}
}
\label{Fig: CH3OH map}
\end{figure*}

\begin{figure*}[htbp]
\includegraphics[bb= 150 170 500 400, scale=1.1]{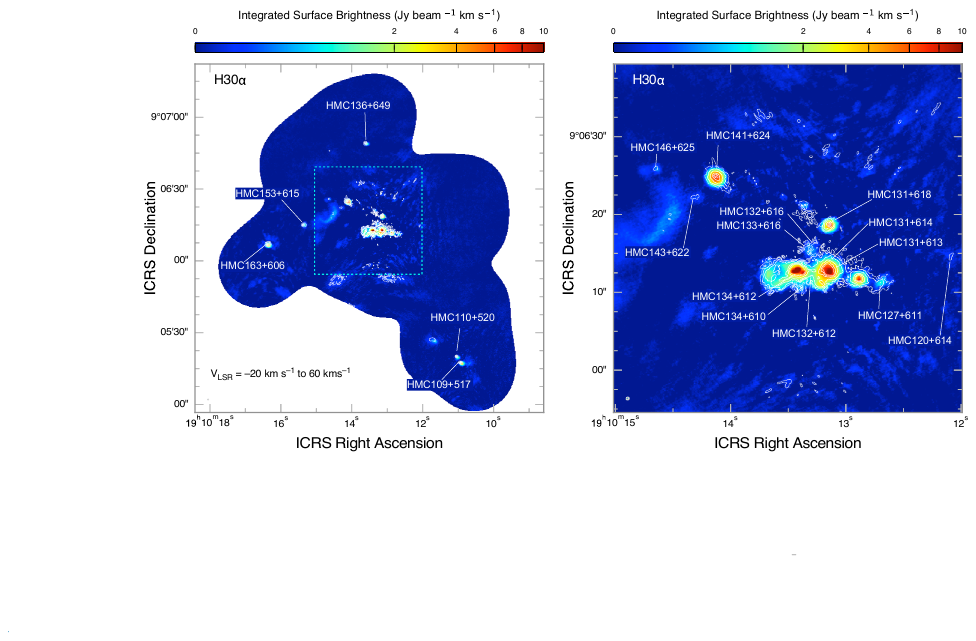}
\caption{
Same as Fig.~\ref{Fig: CH3CN map}, but for the H30$\alpha$ emission integrated from $-$20 to 60 km\,s$^{-1}$.
{Alt text: Two-panel integrated-intensity map of H30$\alpha$ emission integrated over $V_{\rm LSR}$ = -20 to 60 km~s$^{-1}$. Layout matches Fig. 5: left panel shows the full field and right panel zooms into the central region. White contours show 1.3~mm continuum emission at 5, 10, 20, 40, 60, and 80\% of the peak.}
}
\label{Fig: H30alpha map}
\end{figure*}

We show the integrated intensity (moment 0) maps of CH$_{3}$CN ($J_K=12_3-11_3$), $^{13}$CS ($J=5-4$), SO ($N_J=6_5-5_4$), SiO ($J=5-4$), CH$_{3}$OH ($J_K=4_3-3_1$), and H30$\alpha$ emissions from Figures~\ref{Fig: CH3CN map} to \ref{Fig: H30alpha map}, respectively.
The panels on the left display maps over the entire observed area, while the panels on the right offer detailed views of the central area, including the UCHII ring. 

The molecular line emissions were integrated over the velocity range from $-$4 to 30~km\,s$^{-1}$.
This integration range was determined by examining the CH$_{3}$CN line velocities\footnote{The notation $-$4~km\,s$^{-1}$, for example, denotes the velocity channel centered at this value. Since each velocity channel has a width of 2~km\,s$^{-1}$, the velocity-integrated specific intensity corresponds to the average over $-$5 to $-$3~km\,s$^{-1}$. This convention is adopted throughout this paper.} across the entire observed area and confirming that the emission is contained within this range.
This velocity range was then applied as the integration range for all molecular lines in Figures~\ref{Fig: CH3CN map} to \ref{Fig: CH3OH map}.
For the detailed maps of each selected CH$_{3}$CN source (Figures~\ref{Fig109+517}--\ref{Fig131+618}, \ref{Fig131+613}--\ref{Fig163+606}, and Figures~\ref{Fig:Other_mol_lines_map01} and \ref{Fig:Other_mol_lines_map02}), we adopted velocity integration ranges specific to each source, as shown in Figure~\ref{Fig: CH3CN line profiles} and listed in Table~\ref{Table: HMC candidates}.
The emission velocity of the H30$\alpha$ line was also examined across the entire area, confirming that it lies within $-$20 to 60~km\,s$^{-1}$.
The H30$\alpha$ line map in Figure~\ref{Fig: H30alpha map} was integrated over this range.

Figure~\ref{Fig: CH3CN map} shows the integrated intensity maps of the CH$_3$CN emission.
The emission features are compact and appear to be well separated from one another, frequently coinciding with the 1.3~mm continuum sources.
In the next section, we identify 18 HMCs based on this map.
The positions of these 18 HMCs are indicated, together with their names, on the CH$_3$CN maps.
Details of each HMC with additional maps of other molecular lines are given in Appendix~\ref{Individual HMCs}.

%:%%% Figure 3:CH3CN Integrated Intensity %%%
%%% Figure 3:CH3CN Integrated Intensity %%%

Figure~\ref{Fig: 13CS map} shows the integrated intensity maps of the $^{13}$CS emission.
The spatial distribution of $^{13}$CS closely resembles that of CH$_3$CN, although the $^{13}$CS emission appears to be slightly more extended.
This is reasonable because CS traces both dense, low-temperature gas as well as dense, high-temperature gas that is sampled by CH$_3$CN.
The $^{13}$CS emission is optically thin \citep[e.g.,][]{LeGal2019} and primarily traces regions with densities above $n$(H$_2$)~$\sim10^6$~cm$^{-3}$, corresponding to the critical density of the $J=5$ level of CS at 100~K.
$^{13}$CS features without accompanying CH$_3$CN emission can therefore be interpreted as cold cores.

%:%%% Figure 4:13CS Integrated Intensity %%%
%%% Figure 4:13CS Integrated Intensity %%%

The integrated intensity maps of the SO line are shown in Figure~\ref{Fig: SO map}.
The SO emission is detected toward many of the 18 HMCs, with particularly strong features associated with the UCHII regions B, G2, and J1.
SO emission is known to partially trace outflows, as observed in MCN-a \citep{Miyawaki2022b}, which is referred to as HMC141+624 in this paper.

%:%%% Figure 5:SO Integrated Intensity %%%
%%% Figure 5:SO Integrated Intensity %%%

The integrated intensity maps of the SiO line are shown in Figure~\ref{Fig: SiO map}.
SiO emission is detected toward many of the 18 HMCs.
As in the case of SO, the SiO emission is particularly strong toward the three 3.6~cm continuum sources B, G2, and J1.
Since SiO lines are known to originate from shocked regions often excited by outflows, we use the SiO and SO lines as outflow tracers to characterize the HMCs in \S\ref{correlarion logS-logS}.

%:%%% Figure 6:SiO Integrated Intensity %%%
%%% Figure 6:SiO Integrated Intensity %%%

%\subsubsection{CH$_3$OH emission}\label{Result-CH3OH}% outflow
Figure~\ref{Fig: CH3OH map} shows the integrated intensity maps of the CH$_3$OH line.
The overall distribution is similar to that of other molecular emission lines, although some HMCs exhibit notable differences in CH$_3$OH emission.

%:%%% Figure 7:CH3OH Integrated Intensity %%%
%%% Figure 7:CH3OH Integrated Intensity %%%

%\subsubsection{H30$\alpha$ emission}\label{Result-H30alpha} %HCHII or UCHII
Figure~\ref{Fig: H30alpha map} shows the integrated intensity maps of the H30$\alpha$ line.
The distribution of H30$\alpha$ emission, particularly the diffuse component, is consistent with that of the 3.6~cm continuum map \citep{DePree1997}.
In addition, many compact H30$\alpha$ sources are associated with the 1.3~mm continuum sources.

%:%%% Figure 8:H30alpha Integrated Intensity %%%
%%% Figure 8:H30alpha Integrated Intensity %%%

The H30$\alpha$ line is a reliable tracer of ionized gas, and we use its luminosity as an indicator of ionized gas in \S\ref{correlarion logS-logS}.
However, special care must be taken because, in many cases, emission lines of COMs, such as CH$_3$OCH$_2$OH, coincide with the H30$\alpha$ frequency.
For example, MCN-a (HMC141+624; UCHII J1) exhibits moderately strong emission in Figure~\ref{Fig: H30alpha map}, but shows weak 3.6~cm continuum emission, and only COM lines were detected at the H30$\alpha$ frequency \citep{Miyawaki2022b}.
We therefore examine the line width, intensity, and spatial distribution of the emission for each HMC to assess whether it represents a genuine H30$\alpha$ line or a COM line.
H30$\alpha$ lines from HCHII regions typically have FWHM line widths ($\Delta V_{\rm FWHM}$) greater than 30~km\,s$^{-1}$, which serves as a reliable diagnostic.
To confirm that spectral features at the H30$\alpha$ frequency indeed correspond to a recombination line, we also consider information from the H50$\beta$, H29$\alpha$, H26$\alpha$, and H32$\beta$ lines.

Many H30$\alpha$ sources are associated with the compact 3.6~cm continuum sources, including UCHII A, B, C, and G2 \citep{DePree1997}.
They exhibit H30$\alpha$ lines with large widths ($\Delta V_{\rm FWHM} \ga 30$ km\,s$^{-1}$) and broad wings.
In addition, H50$\beta$ lines have been detected from these sources, confirming that the emission lines observed at the H30$\alpha$ frequency toward them indeed correspond to the recombination line H30$\alpha$.

\subsection{Identification of Hot Molecular Cores}

\label{Identification Hot Molecular Cores}

We identified 18 HMCs in the CH$_{3}$CN ($J_K=12_3-11_3$) map (Figure~\ref{Fig: CH3CN map}) based on their spatial and velocity structures.
The integrated CH$_{3}$CN map has a typical noise level of 0.07~Jy\,beam$^{-1}$\,km\,s$^{-1}$ toward and around the UCHII ring, and 0.02~Jy\,beam$^{-1}$\,km\,s$^{-1}$ in the outer regions.
Accordingly, we adopted detection thresholds of 0.3~Jy\,beam$^{-1}$\,km\,s$^{-1}$ and 0.1~Jy\,beam$^{-1}$\,km\,s$^{-1}$ for the inner and outer regions, respectively, corresponding to the $>$4$\sigma$ noise levels, and identified isolated, simply shaped emission blobs.

We then examined the velocity channel maps and confirmed that the emission features are smoothly connected across contiguous channels.
If emission features did not connect smoothly across contiguous velocity channels—even when they appeared as a single, simple blob in the integrated intensity map—we identified them as two distinct HMCs.
Only HMC131+613 and HMC131+614 fall into this category (see Figure~\ref{Fig131+613-4}).

As a result, 18 emission blobs were identified as HMCs.
Among these, five are located away from the UCHII ring and can be clearly distinguished (left panel of Figure~\ref{Fig: CH3CN map}), while the remaining 13 are concentrated toward and around the UCHII ring at the center of W49N (right panel of Figure~\ref{Fig: CH3CN map}).
For the 13 HMCs in this crowded region, we confirmed their identification using the detailed maps presented in Appendix~\ref{Individual HMCs}.

To determine the peak positions, specific intensities, major and minor axis lengths, position angles, and line luminosities of the 18 HMCs, we performed 2D Gaussian fitting using the CASA routine.
The line luminosities were derived from the CH$_{3}$CN map integrated over the characteristic line-of-sight velocity range of each HMC, as listed in Table~\ref{Table: HMC candidates}.

For the specific intensity and line luminosity measurements, the overall uncertainty is estimated to be 5--10\%, reflecting systematic calibration uncertainties.
As discussed in \S2.1 and \S3.1, all detected HMCs have sizes (FWHM diameters) smaller than $\sim$3$''$---nearly an order of magnitude smaller than the MRS of 25$''$---and thus the missing-flux effect is negligible.
The angular sizes of the HMCs correspond to physical scales ranging from $\sim$5000~au to 10000~au.

The resulting parameters are listed in Table~\ref{Table: HMC candidates}. 
As with the continuum sources, we named each HMC according to its peak position in right ascension and declination. 
The positional uncertainties are smaller than the pixel size of 0\farcs05.  

We employed the same method to derive the positions and other parameters of the HMCs for $^{13}$CS ($J=5-4$), SO ($N_J=6_5-5_4$), SiO ($J=5-4$),
CH$_{3}$OH ($J_K=4_3-3_2$), HNCO ($J_{K_a,K_c}=10_{0,10}-9_{0,9}$),
HC$_{3}$N ($J=24-23$), OCS ($J=19-18$), H$_{2}$CO ($J_{K_a,K_c}=3_{2,2}-2_{2,1}$), DCN ($J=3-2$), C$^{18}$O ($J=2-1$), and H30$\alpha$.
These parameters are listed in Tables~\ref{Table:13CS_HMC_properties}--\ref{Table:C18O_HMC_properties} in Appendix~\ref{Additional tables and maps}.
The uncertainties in the measured specific intensities and luminosities are 5--10\%, for the same reason as for the CH$_{3}$CN parameters.

%:%%% Table 5: HMC candidates%%%
\begin{table*}[htbp]
\caption{HMC candidates and their CH$_{3}$CN ($J_K=12_3-11_3$) parameters}
\label{Table: HMC candidates}
\begin{center}
\scalebox{0.75}[0.75]
{
\begin{tabular}{ccccccccl}
\hline\hline\noalign{\vspace{1pt}}
HMC & \multicolumn{2}{c}{Position}  & Brightness &  FWHM Size &PA & Integration &Line &\hfil Remarks\\

Name & R.A. & DEC &at Fitted Peak & major $\times$ minor & & Range &Luminosity &\\
  &19$^{\rm h}$ 10$^{\rm m}$ & \hfil +9\degree &  (Jy\,beam$^{-1}$ km~s$^{-1}$) &   ($''\,\times\,''$) & (deg) & (km~s$^{-1}$) &(L$_\odot$)&\\

\noalign{\vspace{1pt}}\hline\noalign{\vspace{1pt}}
HMC109+517 & 10\fs943 & 05$'$ 17\farcs39 & 0.970 & 1.06 $\times$ 0.95 & 63.7 &2 to 16&  2.73$\times$10$^{-4}$ &western edge of R3, weak 3.6~cm peak\\
HMC110+520 & 11\fs061 & 05$'$ 20\farcs14 & 0.310 & 1.04 $\times$ 0.98 & 157.6 &6 to 14& 8.82$\times$10$^{-5}$&UCHII R\\
HMC120+614 & 12\fs095 & 06$'$ 14\farcs62 & 0.189 & 1.00 $\times$ 0.81 & 54.1 &8 to 14& 4.27$\times$10$^{-5}$ &marginal 1.3~mm, no 3.6~cm\\
HMC127+611 & 12\fs704 & 06$'$ 11\farcs23 & 1.684 & 1.08 $\times$ 0.99 & 87.1 &8 to 20& 5.02$\times$10$^{-4}$ &east of UCHII~A, no 3.6~cm peak\\
HMC131+618 & 13\fs136 & 06$'$ 18\farcs77 & 0.675 & 0.87 $\times$ 0.85 & 16.7 &0 to 10& 1.39$\times$10$^{-4}$ &UCHII~C\\
HMC131+613 & 13\fs159 & 06$'$ 13\farcs13 & 4.297 & 1.98 $\times$ 1.70 & 114.6 &12 to 30& 4.04$\times$10$^{-3}$ &0\farcs5 N of UCHII~B, cavity, ring or shell?\\
HMC131+614 & 13\fs166 & 06$'$ 14\farcs14 & 1.644 & 2.10 $\times$ 1.36 & 169.1 &4 to 12& 1.31$\times$10$^{-3}$ &2$\arcsec$ N of UCHII~B, 2$''$ SE of C1\\
HMC132+612 & 13\fs287 & 06$'$ 12\farcs66 & 1.648 & 1.30 $\times$ 1.03 & 101.6 &4 to 12&6.16$\times$10$^{-4}$ &0\farcs5 NE of UCHII E\\
HMC133+619 & 13\fs314 & 06$'$ 19\farcs85 & 0.177 & 0.95 $\times$ 0.73 & 110.6 &2 to 8& 3.42$\times$10$^{-5}$ &1$\arcsec$ SW of UCHII~F\\
HMC133+616 & 13\fs325 & 06$'$ 16\farcs15 & 1.578 & 1.68 $\times$ 1.10 & 104.4 &$-$2 to 8& 8.13$\times$10$^{-4}$ &1$\arcsec$ NE of UCHII B1$^{1)}$, triple?\\
HMC134+610 & 13\fs408 & 06$'$ 10\farcs56 & 1.268 & 1.61 $\times$ 1.25 & 175.6 &$-$4 to 8& 7.12$\times$10$^{-4}$ &2$\arcsec$  S of UCHII~G1 \& G2\\
HMC134+612 & 13\fs434 & 06$'$ 12\farcs58 & 1.640 & 1.32 $\times$ 0.96 & 88.7 &$-$4 to 6& 4.41$\times$10$^{-3}$ &0\farcs4 S of UCHII~G2, 0\farcs9 E of UCHII~G1\\
HMC136+649 & 13\fs613 & 06$'$ 49\farcs06 & 1.650 & 1.00 $\times$ 0.94 & 82.8 &14 to 26&4.33$\times$10$^{-4}$ &1\farcs5 SW of  UCHII ~GG\\
HMC141+624 & 14\fs132 & 06$'$ 24\farcs79 & 8.365 & 1.41 $\times$ 1.34 & 22.4 &4 to 24& 1.27$\times 10^{-4}$ &UCHII~J1, weak 3.6~cm peak\\
HMC143+622 & 14\fs311 & 06$'$ 22\farcs16 & 0.376 & 1.06 $\times$ 0.92 & 99.1 &$-$4 to 4& 1.02$\times$10$^{-4}$ &4$\arcsec$ W of UCHII~L, 4$''$ SE of UCHII~J1\\
HMC146+625 & 14\fs650 & 06$'$ 25\farcs93 & 0.719 & 0.95 $\times$ 0.90 & 132.0 &6 to 20& 1.72$\times$10$^{-4}$ &1\farcs5 W of UCHII~M\\
HMC153+615 & 15\fs364 & 06$'$ 15\farcs22 & 1.229 & 1.11 $\times$ 1.00 & 140.0 &12 to 22& 3.81$\times$10$^{-4}$ &0\farcs3 N of UCHII~N\\
HMC163+606 & 16\fs379 & 06$'$ 06\farcs52 & 0.128 & 0.96 $\times$ 0.90 & 125.7 &$-$4 to 0&3.08$\times$10$^{-5}$ &$-$1\farcs5 S of UCHII~O\\
\noalign{\vspace{1pt}}\hline\noalign{\vspace{1pt}}

Average &&&&1.25 $\times$ 1.04&&&8.15$\times$10$^{-4}$&\\
\noalign{\vspace{1pt}}\hline
\end{tabular}
}
\end{center}
$^{1)}$ This UCHII B1 is the source defined in \citet{DePree1997}, and it is distinct from the B1 in UCHII B defined in \citet{DePree2000,DePree2004,DePree2020}.\\
\end{table*}
% Table 5

%:%%% Dust core positions and HMC

We examined the positional correspondence between the HMC (CH$_{3}$CN) peaks and 1.3~mm continuum sources and summarize the results in Table~\ref{Table: HMC-Continuum}.
Some of the HMCs were previously identified by \citet{Wilner2001}, and 17 out of 18 HMCs correspond to the cores identified by \citet{Nony2024}.
This correspondence is also noted in Table~\ref{Table: HMC-Continuum}. 
Ten of the 18 HMCs show good agreement between the CH$_{3}$CN and 1.3~mm peak positions, where ``good'' agreement means that the two peaks coincide within 0\farcs1 (0.005~pc or 1100~au) or within two pixels.
Five HMCs (HMC110+520, HMC131+618, HMC134+610, HMC153+615, and HMC163+616) exhibit notable shifts (0\farcs13--0\farcs4) between the two peaks.
These positional shifts may indicate the presence of multiple hot cores if the 1.3~mm continuum peak traces another HMC, or they may represent unresolved shells whose original centers have been ionized by the stars that now produce the free-free 1.3~mm continuum.

%:%%% Table 6: HMC and Continuum Correspondence%%%
\begin{table*}[htbp]
\caption{HMC-1.3 mm Continuum correspondence}\label{Table: HMC-Continuum}
\begin{center}
\scalebox{0.8}[0.8]
{
\begin{tabular}{cccccc}
\hline\hline\noalign{\vspace{1pt}}
HMC & Continuum& Offset$^{1)}$ &CH$_{3}$CN-1.3 mm & Wilner et al. & Nony et al. \\
Name & Source & ($''\,\times\,''$) &Coincidence & (2001) $^{2)}$ & (2024)$^{3)}$ \\
\noalign{\vspace{1pt}}\hline\noalign{\vspace{1pt}}

HMC109+517 &C109+517& $0\farcs00\times0\farcs09$ &good&&N4 \\
HMC110+520 &C110+520& $-0\farcs10\times-0\farcs07$ &shifted ($0\farcs13$)&&N9  \\
HMC120+614 &[C120+614]& $-0\farcs09\times0\farcs04$ &good&& N36 \\
HMC127+611 &C127+611& $0\farcs04\times-0\farcs10$ &good&& N44 \\
HMC131+618 &C131+618& $0\farcs21\times-0\farcs07$ &shifted ($0\farcs22$)&& N7 \\
HMC131+613 &C131+612& $-0\farcs09\times-0\farcs44$ &shifted ($0\farcs45$) & MCN-f & (N32, N56, N83)$^{4)}$  \\
HMC131+614 &&&no peak$^{5)}$ & MCN-f & N28\\
HMC132+612 &C132+612& $-0\farcs12\times0\farcs02$ &good& MCN-e & N20 \\
HMC133+619 &C133+619& $0\farcs07\times0\farcs01$ &good&& N18 \\
HMC133+616 &[C133+616]& $0\farcs06\times-0\farcs08$ &good& MCN-d & N54 (N27, N43) \\
HMC134+610 & & & shifted ($0\farcs3$)$^{6)}$& MCN-c &  \\
HMC134+612 &C134+612 & $-0\farcs12\times0\farcs36$ & no peak$^{4)}$ & MCN-b & (N1, N10)$^{7)}$ \\
HMC136+649 &C136+649& $0\farcs07\times0\farcs02$ &good&& N6   \\
HMC141+624 &C141+624& $-0\farcs01\times0\farcs10$ &good& MCN-a& N8 \\
HMC143+622 &[C143+622]& $-0\farcs04\times0\farcs09$ &good&& N24 (N31) \\
HMC146+625 &[C146+625]& $0\farcs07\times-0\farcs05$ &good&& N19  \\
HMC153+615 &C153+614& $0\farcs04\times-0\farcs25$ &shifted ($0\farcs25$)&&N25 \\
HMC163+606 &C163+606& $-0\farcs16\times0\farcs36$ &shifted ($0\farcs4$)&& (N12) \\
% certain -- probable -- marginal -- impossible

\noalign{\vspace{1pt}}\hline
\end{tabular}
}
\end{center}
$^{1)}$ Offset of the fitted continuum peak in $\Delta\alpha\times\Delta\delta$ with respect to the fitted CH$_{3}$CN peak.\\
$^{2)}$ Name of the HMCs identified by \citet{Wilner2001}.\\
$^{3)}$ The 1.3~mm continuum cores identified by \citet{Nony2024}. 
Those located close to the corresponding HMCs are in parentheses.\\
$^{4)}$ The HMC peak is located within 0\farcs5 of the three cores.\\
$^{5)}$ No 226 GHz continuum peak is visible in the vicinity of the CH$_{3}$CN peak position.\\
$^{6)}$ A marginal 1.3~mm continuum peak of $\sim$20~mJy\,beam$^{-1}$ is located 0\farcs3 west of the CH$_{3}$CN peak.\\
$^{7)}$ The CH$_3$CN emission peak is located 0\farcs4 south of N1 and 0\farcs4 southwest of N10.\\
\end{table*}
% Table 6

\subsection{Rotation temperature of CH$_3$CN}\label{Rotation temperature}

Two sets of rotational transitions of CH$_3$CN, $J_K=12_K-11_K$ and $J_K=24_K-23_K$, were detected in the present data.
From the intensities of these lines, which originate from different energy levels, we estimated the rotation temperature of CH$_3$CN.
The analysis was performed using the rotational diagram method introduced by \citet{Hollis1982} and subsequently refined by several authors \citep[e.g.,][]{Loren1984, Turner1991, Goldsmith1999}.

Details of the rotation temperature derivation are given in \citet{Miyawaki2022b}.
In deriving the rotation temperatures, we assumed that all CH$_3$CN lines are optically thin. This assumption is supported by the observed CH$_3$CN-to-CH$_3$$^{13}$CN ($J_K=12_3-11_3$) intensity ratio of $\sim$40.
Furthermore, we do not find that the rotation temperatures obtained from the low-$K$ components are systematically higher than those derived from the high-$K$ components, which would be expected if significant optical depth effects were present \citep{Araya2005}.
We therefore conclude that optical depth effects are negligible.

The rotation temperatures derived from the CH$_3$CN ($J_K=12_K-11_K$)
transitions for the 18 HMCs are listed in Table~\ref{Table: Rotation Temperature}.
Rotation temperatures derived from the CH$_3$CN ($J_K=24_K-23_K$)
transitions for seven HMCs are also presented in Table~\ref{Table: Rotation Temperature}.
For most HMCs, the 1$\sigma$ uncertainties in the derived temperatures are within 20\%.
The average rotation temperatures are 176$\pm52$~K and 222$\pm25$~K for the
$J_K=12_K-11_K$ and $J_K=24_K-23_K$ transitions, respectively.
No significant correlations were found between HMC size and rotation temperature,
or between line luminosity and rotation temperature.

%:%%% Table 8 Rotation Temperature  derived from CH$_3$CN%%%
{\renewcommand{\arraystretch}{1.3}
\begin{table}[htbp]
\caption{Rotation Temperature  derived from CH$_3$CN}
\begin{center}
\scalebox{0.7}[0.7]
{
\begin{tabular}{ccccccccccl}

\hline\hline
Name  & $J=12_K-11_K$ (220~GHz) &  $J=24_K-23_K$ (440~GHz) &    \\
 & (K) & (K)  & &  &   \\ 

 \hline
 
HMC109+517 &  171.66$^{+10.30}_{-9.19}$ &    &  & \\
HMC110+520 &  131.55$^{+8.21}_{ -7.30}$ &    &  &  \\
HMC120+614 &  238.64$^{+32.64}_{ -25.63}$ &    &  &  \\
HMC127+611 &  150.09$^{+26.19}_{ -19.41}$ &    &  &  \\
HMC131+618 &  171.06$^{+18.84}_{ -15.44}$ &   236.35$^{+48.42}_{-34.35}$  \\
HMC131+613 &  108.01$^{+6.64}_{ -5.92}$ &   221.24$^{+22.85}_{-18.40}$ \\
HMC131+614 &  153.06$^{+12.19}_{ -10.51}$ &   245.60$^{+56.63}_{-38.76}$   \\
HMC132+612 &  218.01$^{+14.65}_{ -12.91 }$ &  252.95$^{+83.64}_{-50.34}$   \\
HMC133+619 &  181.86$^{+31.40}_{ -23.34}$ & & & \\
HMC133+616 &  131.62$^{+4.90}_{ -4.56}$ &  199.82$^{+25.11}_{-20.06}$ \\
HMC134+610 & 116.41$^{+8.52}_{ -7.44}$ &  182.99$^{+20.80}_{-16.95}$   \\
HMC134+612 & 227.50$^{+11.49}_{ -10.44}$ &   211.88$^{+36.83}_{-27.33}$   \\
HMC136+649 &  254.11$^{+36.77}_{ -28.52}$ &  &  &  &  \\
HMC141+624  & 305.78$^{+137.24}_{ -72.32}$ &    \\
HMC143+622 &  153.28$^{+13.92}_{ -11.78}$ &    &  &  \\
HMC146+625 &  181.06$^{+15.57}_{ -13.78}$ &    &  &  \\
HMC153+615 & 152.87$^{+12.27}_{ -10.57}$ &    &  & \\
HMC163+606 & 116.70$^{+6.31}_{ -5.56}$ &   &  &  \\[1pt]

\hline

\end{tabular}
}
\end{center}
\label{Table: Rotation Temperature}
\end{table}
}
% Table 8 Rotation Temperature  derived from CH$_3$CN%%%

\section{Discussion}

\subsection{Correlations among line luminosities}
\label{correlarion logS-logS}
% Figure 33: Line Luminosity Histograms %%%

In the preceding sections, we analyzed the characteristics of 18 hot molecular cores (HMCs), as presented in Figures~\ref{Fig109+517}--\ref{Fig163+606}, Tables~\ref{Table: HMC candidates}--\ref{Table: HMC-Continuum}, and  
Tables~A1--A12.
The twelve molecular transitions and one hydrogen recombination line, along with the 1.3~mm continuum emission, exhibit distinct behaviors, which may reflect both the evolutionary stages of the embedded stars.
In this subsection, we will discuss their statistical properties.

\begin{figure}[htbp]
\includegraphics[bb= 0 0 300 400, scale=0.38]{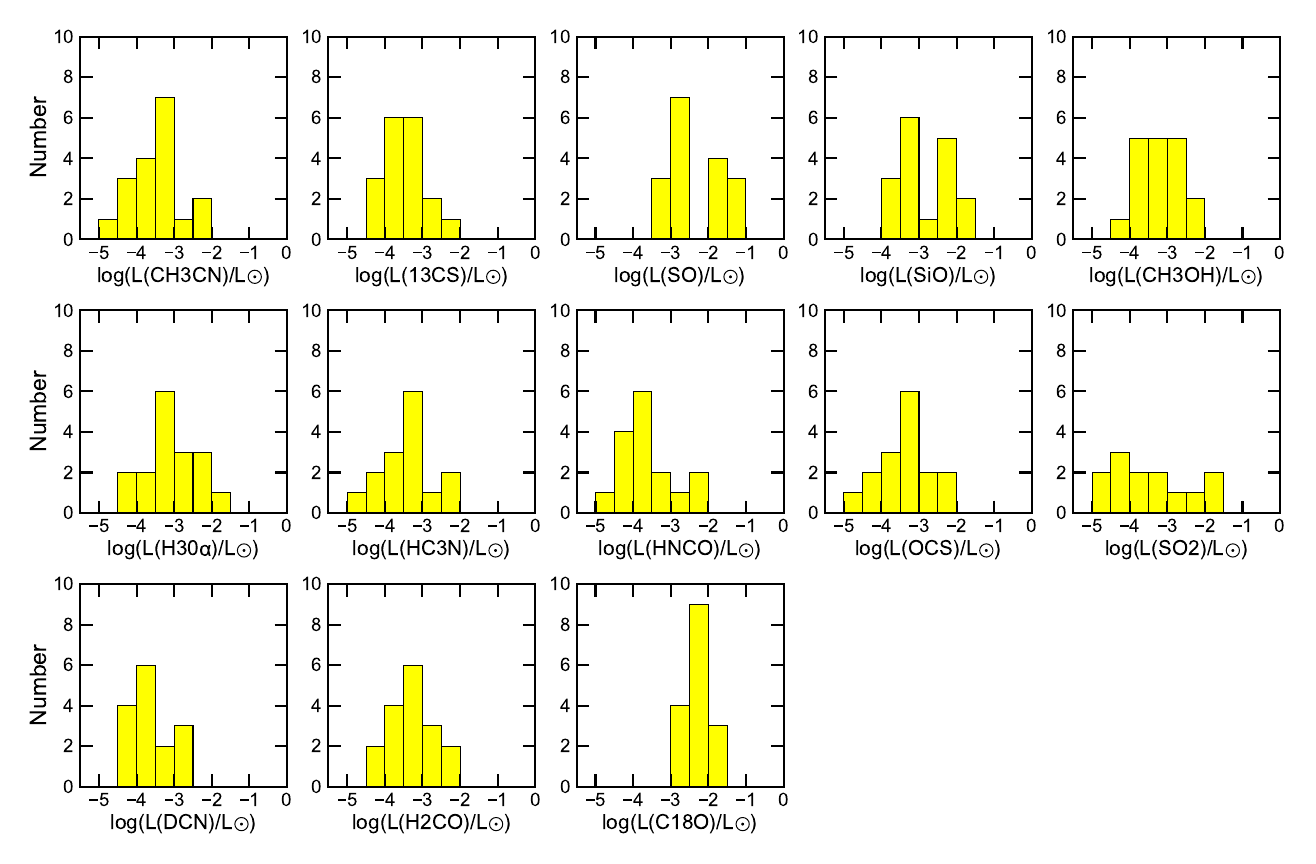}
\caption{Line Luminosity Histograms.
{Alt text: Grid of histograms showing the distribution of line luminosities for each analyzed transition: CH$_3$CN, $^{13}$CS, SO, SiO, CH$_3$OH, H30$\alpha$, HC$_3$N, HNCO, OCS, SO$_2$, DCN, H$_2$CO, and C$^{18}$O. Each panel plots the number of HMCs versus log10[L(line)/L$_\odot$] for that line.}
}
\label{Figs:L-histograms}
\end{figure}

Figure~\ref{Figs:L-histograms} presents histograms of the line luminosities for the observed molecular lines and H30$\alpha$.
For the brightest HMCs, the luminosity emitted by each line reaches up to 10$^{-2}$~L$_\odot$, although it can exceed 10$^{-2}$~L$_\odot$ for lines that tend to be spatially extended, such as SO, SiO, and C$^{18}$O.
The lower end of the luminosity distribution, around $\sim$10$^{-4}$~L$_\odot$, is limited by the methodology and sensitivity; there may exist smaller and/or less luminous HMCs, and the actual number of HMCs could be higher in that range.
If an HMC we identified actually consists of multiple unresolved HMCs, the line luminosity associated with each constituent would be smaller.

%:%%% Figure 9: Line Luminosity Histograms %%%
%%% Figure 9: Line Luminosity Histograms %%%

\begin{figure}[htbp]
\includegraphics[bb= 50 130 600 450, scale=0.32]{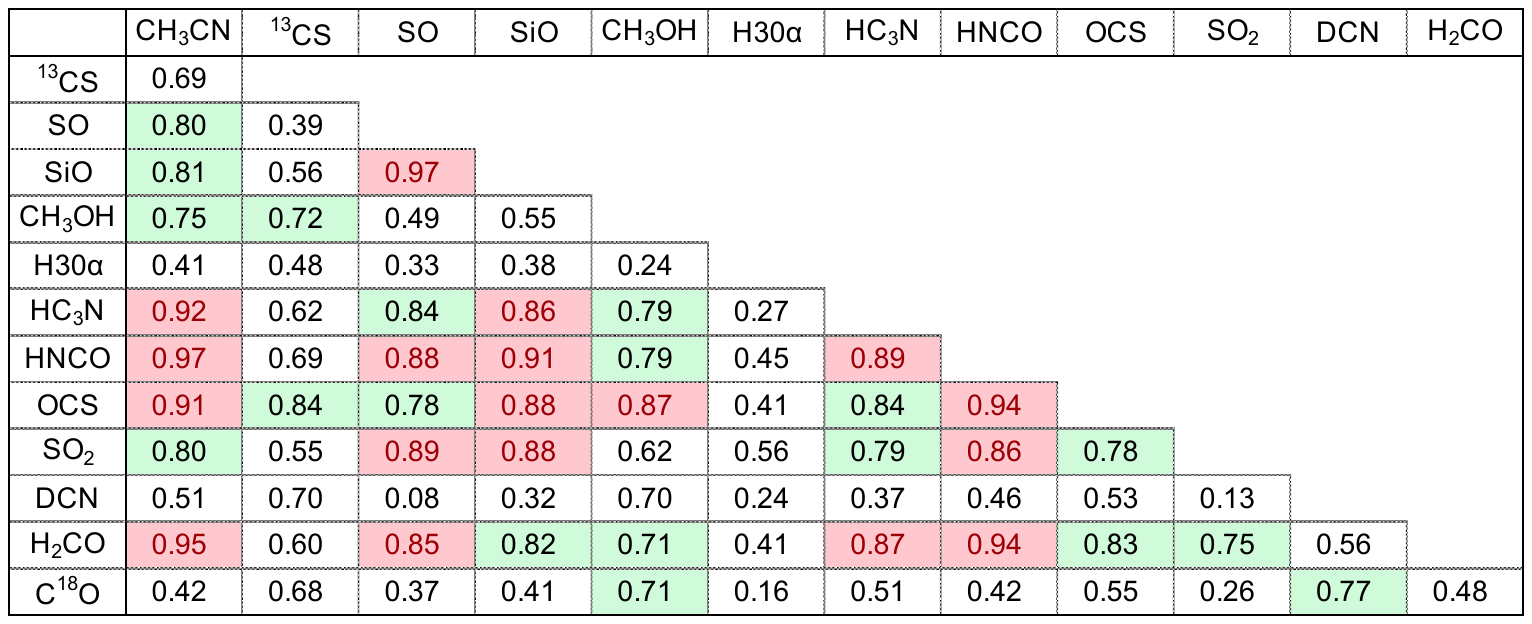}
\caption{Correlation coefficient matrix for line luminosities.
Correlations with $r \ge 0.85$ are shown in red, and those with $0.85 > r \ge 0.70$ are shown in green.
Note that the $\pm1\sigma$ ranges of the correlation coefficients for values of 0.9, 0.8, and 0.7 are 0.82--0.94, 0.66--0.89, and 0.51--0.82, respectively.
{Alt text: Correlation-coefficient matrix for the set of line luminosities. Each cell reports the correlation between two lines; cells with $r \ge 0.85$ are highlighted in red and those with $0.85 > r \ge 0.70$ are highlighted in green to emphasize strong and moderate correlations.}
}
\label{Correlation Matrix}
\end{figure}

We calculated correlation coefficients for all pairs of the analyzed
molecular and recombination line luminosities.
Figure~\ref{Correlation Matrix} presents these coefficients in matrix
form, allowing a clear visualization of the correlations among the line
luminosities.
It shows that some line luminosities are well correlated, whereas others
are not.
Although the central stellar luminosity or the total gas column density
is clearly an important factor contributing to these correlations, the
presence of both strong and weak correlations provides insight into the
chemical properties of the emitting regions.
\begin{enumerate}
\item The line luminosities of CH$_{3}$CN, HC$_{3}$N, HNCO, OCS, and H$_{2}$CO are strongly correlated with each other ($r \ge 0.85$).
\item The CH$_{3}$CN luminosity is moderately correlated ($0.70 \le r < 0.85$) with those of SO, SiO, CH$_{3}$OH, and SO$_{2}$.
\item The line luminosities of SO and SiO show a strong correlation ($r = 0.97$).
\item The line luminosities of $^{13}$CS, DCN, and C$^{18}$O are generally weakly correlated ($r < 0.70$) with other line luminosities.
\item The H30$\alpha$ luminosity shows weak correlation with all molecular line luminosities.
\end{enumerate}

%:%%% Figgure 10: Correlations Matrix %%%
% Figgure 10: Correlations Matrix %%%

Based on these line luminosity correlations, we classify the analyzed molecular lines into the following categories:
\begin{description}
\item A (Hot and dense gas): CH$_3$CN, HC$_3$CN, HNCO, OCS, and H$_2$CO 
\item B (Outflow gas): SiO and SO 
\item C (Envelope gas): SO$_2$ and CH$_3$OH
\item D (Extended gas): $^{13}$CS, DCN, and  C$^{18}$O
\item E (Ionized gas): H30$\alpha$ (or COMs for low luminosity cases)
\end{description}

The molecular emissions in Category~A are characteristic of an HMC, a dense ($n({\rm H}_2) \ga 10^6~{\rm cm}^{-3}$) and hot ($T \ga 100~{\rm K}$) region around a high-mass mprotostellar object (HMPO).
We therefore refer to them as ``hot and dense gas.''
A previous study of the specific object HMC141+624 (MCN-a) showed that these lines are emitted from a possibly infalling torus of radius $\sim 7000~{\rm au}$ around a massive protostar with an accretion disk \citep{Miyawaki2022b}.

Category~B molecules are well-known outflow tracers; they are released from dust grains in shocked regions of outflows or in boundary layers impacted by them. We refer to this component as “outflow gas.”

The line luminosities in Category~C are correlated with those of Categories~A and~B, but not as consistently as the correlations within each of the latter two categories. 
These emissions may arise from the outer region of an HMC, although in the specific case of HMC141+624 no significant difference in radius was reported \citep{Miyawaki2022b}. 
We refer to them as ``envelope gas.''

The line luminosities of Categories~D and~E show only weak correlations with other line luminosities, as well as within each category.
The emission regions of Category~D are often extended, tracing the region from the envelope of an HMC to more extended molecular gas.
We refer to them as ``extended gas.''

Category~E, H30$\alpha$, represents ionized gas primarily associated with HCHII/UCHII regions located at or near the centers of HMCs.
In the latter case, the H30$\alpha$ emission peak does not coincide with the CH$_3$CN peak, and part of the HMC is ionized by the adjacent HII region.
Emission from complex organic molecules (COMs) within the HMC may sometimes contaminate the recombination line at its frequency, as discussed in \S\ref{Line emission}, but the relatively low level ofcorrelation between H30$\alpha$ and other luminosities suggests that the contamination has only a minor effect.

\begin{figure}[htbp]
\includegraphics[bb= 0 0 500 580, scale=0.3]{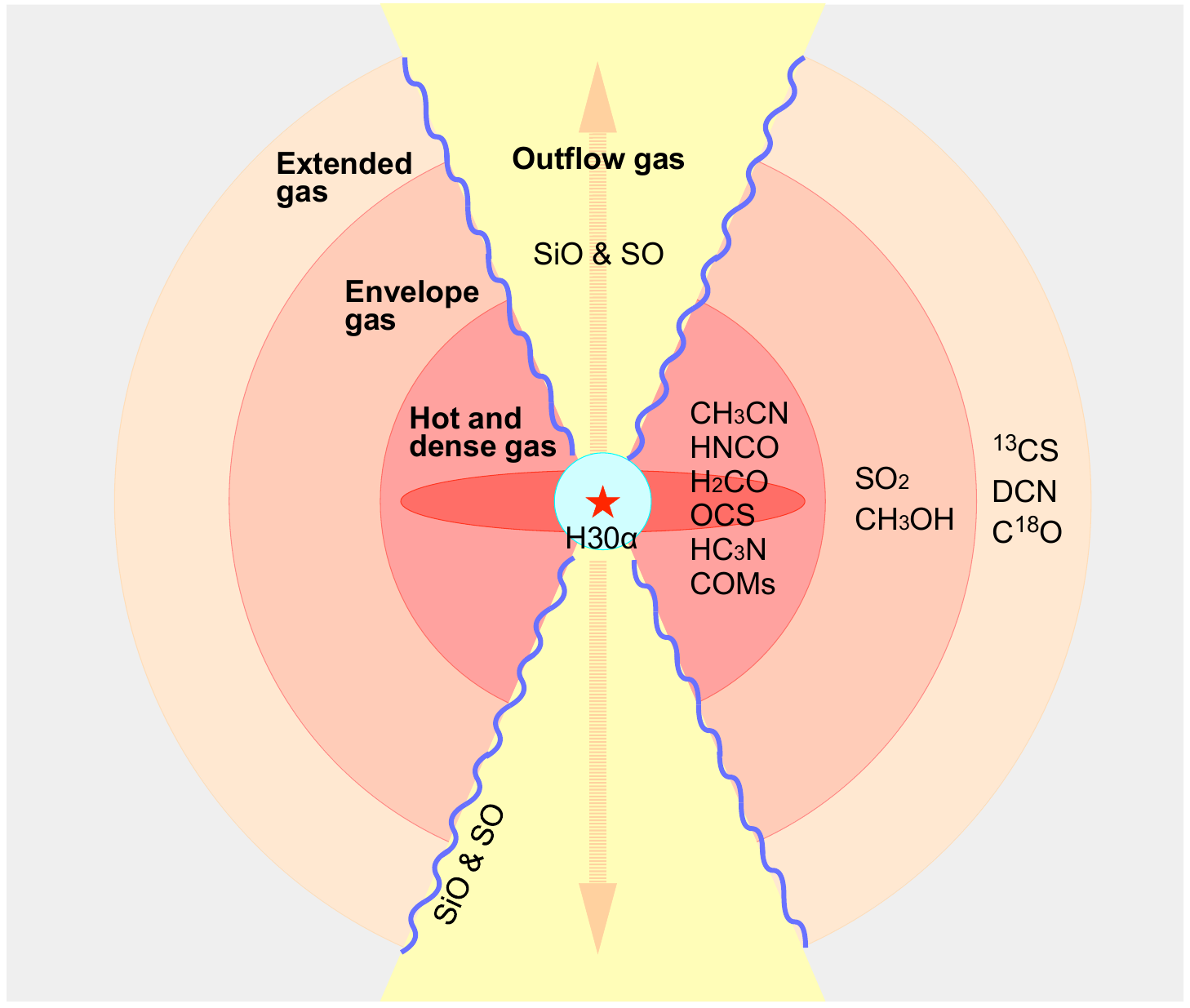}
\caption{Classification of the observed lines and their emitting regions.
{Alt text: Diagram classifying the observed emission lines into groups and linking each group to a characteristic emitting region, such as hot-core gas, outflow/shocked gas, envelope gas, more extended molecular gas, and ionized gas traced by H30$\alpha$.}
}
\label{Fig:schematic_fig}
\end{figure}

Figure~\ref{Fig:schematic_fig} illustrates our classification of the observed emission lines and their emitting regions.
The diagram is broadly consistent with previous molecular line studies of HMCs \citep[e.g.,][]{Shimonishi2021}.
Correlations in the emission line intensities across the HMCs show that the variation in emission regions around an HMC, depending on the molecular species, is a general phenomenon.
This can be interpreted, as previously suggested, as resulting from differences in the physical and chemical conditions of the emission regions for each line.
This study reveals that the categorized regions of molecular emissions are universally present within an HMC, regardless of the central star's luminosity or evolutionary stage.

%:%%% Figure 11: Schematic illustration of emission line regions %%%
% Figure 11: Schematic illustration of emission line regions %%%

\subsection{Origin of the 226 GHz continuum}\label{Origin of 226 GHz continuum}

Following the discussion of individual HMCs presented in Appendix~\ref{Individual HMCs}, we summarize in Table~\ref{Table: 226 GHz likelihood} the likelihood that the 1.3~mm continuum emission at the listed CH$_3$CN peaks is dominated by thermal dust emission from the hot cores.
The last column 7 presents the estimated likelihood, while Columns~2--6 provide the basis for this assessment, as discussed in the preceding section, Appendix~\ref{Individual HMCs}, and the following paragraphs.

%:%%% Table 7: Likelihood of 230 GHz by dust emission %%%
\begin{table*}[htbp]
\caption{Likelihood of 1.3~mm continuum dominated by dust emission}\label{Table: 226 GHz likelihood}
\begin{center}
\scalebox{0.8}[0.8]
{
\begin{tabular}{ccccccc}
\hline\hline\noalign{\vspace{1pt}}
Name &CH$_{3}$CN-1.3~mm& 1.3~mm &CH$_{3}$CN-3.6~cm&1.3~mm-H30$\alpha$&Steeply &Dust dominated\\
&Peak&Excess at&Peak&Peak&Rising SED$^{3)}$&226 GHz\\
&Coincidence&CH$_{3}$CN Peak&Coincidence$^{1)}$&Coincidence$^{2)}$&($\alpha>2$)&Continuum\\
{[}1] & [2] & [3] & [4] & [5] & [6] & [7] \\
\noalign{\vspace{1pt}}\hline\noalign{\vspace{1pt}}

HMC109+517 &Yes (Group 1) &--&No&Yes&No?&Probable\\
HMC110+520 &No (Group 2) &No&No&Yes&No&No\\
HMC120+614 & Yes (Group 1) &--&No&No&Yes&Certain\\
HMC127+611 & Yes (Group 1) &--&No&Yes (COMs?)&No&Certain\\
HMC131+618 &No (Group 2) &No&No&Yes&No&No\\
HMC131+613 &No (Group 2) &Yes&No&Yes&Yes&Probable\\
HMC131+614 &No (Group 2) &Marginal&No&No&n/a&Probable\\
HMC132+612 &Yes (Group 1) &--&No&No&No&Probable\\
HMC133+619 &Yes (Group 1) &--&No&No&Yes&Certain\\
HMC133+616 &Yes (Group 1) &--&Yes&No&Yes&Certain\\
HMC134+610 &No (Group 2) &Probable&No&No&n/a &Probable\\
HMC134+612 &No (Group 2) &Probable&No&Yes&No&Probable\\
HMC136+649 &Yes (Group 1) &--&No&No&Yes&Certain\\
HMC141+624 &Yes (Group 1) &--&Yes&Yes (COMs)&Yes&Certain\\
HMC143+622 &Yes (Group 1) &--&No&Yes (COMs)&Yes&Certain\\
HMC146+625 &Yes (Group 1) &--&No&Yes (COMs)&n/a&Certain\\
HMC153+615 &No (Group 2) &Yes&No&Yes&Yes&Certain\\
HMC163+606 &No (Group 2) &Yes&No&Yes&No&?\\

\noalign{\vspace{1pt}}\hline
\end{tabular}
}
\end{center}
$^{1)}$ ``No'' suggests a dust origin for the 1.3~mm continuum.\\
$^{2)}$ ``No'' suggests a dust origin for the 1.3~mm continuum when the CH$_{3}$CN and 1.3~mm emission peaks coincide (Group 1). This column does not mean much for the HMC with its CH$_{3}$CN peak not coincident with the 1.3~mm continuum (Group 2).\\
$^{3)}$ ``Yes'' indicates that the 96-226GHz spectral index ($\alpha=\log(F_{226}/F_{93})/\log(226/93)$ ) is larger than 2, suggesting a dust origin for the 226GHz continuum.\\
\end{table*}
% Table 7: Likelihood of 230 GHz by dust emission %%%

First, we divided the HMCs into two groups: those in which the CH$_{3}$CN and 1.3~mm continuum peaks coincide (Group~1), and those in which the two peaks are misaligned or no corresponding 1.3~mm peak is present (Group~2).
Group~1, corresponding to the HMCs judged as ``good'' in the fourth column of Table~\ref{Table: HMC-Continuum}, contains 10 HMCs.
Group~2 comprises 8 HMCs.

Group~1 indicates that the CH$_3$CN emission and the corresponding 1.3~mm continuum are likely to originate from the same source, at least at the current angular resolution of ALMA.
In such cases, a large fraction of the 1.3~mm continuum can be attributed to thermal dust emission if the signatures of HII regions are weak or if the SED exhibits a steep rise ($\alpha>2$) toward 1.3~mm and higher frequencies (“Yes” in Column~6).

To evaluate the signatures of HII regions, we used the 3.6~cm continuum distribution and the H30$\alpha$ intensity.
Specifically, the 1.3~mm continuum is considered to be dominated by dust emission if the 3.6~cm continuum is absent or its peak is misaligned with the HMC peak (“No” in Column~4), and if the H30$\alpha$ intensity is weak or lines of COMs are seen instead (“No” or “Yes (COMs)” in Column~5).
Among the 10 Group~1 HMCs, seven meet all these criteria and are therefore classified as “certain” cases of thermal dust emission.
In the case of HMC141+624 (MCN-a), we also classify it as a “certain” case, as discussed in \citet{Miyawaki2022b}, even though its 3.6~cm peak (UCHII J1) coincides with the 1.3~mm peak.
For the remaining two HMCs, HMC109+517 can be classified as “probable,” while the status of HMC163+606 remains unknown (marked as “?”).
These classifications are consistent with those in Table~\ref{Table: 226-8.3 correspondence}.

The CH$_{3}$CN sources in Group 2 may be different from their nearby 1.3~mm continuum sources.
We therefore examined whether any excess 1.3~mm continuum emission is present directly at the CH$_{3}$CN peaks (Column~3 in Table~\ref{Table: 226 GHz likelihood}) by visually inspecting the maps, and found that three of the eight HMCs, HMC131+613, HMC153+615, and HMC163+606, show such excesses.
For these three HMCs, we applied the same criteria as for Group~1 regarding HII region signatures and SED characteristics to assess whether the excess 1.3~mm continuum is dominated by thermal dust emission.
As a result, HMC153+615 ($\alpha>2$) is judged as a “certain” case, HMC131+613 ($\alpha>2$) as a “probable” case, and HMC163+606 as “not the case.”

The cases of HMC134+610 and HMC134+612 are difficult to judge, as they lie in the complex region in or around UCHII~G.
Careful inspection of the 1.3~mm continuum map reveals signs of excess emission at their positions compared with the 3.6~cm continuum and H30$\alpha$ distributions, leading us to judge them as “probable” cases.
HMC131+614 is a similar case: the 1.3~mm continuum shows a marginal excess at the HMC position compared to the 3.6~cm and H30$\alpha$ distributions, and is therefore also judged as “Probable.”
For the remaining two sources, HMC110+520 and HMC131+618, there is no evidence that their 1.3~mm continuum is dominated by thermal dust emission; hence they are judged as “No.”

\subsection{Fractional abundance of CH$_{3}$CN}\label{Mass}

We estimate the fractional abundance of CH$_{3}$CN with respect to H$_{2}$ for the Group~1 HMCs, toward which the CH$_{3}$CN and 1.3~mm continuum peaks coincide.
In this case, the 1.3~mm continuum emission is dominated by thermal dust emission, and the peak specific intensity is proportional to the dust column density and hence to the molecular hydrogen column density. This allows a direct comparison with the CH$_{3}$CN column density to derive its fractional abundance at the peaks of these HMCs.

The molecular hydrogen column density $N(\mathrm{H}_2)$ is derived from the 1.3~mm specific intensity $I_\nu$ using the following formula: $N(\mathrm{H}_2)=C_\nu I_\nu/[2\pi ab \, m_\mathrm{H} \, \mu \, B_\nu(T)]$ \citep{Hildebrand1983}\footnote{\citet{Hildebrand1983} used the total hydrogen column density, $N(\mathrm{H}+\mathrm{H}_2)$, in this formula. Because $N(\mathrm{H}_2)$ in this paper refers to the molecular hydrogen column density, we divide their original formula by a factor of 2.}.
Here, $C_\nu$ is the conversion factor inversely proportional to the dust optical depth, and $a$ and $b$ are the semi-major and semi-minor axes of the beam, respectively. $m_\mathrm{H}$ and $\mu$ denote the hydrogen atom mass and the gas-to-hydrogen mass ratio ($=1.36$), respectively. We adopt the rotation temperature derived from CH$_{3}$CN ($J_K=12_K-11_K$) (see Table~\ref{Table: Rotation Temperature}) for the Planck function $B_\nu(T)$.

The column density of CH$_{3}$CN in the $J=12$ level is obtained by summing the column densities of the available $K$ components, derived from the velocity-integrated specific intensities at the peak position of each HMC. 
Assuming that all CH$_{3}$CN levels are thermalized at the rotation temperatures derived from the $J_K=12_K-11_K$ transitions (listed in Table~\ref{Table: Rotation Temperature}), we compute the total CH$_{3}$CN column density using the corresponding partition function.

The CH$_{3}$CN column density at the peak of each HMC is then divided by the corresponding molecular hydrogen column density to obtain the fractional abundance $X(\mathrm{CH_{3}CN})$, which is listed in Table~\ref{X(CH3CN)}.
Note that these column densities have an uncertainty of about a factor of three, arising from uncertainties in the value of $C_\nu$ measured at 400~$\mu$m \citep[$C_{400\,\mu m}=27$~g\,cm$^{-2}$,][]{Keene1982} and in the frequency dependence index $\beta$ used to extrapolate this value to 1.3~mm, i.e., $C_{1300\,\mu\mathrm{m}}=(1300/400)^\beta C_{400\,\mu\mathrm{m}}$, where we adopt $\beta=2$.
As discussed in \S2.1, \S2.2, and \S4.2, the uncertainty due to missing flux is negligible, and the flux calibration uncertainty of $\sim$10\% is also minor here.

%:%%% Table: X(CH3CN)
\begin{table}[htbp]
\caption{Fractional abundance of CH$_{3}$CN}
\label{X(CH3CN)}
\begin{center}
\scalebox{0.8}[0.8]
{\begin{tabular}{ccc}
\hline\hline\noalign{\vspace{1pt}}

Name & X(CH$_{3}$CN) & Mass (M$_\odot$)\\

\noalign{\vspace{1pt}}\hline\noalign{\vspace{1pt}}

HMC109+517 & 7.8$\times$10$^{-9}$&  415\\
HMC120+614 & 2.1$\times$10$^{-8}$&  498\\
HMC127+611 & 1.3$\times$10$^{-8}$& 832\\
HMC132+612 & 1.5$\times$10$^{-8}$&  504\\
HMC133+619 & 1.9$\times$10$^{-9}$& 1200\\
HMC133+616 & 1.5$\times$10$^{-8}$& 489\\
HMC136+649 & 4.4$\times$10$^{-8}$& 88\\
HMC141+624 & 2.7$\times$10$^{-7}$& 804\\
HMC143+622 & 9.9$\times$10$^{-9}$& 147\\
HMC146+625 & 4.2$\times$10$^{-8}$& 152\\

\noalign{\vspace{1pt}}\hline
\end{tabular}}
\end{center}
\end{table}

The derived fractional abundance of $\mathrm{CH_3CN}$ varies over three orders of magnitude, from $X(\mathrm{CH_3CN})=1.9\times10^{-10}$ to $2.7\times10^{-7}$.
This variation within a single massive star-forming region indicates that local physical conditions, particularly temperature, are the dominant factors controlling the fractional abundance.
The range of $X(\mathrm{CH_3CN})$ is consistent with previous studies \citep[e.g.,][]{Hernandez2014, Pols2018, He2021}.
\citet{Hernandez2014} estimated that the typical values of $X(\mathrm{CH_3CN})$ lie between 10$^{-9}$ and 10$^{-7}$, with relatively high values of (1.8$\times$10$^{-7}$  found toward an HMC in IRAS~17233-3606 , based on a survey of 17 HMCs in various Galactic massive star-forming regions.
\citet{Pols2018} find that $X(\mathrm{CH_3CN})$ in the hot core Sgr~B2(M) spans a wide range of $\sim10^{-11}$--$10^{-7}$, with significant spatial variation across the source.
\citet{He2021} report that $X(\mathrm{CH_3CN})$ in massive star-forming clumps is typically $\sim10^{-10}$--$10^{-8}$ and tends to decrease with increasing luminosity-to-mass ratio $L/M$, which they adopt as an indicator of evolutionary stage.

Table~\ref{X(CH3CN)} also includes the masses of the HMCs derived from the molecular hydrogen column density and their 1.3~mm FWHM sizes.
The masses range from 88~M$_\odot$ to 1200~M$_\odot$, with a mean value of 510~M$_\odot$.
The uncertainty in the mass estimates is also about a factor of three.

\begin{figure}[htbp]
\includegraphics[bb= 0 0 400 270, scale=0.7]{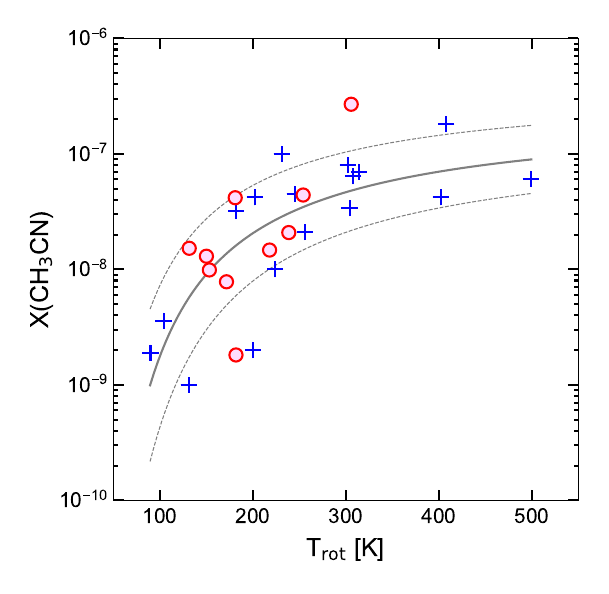}
\caption{The fractional abundance of CH$_{3}$CN with respect to H$_{2}$ is plotted against its rotation temperature, $T_\mathrm{rot}$. 
Red circular symbols represent the values for the HMCs in this study, while blue plus symbols indicate those taken from \citet{Hernandez2014}.
The solid curve represents the fitted relation, $X(\mathrm{CH_3CN})=X\mathrm{_\infty}\exp(-E/kT_\mathrm{rot})$, with the $\pm1\sigma$ ranges shown by broken lines.
Alt text: Scatter plot showing the fractional abundance of CH3CN relative to H2 as a function of rotational temperature for hot molecular cores. 
Red open circles represent sources from this study and blue plus symbols indicate data from Hern\'andez-Hern\'andez et al. (2014). 
The CH3CN abundance increases systematically with rotational temperature, suggesting enhanced production and desorption in warmer hot cores. 
The solid curve shows the fitted exponential relation, while dashed curves indicate the $\pm1\sigma$ uncertainty range.
}
\label{Abundance plot}
\end{figure}

The fractional abundances of the 10 HMCs are plotted against their rotation temperatures in Figure~\ref{Abundance plot}, together with those of the 17 sources studied by \citet{Hernandez2014}, for which rotation temperatures are available.
The data points from the current study agree well with those reported by \citet{Hernandez2014}.

%:%%% Figure 12: X(CH3CN) %%%
%%% Figure 12: Abundance %%%

We fitted the data points assuming the relation $X(\mathrm{CH_3CN})=X_\mathrm{\infty}\exp(-E/kT_\mathrm{rot})$ between the CH$_{3}$CN fractional abundance and the rotation temperature, where $X_\mathrm{\infty}$ is the fractional abundance in the high-temperature limit ($T_\mathrm{rot}\gg100~\mathrm{K}$), and $E$ denotes the characteristic energy of warm-up chemistry to produce gas-phase CH$_{3}$CN, including the desorption energy from evaporating grain mantles.
The fit resulted in the relation: 
\begin{equation}
X(\mathrm{CH_3CN})=(2.5^{+1.6}_{-1.0}\times10^{-7})\exp[-(490\pm90)/T_\mathrm{rot}],
\end{equation}
where $T_\mathrm{rot}$ is measured in K.
The characteristic temperature of 490~K corresponds to an energy of $E=0.042\pm0.008~\mathrm{eV}$, which is consistent with an effective energy scale governing the release of COMs from grain mantles, where thermal desorption is expected to dominate over non-thermal processes \citep[e.g.,][]{Minissale2022}.

The value of $X_\mathrm{\infty}=2.5\times10^{-7}$, with a systematic uncertainty of a factor of three, may represent the upper limit of the fractional abundance of CH$_{3}$CN in our Galaxy, considering that the relation is applicable not only to the W49A region but also to other Galactic massive star-forming regions.
\citet{Pols2018} obtained a value as large as $X(\mathrm{CH_3CN})=5\times10^{-7}$ in the southern part of Sgr~B2(M).
\citet{Hernandez2014} reported a value of $1.8\times10^{-7}$ toward IRAS~17233-3606.
In the current study, only HMC141+624 (MCN-a) shows a fractional abundance as large as this.

The fitted relation suggests that a fractional abundance as large as 10$^{-7}$ is achieved when the temperature exceeds $\sim$300~K.
Various chemical, physical, and evolutionary models have been proposed to explain such a high fractional abundance of $\mathrm{CH_3CN}$.

\citet{Gieser2019} used the {\tt MUSCLE} chemical evolution code to model the molecular emission of AFGL~2591 VLA3, showing that $X(\mathrm{CH_{3}CN})$ reaches $\sim10^{-7}$ within radii $r \lesssim 1000~\mathrm{au}$, where the molecular hydrogen density is $\gtrsim10^{7}\mathrm{cm}^{-3}$ and the temperature exceeds $\sim$100--150K, at a chemical age of $\sim4600$~yr in their best-fit hot-core model.
They attribute the high fractional abundance of CH$_3$CN ($\sim10^{-7}$) to the combined effects of thermal desorption of ice mantles at $T \gtrsim$100--150~K and subsequent warm gas-phase reactions that efficiently reform and sustain CH$_3$CN in the hot-core region.

Giani et al. (2023) show that a high fractional abundance of CH$_3$CN ($\sim10^{-7}$) can be achieved in dense ($n \gtrsim 10^{6}$--$10^{7}$~cm$^{-3}$) and warm ($T \gtrsim 100$~K) gas, where efficient, barrierless ion-molecule reactions, particularly involving CH$_3^+$ and HCN, rapidly form and sustain CH$_3$CN.
Garrod et al. (2021) show that high CH$_3$CN abundances ($\sim10^{-7}$) arise from the buildup of CH$_3$CN in grain mantles via radical chemistry in both surface and bulk ices during the warm-up phase, followed by thermal desorption and subsequent gas-phase processing in hot-core conditions.
Giani et al. (2023) argue that the high CH$_3$CN abundance can be produced predominantly by efficient gas-phase ion-molecule reactions, whereas Gieser et al. (2019) attribute it mainly to thermal desorption of ice mantles followed by gas-phase processing, and Garrod et al. (2021) propose a more comprehensive scenario in which CH$_3$CN is first built up in grain mantles via radical chemistry in surface and bulk ices during the warm-up phase, and subsequently released by thermal desorption and further processed in the gas phase.

These theoretical calculations suggest that a minimum temperature of 100--150~K is required for the fractional abundance of CH$_{3}$CN to reach $\sim10^{-7}$.
However, Equation~(1), together with the data points shown in Figure~\ref{Abundance plot}, indicates that the fractional abundance remains at $\lesssim10^{-8}$ within this temperature range, and that a higher abundance of $\sim10^{-7}$ or more is achieved only at temperatures exceeding $\sim200$~K.
While previous observational and chemical studies have generally shown that the abundance of CH$_3$CN increases with temperature due to warm-up chemistry and ice-mantle desorption, our exponential fit provides a simple empirical parametrization of this trend as a continuous function of rotational temperature; this expression also offers a quantitative empirical relation describing the increase of the CH$_3$CN abundance at higher temperatures.

\subsection{Implications for massive star formation}
\label{Evolutionary timescale}

\subsubsection{HMCs with and without embedded HII regions}

Table~\ref{Table:HMCs_summary} summarizes the characteristics of the 18 HMCs 
identified in this study.
The qualitative strengths of molecular line categories~A through D are also listed.
The classifications “Prominent~($\surd\surd\surd$),” “Evident~($\surd\surd$),” 
and “Subtle~($\surd$)” are primarily based on the line luminosities---greater 
than $10^{-3}$~L$_\odot$, between $10^{-4}$ and $10^{-3}$~L$_\odot$, and less 
than $10^{-4}$~L$_\odot$, respectively---and are supplemented by visual 
inspection of the spatial distributions of the molecular emission in the 
maps.
The presence or absence of H$_{2}$O, OH, and CH$_{3}$OH masers is also 
indicated.

%:%%% Table: 9 HMCs_summary %%%
\begin{table*}[ht]
\caption{Summary of HMC properties}
\begin{center}
\scalebox{0.8}[0.8]
{
\begin{tabular}{lccccccccccc}

\hline\hline
\hfil Name & HII (H30$\alpha$) & 1.3 mm & Hot and dense$^{2)}$ & Outflow$^{2)}$ & Envelope$^{2)}$ & Extended$^{2)}$ & Case \\
& or COMs$^{1)}$ & Thermal Dust & CH$_3$CN, etc.  & SiO \& SO & SO$_{2}$ \& CH$_{3}$OH & $^{13}$CS, DCN \& C$^{18}$O &\\
\hline
HMC109+517 & HCHII & Probable & $\surd\surd$ & $\surd\surd$ & $\surd\surd$ & $\surd\surd$ &ii\\
HMC110+520 & HCHII & No & $\surd$ & $\surd\surd$ & $\surd$ & $\surd\surd$ &ii\\
HMC120+614 & COMs & Certain & $\surd$ & $\surd\surd$ & $\surd$ & $\surd$ &i\\
HMC127+611 & HCHII+COMs & Certain & $\surd\surd$ & $\surd\surd$ & $\surd\surd$ & $\surd\surd$ &ii\\
HMC131+618 & HCHII & No & $\surd\surd$ & $\surd$ & $\surd$ & $\surd\surd$ &ii\\
HMC131+613 & HCHII & Probable & $\surd\surd\surd$ & $\surd\surd\surd$? & $\surd\surd\surd$ & $\surd\surd\surd$ &ii\\
HMC131+614 & Background HII? & Probable & $\surd\surd\surd$? & $\surd\surd\surd$? & $\surd\surd\surd$ & $\surd\surd$? &i\\
HMC132+612 & Background HII? & Probable & $\surd$ & $\surd\surd\surd$ & $\surd$ & $\surd$&i\\
HMC133+619 & COMs & Certain & $\surd$ & - & $\surd$ & $\surd\surd$ &i\\
HMC133+616 & COMs & Certain & $\surd\surd$ & ? & $\surd$ & $\surd\surd$ &i\\
HMC134+610 & COMs & Probable & $\surd\surd$ & ? & $\surd\surd$ & $\surd$ &i\\
HMC134+612 & ? & Probable & $\surd\surd\surd$ & $\surd\surd\surd$ & $\surd$ & $\surd$ &iv\\
HMC136+649 & COMs & Certain & $\surd\surd$ & $\surd\surd$ & $\surd\surd$ & $\surd\surd$ &i\\
HMC141+624 & COMs &  Certain & $\surd\surd$ & $\surd\surd\surd$ & $\surd\surd\surd$ & $\surd\surd\surd$ &i\\
HMC143+622 & COMs & Certain & $\surd\surd$ & $\surd\surd\surd$ & $\surd\surd$ & $\surd\surd$ &i\\
HMC146+625 & COMs & Certain & $\surd\surd$ & $\surd\surd$ & $\surd\surd$ & $\surd\surd$ &i\\
HMC153+615 & HCHII+COMs & Certain & $\surd\surd$ & $\surd\surd$ & $\surd\surd$ & $\surd\surd$ &ii\\
HMC163+606 & ? & ? & $\surd$ & $\surd\surd$ & $\surd$ & $\surd\surd$ &iv\\[1pt]

\hline

\end{tabular}
}
\end{center}
$^{1)}$ Complex organic molecules\\
$^{2)}$ Prominent ($\surd\surd\surd$), Evident ($\surd\surd$), Subtle ($\surd$), or Undeterminable (?)\\
% Certain -- Probable -- Marginal -- Impossible
$^{3)}$ Associated ($\surd$), or Possibly associated ($\surd$?) \\

\label{Table:HMCs_summary}
\end{table*}
% Table: 9 HMCs_summary %%%

Based on the discussion in the previous sections, the observed HMCs summarized in Table~\ref{Table:HMCs_summary} can be classified into the following four cases.
The latter two may also represent intermediate stages between the first two.

\begin{enumerate}
\renewcommand{\labelenumi}{\roman{enumi}.}

\item
HMCs in which the 1.3 mm continuum is dominated by thermal dust emission and no HCHII/UCHII region has yet developed at the center: HMC120+614, HMC131+614, HMC132+612, HMC133+619, HMC133+616, HMC134+610, HMC136+649, HMC141+624, HMC143+622, and HMC146+625.

\item
HMCs in which the 1.3 mm continuum is dominated by free-free emission and HCHII/UCHII regions have developed at the center: HMC109+517, HMC110+520, HMC131+618, and HMC131+613.

\item
HMCs in which both the free-free and thermal dust emissions may contribute to the 1.3 mm continuum: HMC127+611, and HMC153+615.

\item
HMCs for which neither of the above cases can be distinguished: HMC134+612, and HMC163+606. 
For these sources, nearby HCHII regions are so strong and close to the HMCs that it remains unclear whether the 1.3~mm continuum arises from free-free or thermal dust emission.

\end{enumerate}

Although not statistically significant, it is interesting to note that the number of HMCs without embedded HCHII/UCHII regions exceeds that of those with such regions.
This may suggest that the duration of the HMC phase without an embedded 
HCHII/UCHII region is at least comparable to, or possibly longer than, that of 
the phase with an embedded HCHII/UCHII region.
Once the HMPO begins to ionize its surroundings, it may rapidly 
($\sim$10$^{4}$~yr) ionize both the infalling gas and the disk/torus material 
of the HMC, leading to the Hollow Hot Core stage, accompanied by the 
development of an HCHII/UCHII region.
Thus, an HMC without an embedded HCHII/UCHII region may either not yet be 
massive enough to produce sufficient Lyc photons, or the formation of an HII 
region may be suppressed due to a very high accretion rate 
(\ga10$^{-3}$~M$_{\odot}$\,yr$^{-1}$).
In both cases, the timescale for the formation of an HCHII/UCHII region around 
the HMPO is expected to be comparable to the free-fall timescale of 
$\sim$10$^{5}$~yr, which is compared with the shorter timescale of the Hollow Hot Core of $\sim$10$^{4}$~yr \citep{Nony2024,Stephan2018}

\begin{figure}[htbp]
\includegraphics[bb= 80 120 600 440, scale=0.45]{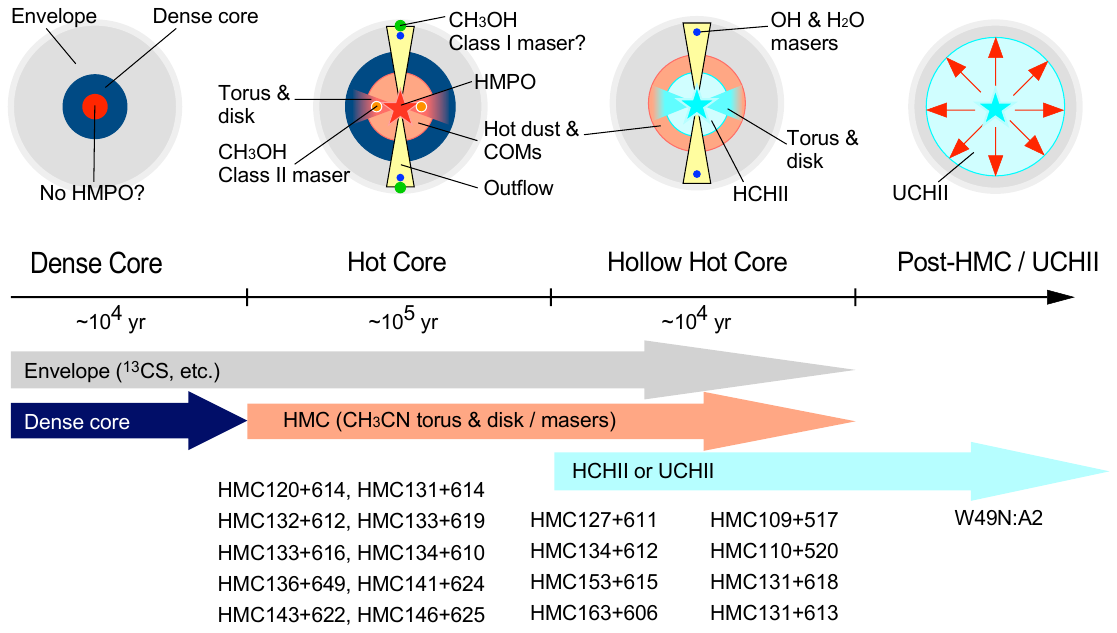}
\caption{A schematic figure showing the HMC evolution.
{Alt text: Schematic evolutionary sequence for hot molecular cores. The diagram illustrates four stages of massive star formation, from a cold dense core (with no hot-core molecular emission) to a hot-core phase with disks/outflows/masers, followed by development of an HCHII/UCHII region that creates a hollow hot core, and later ionized-region growth.}
}
\label{Fig:HMCevolution}
\end{figure}

This evolutionary sequence is schematically shown in 
Figure~\ref{Fig:HMCevolution}, where four stages of massive star formation are 
illustrated.
Our observations identify 10 HMCs in the Hot Core stage, four in the Hollow 
Hot Core stage, and the remaining four likely in a transitional phase between 
the Hot Core and Hollow Hot Core stages.

%:%%% Figure 13: A schematic figure showing the HMC evolution %%%
% Figure 13: A schematic figure showing the HMC evolution %%%

The “Dense Core Stage” denotes a cold, quiescent core with no emission from hot dust or hot core molecules, as represented by CH$_{3}$CN in this paper.
A dense dust core and/or a $^{13}$CS core exists at the center in this stage.

The “Hot Core Stage” represents a warm, chemically rich core exhibiting 
CH$_{3}$CN emission, together with outflow, maser, and torus/disk signatures.
The core is associated with an HMPO that accretes material through the disk 
and torus while driving an outflow.

The “Hollow Hot Core Stage” involves the development of an HCHII region that carves a cavity in the surrounding hot molecular core.
The core consists of either an HCHII or UCHII region, associated with an ionized disk and an outflow \citep[e.g.,][]{DePree2000, Tanaka2016}.
It has a density structure similar to that of the Hot Core but contains a central cavity ionized by an HCHII or UCHII region.
This is the final stage of the HMC \citep[e.g.,][]{Furuya2011, Rolffs2011, Serra2012, Fuente2018}, lying between the Hot Core and HII region phases \citep{Stephan2018}.

Finally, the “Post-HMC Stage” corresponds to the dispersal phase of the envelope and torus/disk as the UCHII region expands.
At this stage, the growth of the HCHII region destroys the disk/torus, and the surrounding envelope dissipates.
In Figure~\ref{Fig:HMCevolution}, we include W49A:N2, which exhibits an expanding ionized disk, as an example of this stage \citep{Miyawaki2023}.

\subsubsection{Implications for models of massive star formation}

As shown in the previous section, this study, based on high-resolution ALMA observations, provides further evidence for the existence of HMCs both with and without embedded HCHII/UCHII regions.
The latter appear to have lifetimes comparable to the free-fall timescale of $\sim$10$^{5}$~yr, which may be longer than that of HMCs with embedded HCHII/UCHII regions ($\sim$10$^{4}$~yr).
This difference in timescales provides an important constraint on models of massive star formation.

There are two major theoretical frameworks for massive star formation: the core accretion model and the competitive accretion model.
In the core accretion model, a gravitationally unstable massive molecular cloud core collapses to form a single (or a few) protostellar object(s), 
which subsequently grow through disk-mediated accretion \citep[e.g.,][]{McKee2002,McKee2003}.
In contrast, the competitive accretion model assumes that a cluster of initially similar-mass protostars forms, and that subsequent gas accretion within the gravitational potential of the cluster allows a few central objects to accumulate large masses \citep[e.g.,][]{Bonnell1997,Bonnell2001,Bonnell2004}.

The core accretion model is essentially a scaled-up version of low-mass star formation.
In this framework, a massive ($\gtrsim$10$^{3}$~M$_\odot$) turbulent core, 
characterized by a large effective sound speed, collapses at an accretion rate of $\sim$10$^{-3}$~M$_\odot$\,yr$^{-1}$ \citep{McKee2002,McKee2003}.
If $\sim$30\% of the accreting gas is deposited onto the stellar surface, a 30~M$_\odot$ star can form in $\sim$10$^{5}$~yr, comparable to the initial free-fall timescale.

In the competitive accretion model, stars initially form as low-mass ``seeds'' ($\sim$0.1~M$_\odot$) and subsequently grow by accreting gas drawn in by the gravitational potential of the entire star cluster 
\citep{Bonnell1997,Bonnell2001,Bonnell2004}.
Because the initial dense, turbulent gas clump fragments into many seed stars, the mass accretion rate of the most massive object is generally lower than in the core accretion scenario.
As a result, the formation of a massive star proceeds over a longer timescale, typically $\gtrsim$3$\times$10$^{5}$~yr.
Since these simulations often assume a 100\% accretion efficiency, the timescale may extend to $\sim$10$^{6}$~yr if a more realistic efficiency (e.g., $\sim$30\%) is adopted.

The observational result that the number of HMCs without embedded HII regions is larger than, or at least comparable to, that of HMCs with HII regions may initially appear to favor the competitive accretion model, which predicts a longer formation timescale.
However, the characteristic timescale of $\sim$10$^{6}$~yr in that model is significantly longer than the free-fall and Kelvin--Helmholtz timescales of $\sim$10$^{5}$~yr, which are also consistent with the duration of the HMC phase without an embedded HII region inferred in this study.

The duration of the HMC phase without an embedded HII region is expected to be comparable to the free-fall timescale ($t_\mathrm{ff}$), unless ionization is suppressed by a sufficiently high mass accretion rate 
($\dot{M}\gtrsim10^{-3}$~M$_\odot$\,yr$^{-1}$).
For an HMC to form, the luminosity of the central protostar (including both photospheric and accretion contributions) must exceed $\sim$10$^{4}$~L$_\odot$ \citep{Osorio1999}, corresponding to a stellar mass of $\sim$10~M$_\odot$ \citep{Nomura2004}.
An HII region is expected to develop once the stellar mass reaches $\sim$20~M$_\odot$, unless ionization is suppressed by ongoing high-rate accretion \citep{Keto2002, Keto2003, Hosokawa2010}.

In this context, HMCs without embedded HII regions can be interpreted as an evolutionary phase in which the central stellar mass lies between $\sim$10 and $\sim$20~M$_\odot$.
The growth across this mass range is expected to occur over approximately one free-fall timescale, $t_\mathrm{ff}$ \citep[e.g.,][]{Bonnell2004}.

Even when the stellar mass exceeds 20~M$_\odot$, the formation of an HII region may still be suppressed if the accretion rate remains sufficiently high ($\dot{M}\gtrsim10^{-3}$~M$_\odot$\,yr$^{-1}$; \citealt{McKee2002}).
Such conditions may be realized in some HMCs in W49A.
For example, \citet{Miyawaki2022b} suggested a mass accretion rate of order 10$^{-2}$~M$_\odot$\,yr$^{-1}$ for HMC141+624 (MCN-a).
However, at such high accretion rates, the stellar mass increases rapidly, and an additional $\sim$10~M$_\odot$ can be accreted in a timescale much shorter than the initial free-fall time ($\sim$10$^{5}$~yr), thereby shortening the duration of the HMC phase without an embedded HII region.

In summary, while the relative numbers of HMCs with and without HII regions may appear to support the competitive accretion scenario, the inferred timescales and physical conditions are also naturally explained within the core accretion framework.
The present results therefore do not uniquely distinguish between the two models, but instead suggest that both mechanisms may operate under different physical conditions.

%:SUMMARY
\section{Summary}\label{Summary}

We investigated the physical and chemical properties of hot molecular cores (HMCs) in W49A using high-resolution archival data from ALMA.
Six ALMA projects covering multiple frequency bands were analyzed, including molecular emission lines of CH$_{3}$CN, $^{13}$CS, SO, SiO, CH$_{3}$OH, HNCO, HC$_{3}$N, OCS, SO$_{2}$, H$_{2}$CO, DCN, and C$^{18}$O, as well as the H30$\alpha$ recombination line.
The data were processed and analyzed primarily using the CASA and CARTA software packages for both continuum and line components.

The main results and discussions are summarized as follows:

\begin{enumerate}

\item
We identified 20 continuum sources, including four marginal cases, on the 1.3~mm continuum map with a beam size of 0\farcs32$\times$0\farcs28.
Ten of the 20 sources coincide with previously identified 3.6~cm UCHII regions within the beam size.
Their SEDs suggest that the 1.3~mm continuum emission from most of them contains a significant contribution from thermal dust.

\item
We identified 18 HMCs in the CH$_{3}$CN ($J_K=12_3-11_3$) map and compared in detail the spatial distributions of the molecular, H30$\alpha$, and continuum emission.
Ten HMCs exhibit coincident peaks in the 1.3~mm continuum and CH$_{3}$CN emission within 0\farcs1, indicating that thermal dust emission is dominant, although some sources also show significant contributions from free-free emission.

\item
Rotation temperatures were estimated for HMCs based on CH$_{3}$CN transitions.
The average rotation temperatures are 176$\pm52$~K and 222$\pm25$~K for the $J=12_K-11_K$ and $J=24_K-23_K$ transitions, respectively.
No clear correlation was found between rotation temperature and either the HMC size or line luminosity.

\item
Statistical analysis of line luminosities provided a classification of emission lines into the following five categories.
\begin{description}
\item [\rm A (Hot and dense gas):] CH$_3$CN, HC$_3$CN, HNCO, OCS, and H$_2$CO 
\item [\rm B (Outflow gas):] SiO and SO 
\item [\rm C (Envelope gas):] SO$_2$ and CH$_3$OH
\item [\rm D (Extended gas):] $^{13}$CS, DCN, and  C$^{18}$O
\item [\rm E (Ionized gas):] H30$\alpha$ (or COMs for low luminosity cases)
\end{description}

\item
The above categories represent various regions around an HMC, revealing structural features such as disks/tori, molecular outflows, envelopes, and ionized regions.
Variations of these indicators among the HMCs may reflect different evolutionary stages.

\item
We derived the fractional abundance of CH$_{3}$CN with respect to H$_{2}$.
The abundance ranges from $\sim10^{-10}$ to $\sim10^{-7}$, indicating a 
strong dependence on local temperature.
The abundance-temperature relation is fitted by an exponential form, 
$X(\mathrm{CH_3CN})=2.5\times10^{-7}\exp[-490/T_\mathrm{rot}]$.
This result suggests that $T\gtrsim200$--300~K is required to achieve high 
abundances of $\sim10^{-7}$, whereas theoretical models based on thermal 
desorption from grain mantles and subsequent gas-phase chemistry typically 
assume temperatures of 100--150~K.

\item
The observed HMCs can be categorized into those with and without HCHII regions, with the lifetime of the latter ($\sim$10$^{5}$~yr) possibly longer than that of the former ($\sim$10$^{4}$~yr).
Although the longer lifetime of the former may appear to favor the competitive accretion model of massive star formation, the inferred timescale and physical conditions of the latter are also consistent with the core accretion model.

\end{enumerate}

In conclusion, CH$_{3}$CN is particularly useful for identifying HMCs, and its comparison with other molecular and recombination lines, as well as continuum emission, provides insight into the internal structures and evolutionary processes of HMCs, reflecting the diverse physical and chemical states associated with massive star formation.

%:ACKNOWLEDGMENT
\begin{ack} 
ALMA is a partnership of ESO (representing its member states), NSF (USA) and NINS (Japan), together with NRC (Canada), NSC and ASIAA (Taiwan), and KASI (Republic of Korea), in cooperation with the Republic of Chile. 
The Joint ALMA Observatory is operated by ESO, AUI/NRAO and NAOJ.
We used the ALMA archival data \#2015.1.01535.S : PI A.B\'aez-Rubio, \#2016.1.00620.S: PI A. Ginsburg, \#2017.1.00318.S: PI R. Galvan-Madrid,  \#2017.1.01499.S, PI: X. Fu, \#2018.1.00520.S PI: D. Wilner,  \#2018.1.00589.S PI: R. Galvan-Madrid,  and \#2018.1.00589.S PI: R. Galvan-Madrid. 
ChatGPT was used to improve the wording of this manuscript.
\end{ack}

%\clearpage
%:APPENDIX
%\FloatBarrier
\begin{appendix}\label{APPENDIX}
\renewcommand{\thefigure}{A\arabic{figure}}
\renewcommand{\thetable}{A\arabic{table}}
\setcounter{figure}{0}
\setcounter{table}{0}

%:Appendix 1
\section{Spectral Line Data Analysis}\label{Line data reduction}

\subsubsection{85 -- 101~GHz}

The spectral line data had a frequency resolution (channel width) of 1129~kHz, corresponding to a velocity resolution of $\sim$3.4~km\,s$^{-1}$.
A data cube for the H50$\beta$ line (99.22520843~GHz) was produced with a velocity resolution of 5~km\,s$^{-1}$.
The synthesized beam size of the line images was 1\farcs97 $\times$ 1\farcs65 (PA = $-64.5\degree$).
The resulting noise level was 8.6~mJy\,beam$^{-1}$ per velocity channel.

\subsubsection{217 -- 235~GHz}
\begin{figure}[htbp]
\includegraphics[bb= 130 300 400 600, scale=0.60]{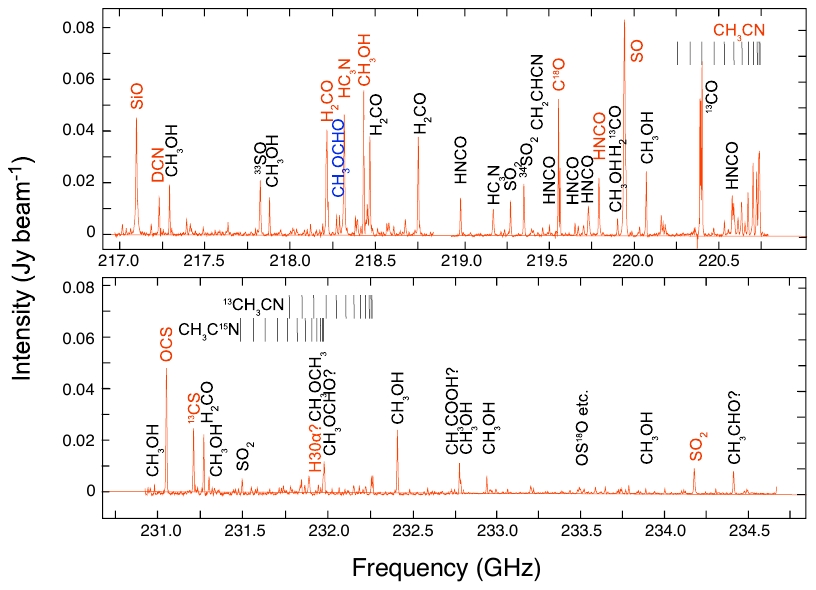}
\caption{Observed line spectra in the 217--235~GHz band toward HMC141+624 (MCN-a).
The specific intensity is averaged over a circular area with a diameter of 2$''$ centered on the CH$_{3}$CN peak.
This figure is reproduced from Fig.~2 of \citet{Miyawaki2022b}, with corrections applied to the identifications of the $^{13}$CS and SO$_2$ lines, which were incorrectly identified in that work.
The lines analyzed in this paper are highlighted in red, and the CH$_{3}$OCHO line used by \citet{Nony2024} to identify HMCs is highlighted in blue.
{Alt text: Spectrum plot of specific intensity versus frequency from 217 to 235 GHz toward HMC141+624 (MCN-a), averaged within a 2-arcsec-diameter aperture centered on the CH$_3$CN peak. Individual molecular lines are labeled; transitions analyzed in this paper are highlighted in red, and the CH$_{3}$OCHO line used by \citet{Nony2024}  is highlighted in blue.}}
\label{Fig: Observed lines}
\end{figure}

The spectral lines in this frequcny band are shown in Figure~\ref{Fig: Observed lines}.
We analyzed the emission lines of CH$_3$CN ($J_K=12_K-11_K$) and others listed in Table~\ref{Table: Observing parameters for emission lines}.
We set up a data cube with a spectral resolution of 2~kms$^{-1}$ for each line.
The original spectral resolution was 976.56~kHz, corresponding to the velocity resolution of $\sim$1.4 km\,s$^{-1}$.
 The spectral line data for this frequency band was previously described in a paper \citep{Miyawaki2022b} that focused on HMC MCN-a. 
 The synthesized beam sizes of the line images were 0\farcs35$\times$ 0\farcs34 (PA=$-87\degree$) and 0\farcs35 $\times$ 0\farcs29 (PA=$-68\degree$) at 219\,GHz and 233\,GHz, respectively.
The resultant noise levels were 4.3\,mJy\,beam$^{-1}$.
The CH$_3$CN ($J_3=12_3-11_3$) line was used to identify HMCs in \S\ref{Hot Molecular Cores}.
The CH$_3$CN ($J_K=12_K-11_K$) lines were used to derive the rotation temperatures of HMCs in \S\ref{Rotation temperature}.

%:Figure A01 %%% Observed Line Spectra %%%
%Figure A01

\subsubsection{241 -- 270~GHz}

We analyzed the spectral lines of H29$\alpha$ (256.30203519~GHz) and CH$_3$CN ($J_K=14_K-13_K$; $K$ = 0-13) in the 250~GHz band.
The synthesized beam size at the H29$\alpha$ line frequency was 0\farcs052 $\times$ 0\farcs033 (PA = $58.0\degree$), and the resulting noise level of the line maps was 2.0~mJy\,beam$^{-1}$.

The H29$\alpha$ data were also used to examine whether the emission line observed at the H30$\alpha$ frequency (231.90092784~GHz) traces ionized gas. This check was necessary because the H30$\alpha$ line could be blended with lines of complex organic molecules (COMs) \citep{Miyawaki2022b}.
The H29$\alpha$ map showed bright UCHII regions that were also visible in the H30$\alpha$ map, indicating that the H30$\alpha$ line is not significantly affected by COM contamination, at least toward the bright UCHII regions.

\subsubsection{342 -- 368~GHz}

We analyzed the H26$\alpha$ (353.62274716~GHz) and H32$\beta$ (366.65253232~GHz) lines from the data obtained by B\'aez-Rubio, A. (\#2015.1.01535.S).
These data cover only the central region of W49N, with a diameter of 16$''$.
Data cubes were produced with a spectral resolution of 2~km\,s$^{-1}$.
The synthesized beam size at 353~GHz was 0\farcs18$\times$0\farcs16 (PA = $-45.8\degree$), and the resulting noise level was 79~mJy\,beam$^{-1}$.
We also examined the spectral lines of CH$_3$CN ($J_K=20_K-19_K$; $K$ = 7-17), but the signal levels were insufficient to determine the rotation temperature.

\subsubsection{441 -- 456~GHz}

We obtained the spectral lines of CH$_3$CN ($J_K=24_K-23_K$; $K$ = 0-9).
Data cubes were produced with a spectral resolution of 2~km\,s$^{-1}$.
The synthesized beam size of the line images at 441~GHz was 0\farcs69$\times$0\farcs56 (PA = $71.8\degree$), and the resulting noise level was 19~mJy\,beam$^{-1}$.
The signal levels of these lines were sufficient to estimate the rotation temperature.

%:Appendix 2
\section{Comparison with previously identified sources}
\label{Nony sources}

\citet{Nony2024} recently identified 129 continuum cores, including 19 HMCs, using the same dataset (\#2016.1.00620.S) employed in this study. 
In contrast, we identified only 20 sources in the 1.3~mm continuum emission---significantly fewer than reported in their study. 
Regarding the HMCs, however, we identified 18 (see \S\ref{Identification Hot Molecular Cores}), a number comparable to that reported by \citet{Nony2024}. 
In the following, we discuss the origin of this discrepancy in the number of 1.3~mm continuum sources.

\begin{figure*}[htbp]
\includegraphics[bb= 80 130 400 480, scale=0.75]{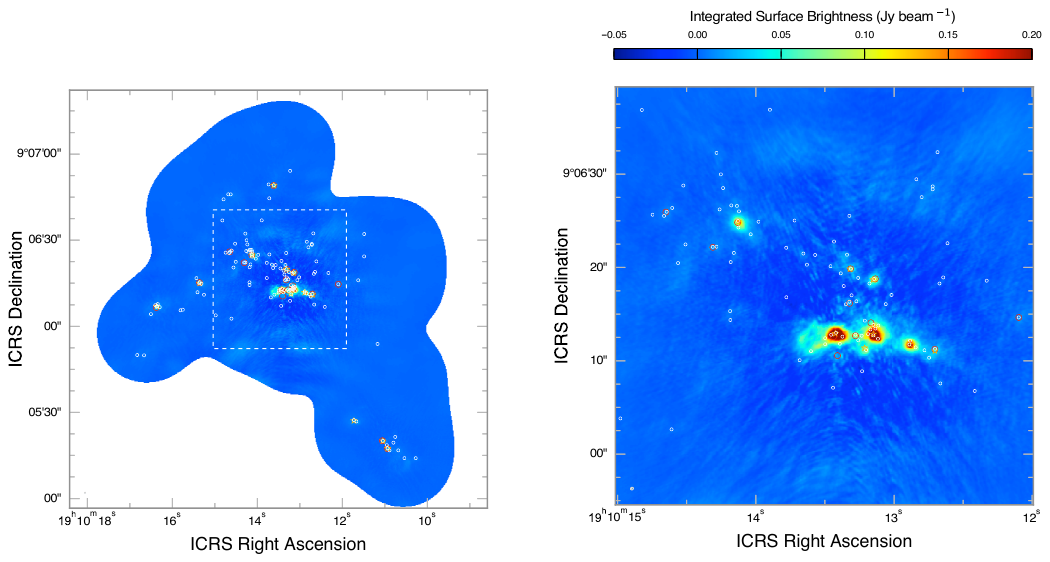}
\caption{Comparison of the 1.3~mm continuum sources (yellow stars) and HMCs (red open circles) with the cores (white open circles) identified by \citet{Nony2024} at the same frequency. The left panel shows the entire region, while the right panel provides a zoomed-in view of the center, corresponding to the area enclosed by the dashed white rectangle in the left panel.
{Alt text: Two-panel sky map comparing source positions. Yellow stars mark 1.3~mm continuum sources, red open circles mark HMCs, and white open circles mark the ‘core’ positions from \citet{Nony2024} . The left panel shows the full observed region; the right panel is a zoom into the central area outlined by a dashed white rectangle in the left panel.}
}
\label{Fig: 226G Comparison}
\end{figure*}

The main difference in the identification of continuum sources between our study and \citet{Nony2024} lies in the selection method: we adopted a threshold of 20~mJy\,beam$^{-1}$, whereas they employed the {\tt getsf} algorithm \citep{Menshchikov2021}.
This methodological difference has two implications.
First, they reported 75 cores with peak brightnesses below 20~mJy\,beam$^{-1}$.
Second, they resolved multiple cores within what we classified as a single source (source confusion).
This difference is illustrated in Figure~\ref{Fig: 226G Comparison}, which compares the 1.3~mm continuum sources and HMCs identified in this study with the cores reported by \citet{Nony2024} at the same frequency.

%:%%% Figure A02:Comparison%%%
% Figure A02

Because of our threshold of 20~mJy\,beam$^{-1}$, all our 20 sources are listed as cores by \citet{Nony2024}.
In this sense, the number of 1.3~mm continuum sources we listed represents a lower limit, and weaker sources are naturally present.
Many of the weaker sources ($\la$20~mJy\,beam$^{-1}$) identified by \citet{Nony2024}, however, have no corresponding prominent peaks on our map and appear much weaker than the spurious features.

Source confusion is another factor that can lead to an underestimation of the number of sources.
There are instances in which \citet{Nony2024} identified multiple cores corresponding to a single 1.3~mm continuum source or HMC listed in this study.
For example, to the north of C131+612 (UCHII B, N2) and around HMC131+613, \citet{Nony2024} detected three cores (N83, N32, and N56, in order of increasing RA).
It is essentially impossible to isolate each of these sources using a threshold method for identification; in such cases, the source size becomes larger---both C131+612 and HMC131+613 have FWHM sizes of 1\farcs5--2$''$.
As we will discuss in \S\ref{HMC131+613}, the HMC physically associated with C131+612 exhibits a shell-like morphology, and algorithmic methods may tend to identify multiple sources in such configurations.

The 1.3~mm sources identified in the current study are on average 2.8 times larger than those reported by \citet{Nony2024}; the average size in this work is 1\farcs13$\pm$0\farcs42, compared to 0\farcs41$\pm$0\farcs15 in their study.
As a result, our 1.3~mm flux densities are typically $\sim$10 times higher than those reported by \citet{Nony2024}.
This discrepancy primarily arises from differences in the source identification methods: setting a fixed threshold for source extraction tends to treat a strong emission feature as a single source with a large flux density, whereas algorithmic extraction tends to divide such a feature into multiple cores, even when the substructures correspond to relatively minor surface brightness variations.
The discrepancy may also be partly attributed to the use of both the 12~m and 7~m ALMA array data in our analysis \citep{Miyawaki2022b}, whereas \citet{Nony2024} used only the 12m array data.
The inclusion of the 7~m data improves sensitivity to extended emission, generally increasing both the apparent size and flux density of the sources.

In contrast to the number of continuum sources, the number of HMCs identified in both studies is nearly the same, although there are differences in the specific HMCs identified: 13 sources are classified as HMCs by both studies, while six HMCs are identified only by \citet{Nony2024}, and five are identified only in the current work.
Seventeen out of the 18 HMCs identified in this study have corresponding 1.3~mm cores detected by \citet{Nony2024}; one source, HMC134+610 (see Appendix~\ref{HMC134+610}), lacks associated continuum emission in both studies.

One reason why both studies identified a similar number of HMCs may also lie in the methodology: \citet{Nony2024} applied the threshold method employed by \citet{Brouillet2022} to determine whether a continuum core is an HMC.
There are likely weaker HMCs that remain undetected due to this threshold method, as is the case for continuum sources.
In this sense, it would be more consistent to apply the same detection method to both continuum and molecular line sources.
As demonstrated by \citet{Nony2024}, the actual number of 1.3~mm sources is larger than that identified in the present study.
Likewise, the number of HMCs would likely increase if the same algorithmic method were applied to the molecular line data.
Therefore, caution must be exercised when comparing the number of identified sources across studies that employ different methodologies.

%: Appendix 3
\section{Individual Hot Molecular Cores}
\label{Individual HMCs} 

We discuss the details of each HMC in this subsection, presenting local maps of the CH$_3$CN, $^{13}$CS, SO, SiO, CH$_3$OH, and H30$\alpha$ lines.
In addition, integrated intensity (moment~0) maps of HNCO ($J_{K_a,K_c}=10_{0,10}-9_{0,9}$), HC$_3$N ($J=24-23$), OCS ($J=19-18$), H$_2$CO ($J_{K_a,K_c}=3_{2,2}-2_{2,1}$), DCN ($J=3-2$), and C$^{18}$O ($J=2-1$) are presented in Figures~\ref{Fig:Other_mol_lines_map01}--\ref{Fig:Other_mol_lines_map02} in the Appendix.

\begin{figure}[htbp]
\includegraphics[bb= 0 0 500 810, scale=0.4]{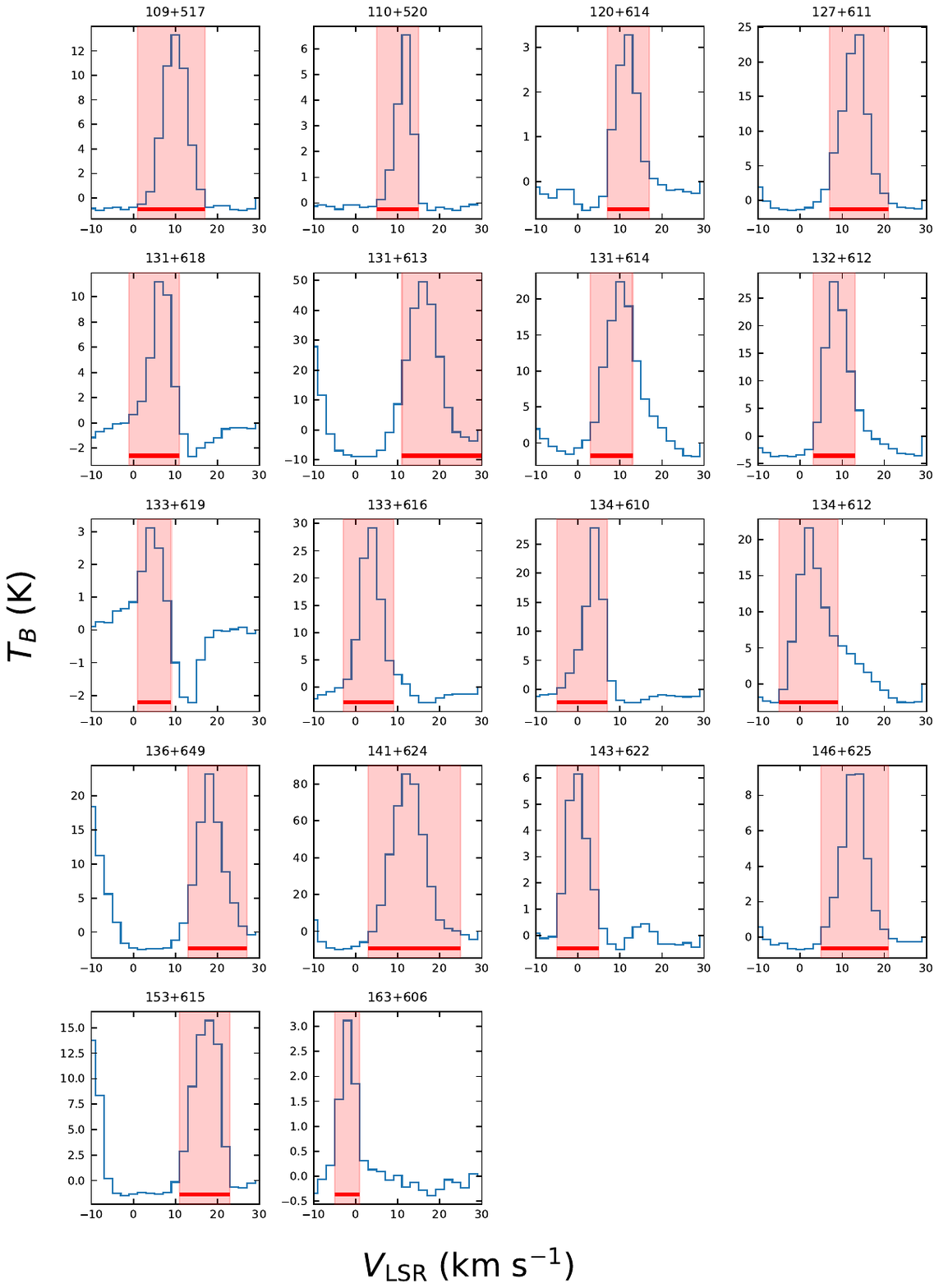}
\caption{CH$_{3}$CN ($J_K=12_3-11_3$) line profiles toward the peak positions of the identified HMCs. 
The shaded regions indicate the velocity ranges used to calculate the integrated intensities of molecular lines.
Alt text: Multi-panel spectra of the CH$_3$CN (J$_K$ = 12$_3$-11$_3$) transition toward the peak positions of the identified hot molecular cores in W49A. 
Each panel shows brightness temperature as a function of LSR velocity for an individual source. 
The shaded regions indicate the velocity ranges used to calculate integrated intensities. 
Most sources exhibit strong emission near the systemic velocity, while the line widths and profile shapes vary among the hot cores.
}
\label{Fig: CH3CN line profiles}
\end{figure}

\begin{figure}[htbp]
\includegraphics[bb= 0 0 500 800, scale=0.4]{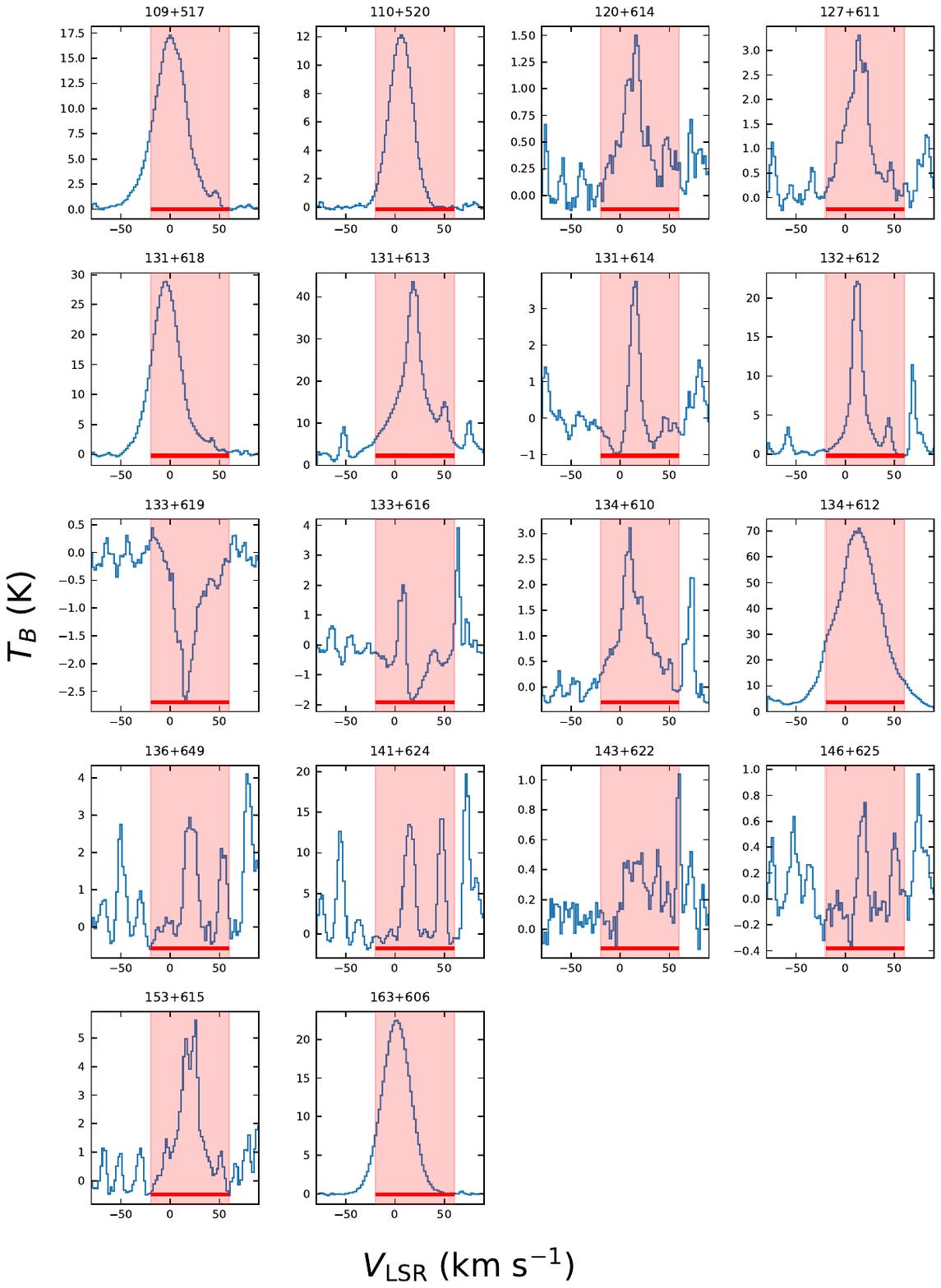}
\caption{H30$\alpha$ line profiles toward the peak positions of the identified HMCs. 
The shaded regions indicate the velocity ranges used to calculate the integrated intensities.
Alt text: Multi-panel H30$\alpha$ recombination-line spectra toward the identified hot molecular cores in W49A. 
Each panel shows brightness temperature as a function of LSR velocity for an individual source. 
The shaded regions indicate the velocity ranges used to calculate integrated intensities. 
Several sources exhibit strong and broad recombination-line emission associated with embedded HII regions, while other sources show weak or undetected H30$\alpha$ emission.
}
\label{Fig: H30alpha line profiles}
\end{figure}

Figures~\ref{Fig109+517}--\ref{Fig131+618} and Figures~\ref{Fig131+613}--\ref{Fig163+606} display local maps for the 18 HMCs. 
The molecular line maps show intensities integrated over the velocity ranges indicated in Figure~\ref{Fig: CH3CN line profiles}, and the H30$\alpha$ maps show intensities integrated over the velocity ranges shown in Figure~\ref{Fig: H30alpha line profiles}.
In these figures, we superimposed markers and contours, if their information is available, as follows. 
A blue circle with a white border shows the local CH$_3$CN peak location from Table~\ref{Table: HMC candidates}.
The thick white dashed contours represent the 5$\sigma$ level of the integrated intensity for each panel.
The magenta contours represent the 226 GHz continuum at 5, 10, 20, 40, 60, and 80\% of the maximum brightness (0.95 Jy\,beam$^{-1}$) over the entire area.
A blue star with a white border indicates the local peak position of the 1.3~mm continuum taken from Table~\ref{Table: 226 GHz continuum sources}.
The white contours show the 3.6~cm continuum emission \citep{DePree1997}.
The red and blue dots represent the OH \citep{Deshpande2013, Zhang2019a, Mendoza2023} and H$_2$O \citep{Sarma2002, McGrath2004, Zhang2013} maser spots\footnote{Spot positions for H$_{2}$O, OH, and CH$_{3}$OH masers were retrieved from the MaserDB database (https://maserdb.net) in May 2022. It should be noted that these positions are time-variable.}, respectively.
The yellow squares mark the CH$_3$OH class~I maser spots \citep{Barrett1971, Haschick1990, Valtts1995, Liechti1996, Rollig1999, Kalenskii2001, Gan2013, Yang2017}, and the yellow asterisks are the CH$_3$OH class~II maser spots \citep{Menten1991, Caswell1995, Slysh1999, Szymczak2000, Malyshev2003, Blaszkiewicz2004, Green2011, Pandian2011, Breen2015, Hu2016, Yonekura2016, Nguyen2022}

%:%%% Figure A03: CH3CN Line Profiles %%%

%:%%% Figure A04:H30 alpha Line Profiles %%%
%%% Figure A04:H30 alpha Line Profiles %%%

\begin{figure}[htbp]
\includegraphics[bb= 80 0 500 600, scale=0.4]{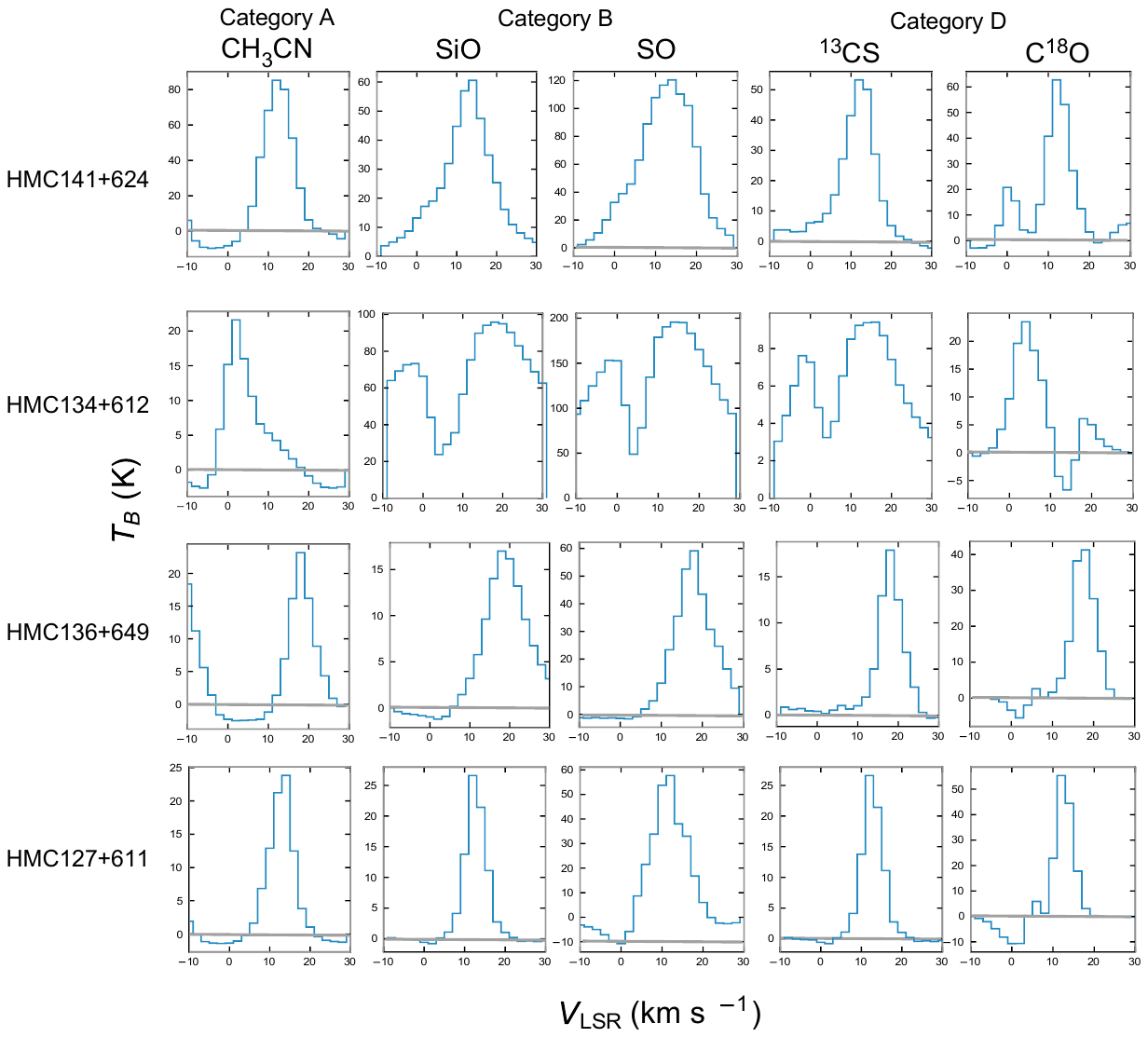}
\caption{Representative molecular-line spectra toward four hot molecular cores (HMC141+624, HMC134+612, HMC136+649, and HMC127+611).
From left to right, the spectra of the Category A hot-core tracer CH$_3$CN, the Category B shock/outflow tracers SiO and SO, and the Category D extended-gas tracers $^{13}$CS and C$^{18}$O are shown on the same $V_{\rm LSR}$ scale. 
The horizontal gray line indicates $T_{\rm B}=0$ K, and the shaded regions mark the velocity ranges used to derive the integrated intensities.
Alt text: Comparison of representative molecular-line spectra toward four hot molecular cores in W49A. 
Rows correspond to HMC141+624, HMC134+612, HMC136+649, and HMC127+611, while columns show the Category A hot-core tracer CH3CN, the Category B shock/outflow tracers
SiO and SO, and the Category D extended-gas tracers 13CS and C18O.
All spectra are displayed on the same LSR velocity scale with the zero brightness-temperature level indicated by a horizontal gray line. 
Shaded regions mark the velocity ranges used for the integrated-intensity measurements. 
SiO exhibits broader high-velocity wings than the other molecular tracers in several sources.
}
\label{Fig:line profiles}
\end{figure}

To illustrate the quality of the spectral data,
Figure~\ref{Fig:line profiles} presents representative line profiles of the Category A compact hot-core tracer (CH$_3$CN), the Category B shock/outflow tracers (SiO and SO), and the Category D extended-gas tracers ($^{13}$CS and C$^{18}$O) toward four representative HMCs.
As indicated by the horizontal zero-level line in Figure~\ref{Fig:line profiles}, all spectra exhibit stable baselines without pronounced negative bowls or artificial absorption features, suggesting that interferometric spatial filtering does not significantly affect the core-scale line measurements.
The SiO spectra frequently exhibit broad high-velocity wings that are substantially wider than those of CH$_3$CN, $^{13}$CS, and C$^{18}$O, supporting their interpretation as localized shocked gas associated with protostellar outflows rather than large-scale expanding shells.

%:%%% Figure A05:Oher Line Profiles %%%
%%% Figure A05:H30 alpha Line Profiles %%%

\subsection{HMC109+517}\label{HMC109+517}

\begin{figure}[htbp]
\includegraphics[bb= 70 100 500 540, scale=0.4]{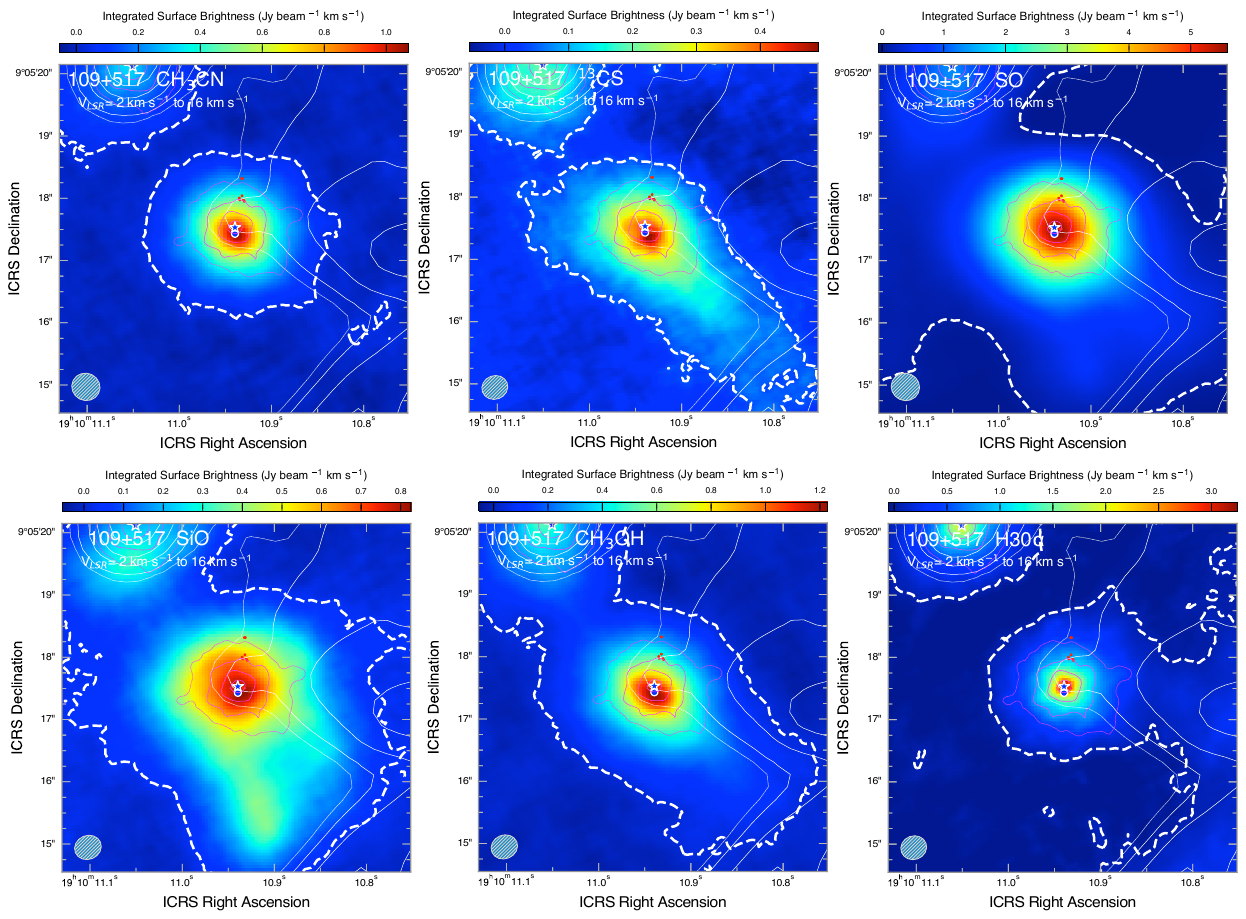}
\caption{Comparing integrated intensity maps of CH$_3$CN, $^{13}$CS, SO, SiO, CH$_3$OH, and H30$\alpha$ emissions for HMC109+517.
The molecular lines were integrated from $V_{\rm LSR} = 2$ to $16$~km\,s$^{-1}$, while the H30$\alpha$ line was integrated from $V_{\rm LSR} = -20$ to $+$60~km\,s$^{-1}$.
See the text for the markers and contours.
The white dashed contours indicate the 5$\sigma$ level.
{Alt text: Six local moment~0 maps centered on HMC109+517 for CH$_3$CN, $^{13}$CS, SO, SiO, CH$_3$OH, and H30$\alpha$. Molecular lines are integrated over $V_{\rm LSR}$ = 2 to 16 km~s$^{-1}$, and H30$\alpha$ over $V_{\rm LSR}$ = -20 to +60 km~s$^{-1}$. Symbols: blue circle marks the CH$_3$CN peak; blue star marks the 1.3~mm continuum peak; magenta contours show 1.3~mm continuum (5-80\%); white contours show 3.6~cm continuum; red/blue dots mark OH/H$_2$O masers; yellow square/asterisk mark CH$_3$OH Class I/II masers.}
}
\label{Fig109+517}
\end{figure}

Figure~\ref{Fig109+517} displays maps of HMC109+517 in CH$_3$CN, $^{13}$CS, SO, SiO, CH$_3$OH, and H30$\alpha$ emission.
All of the line emissions are compact and well-isolated.
This object is located at the western edge of the 3.6~cm continuum source R3 in the W49A~SW region \citep{DePree1997}, near a weak 3.6~cm feature that appears to be connected to the main body of R3.
The 1.3~mm continuum source, C109+517 or N4, essentially overlaps with the molecular emission features.
The peak position of C109+517 coincides with the CH$_3$CN peak within 0.1$''$, corresponding to one-third of the beam radius.

The H30$\alpha$ emission from this HMC has a FWHM line width of 39~km\,s$^{-1}$, suggesting the presence of an HCHII region at or close to its center.
It is not straightforward to determine whether the 1.3~mm continuum arises solely from the free-free emission of the UCHII region.
The absence of a 1.3~mm continuum excess toward the 3.6~cm peak may indicate that thermal dust emission also contributes to the 1.3~mm emission (see Table\ref{Table: 226 GHz likelihood}).

The presence of hot and dense gas, traced by CH$_{3}$CN emission, is evident.
The $^{13}$CS map reveals a weak feature extending in the northeast-southwest direction.
This feature is also apparent in other extended gas tracers, such as DCN and C$^{18}$O (see Figure~\ref{Fig:Other_mol_lines_map01}), with the elongated emission being particularly prominent in C$^{18}$O.
The SiO map shows an elongated structure extending southward from the HMC, which may be associated with an outflow from this HMC.
Several OH maser spots are distributed in the northern part of the HMC, at distances of 0$\farcs$5-1$\farcs$0 from the continuum peak.

%:%%% Figure A06: HMC109+517
%Figure A06

\subsection{HMC110+520}\label{HMC110+520}
\begin{figure}[htbp]
\includegraphics[bb= 50 80 500 520, scale=0.4]{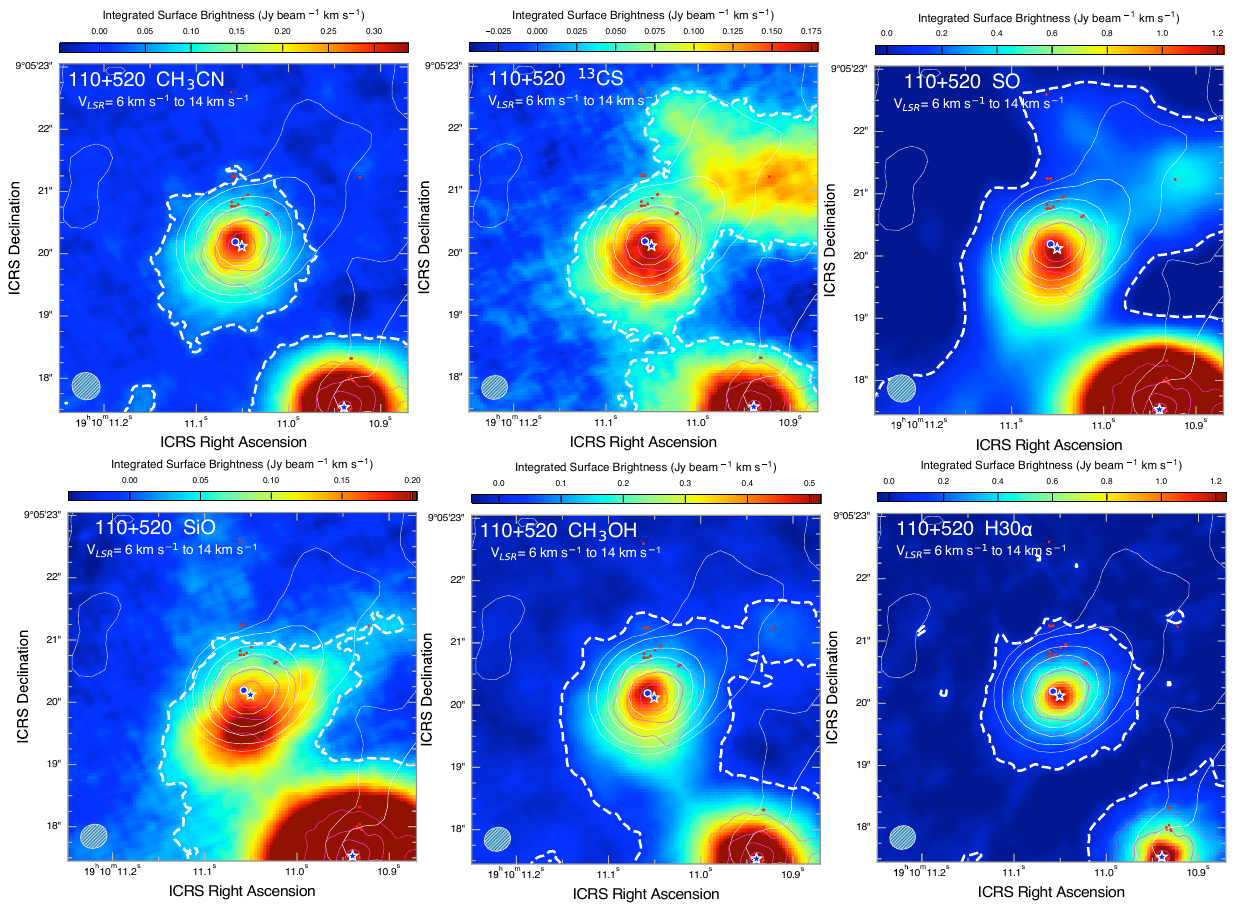}
\caption{Same as Figure~\ref{Fig109+517}, but for HMC110+520, integrated over $V_{\rm LSR} = 6$--14~km,s$^{-1}$ for the molecular lines.
{Alt text: Six local moment~0 maps centered on HMC110+520 for CH$_3$CN, $^{13}$CS, SO, SiO, CH$_3$OH, and H30$\alpha$. Molecular lines are integrated over $V_{\rm LSR}$ = 6 to 14 km~s$^{-1}$, and H30$\alpha$ over $V_{\rm LSR}$ = -20 to +60 km~s$^{-1}$. Symbols: blue circle marks the CH$_3$CN peak; blue star marks the 1.3~mm continuum peak; magenta contours show 1.3~mm continuum (5-80\%); white contours show 3.6~cm continuum; red/blue dots mark OH/H$_2$O masers; yellow square/asterisk mark CH$_3$OH Class I/II masers.}
}
\label{Fig110+520}
\end{figure}

The maps of HMC110+520, corresponding to N9, in the molecular and H30$\alpha$ lines are shown in Figure~\ref{Fig110+520}.
This object is located $\sim3''$ at the northeast of HMC109+517 in the W49A SW region.
The peak position and overall distribution of the 1.3~mm continuum match well with those of the 3.6~cm continuum (UCHII R) and the H30$\alpha$ line.
The latter exhibits a Gaussian profile with a FWHM line width of 28~km\,s$^{-1}$, indicative of its HCHII origin.

The CH$_{3}$CN emission peak is offset by 0$\farcs$13 to the northeast of the continuum peak of C110+520.
The lines of CH$_{3}$OH, HC$_{3}$N, OCS, and H$_2$CO (see also Figure~\ref{Fig:Other_mol_lines_map01}) also peak at the location of the CH$_{3}$CN maximum.
The SiO emission is shifted 0$\farcs$5 to the south of the HMC and the 1.3~mm continuum source, indicative of a southward-directed outflow.
The DCN and C$^{18}$O maps show emission connecting this source with HMC109+517, suggesting a possible interaction between the two HMCs.
Several OH maser spots are distributed around the northern periphery of the HMC.

%:%%% Figure A07: HMC110+520
%Figure A07

\subsection{HMC120+614}\label{HMC120+614}

\begin{figure}[htbp]
\includegraphics[bb= 80 60 500 510, scale=0.4]{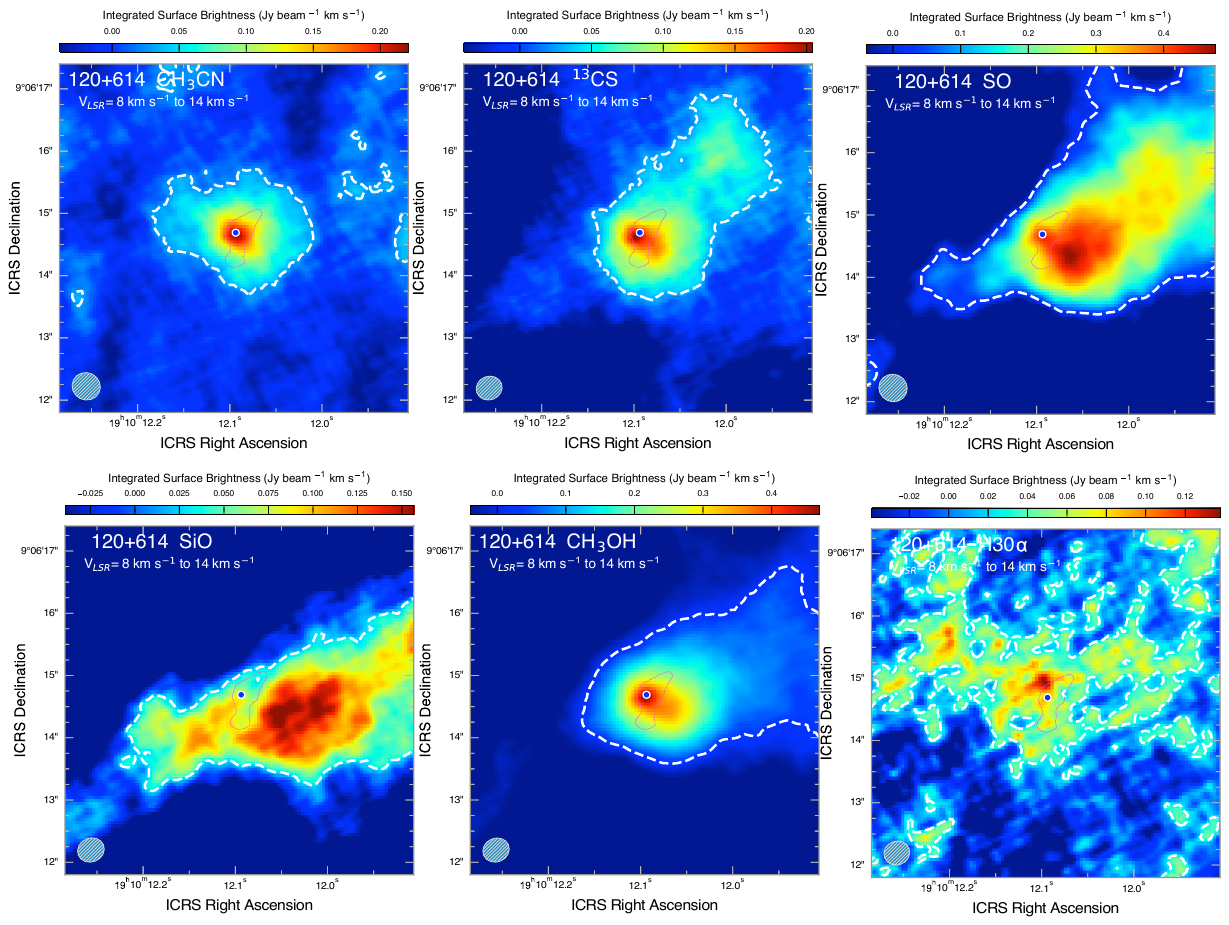}
\caption{Same as Fig~\ref{Fig109+517}, but for HMC120+614, integrated from $V_{\rm LSR} = 8$ to $14$~km\,s$^{-1}$ for molecular lines.
{Alt text: Six local moment~0 maps centered on HMC120+614 for CH$_3$CN, $^{13}$CS, SO, SiO, CH$_3$OH, and H30$\alpha$. Molecular lines are integrated over $V_{\rm LSR}$ = 8 to 14 km~s$^{-1}$, and H30$\alpha$ over $V_{\rm LSR}$ = -20 to +60 km~s$^{-1}$. Symbols: blue circle marks the CH$_3$CN peak; blue star marks the 1.3~mm continuum peak; magenta contours show 1.3~mm continuum (5-80\%); white contours show 3.6~cm continuum; red/blue dots mark OH/H$_2$O masers; yellow square/asterisk mark CH$_3$OH Class I/II masers.}
}
\label{Fig120+614}
\end{figure}

Figure~\ref{Fig120+614} presents maps of molecular lines and H30$\alpha$ emission toward HMC120+614.
The magenta contours indicate the 1.3~mm continuum emission associated with the marginal source [C120+614] (N36; \citealt{Nony2024}), which is elongated in the north-south direction and has a peak intensity of 11~mJy\,beam$^{-1}$, corresponding to a 3$\sigma$ noise level.
No 3.6~cm continuum emission is detected in this region \citep{DePree1997}, suggesting that the weak 1.3~mm continuum likely originates from thermal dust.

A weak emission line is detected at the nominal H30$\alpha$ frequency.
Its spatial distribution is noisy and shows no clear correlation with the molecular emission or the 1.3~mm continuum.
Given the lack of 3.6~cm emission near HMC120+614, we infer that the H30$\alpha$ recombination line is not responsible for this feature.
Weak emission lines of complex organic molecules (COMs) have been detected from this HMC \citep{Madrid2013}, and these COMs may account for the emission observed at the H30$\alpha$ frequency.

The emission features of CH$_3$CN, CH$_3$OH, and H$_2$CO show well-defined distributions with peaks coinciding within 0\farcs1 of the continuum peak.
The maps of $^{13}$CS, HNCO, HC$_3$N, OCS, and H$_2$CO reveal a $\sim$1$''$-long feature elongated in the northeast-southwest direction. 
This may be due to the presence of another emission source located $\sim$0\farcs5 southwest of the CH$_3$CN peak, as suggested by the HC$_3$N and H$_{2}$CO maps, which show a secondary peak. 
The SO and SiO emissions exhibit features extended over more than 2$''$ along the northwest-southeast direction. 
Toward the 1.3~mm and CH$_3$CN peaks, the SiO emission displays a north--south elongated gap that matches well with the 1.3~mm continuum distribution. 
These characteristics suggest that the north-south elongation of the continuum traces a torus and/or disk of HMC120+614, which may be driving a bipolar molecular outflow in the east-west direction.
The C$^{18}$O map indicates that the HMC lies at the eastern edge of a molecular cloud.

%:%%% Figure A08: HMC120+614
% Figure A08

\subsection{HMC127+611}\label{HMC127+611}

\begin{figure}[htbp]
\includegraphics[bb= 70 60 500 500, scale=0.4]{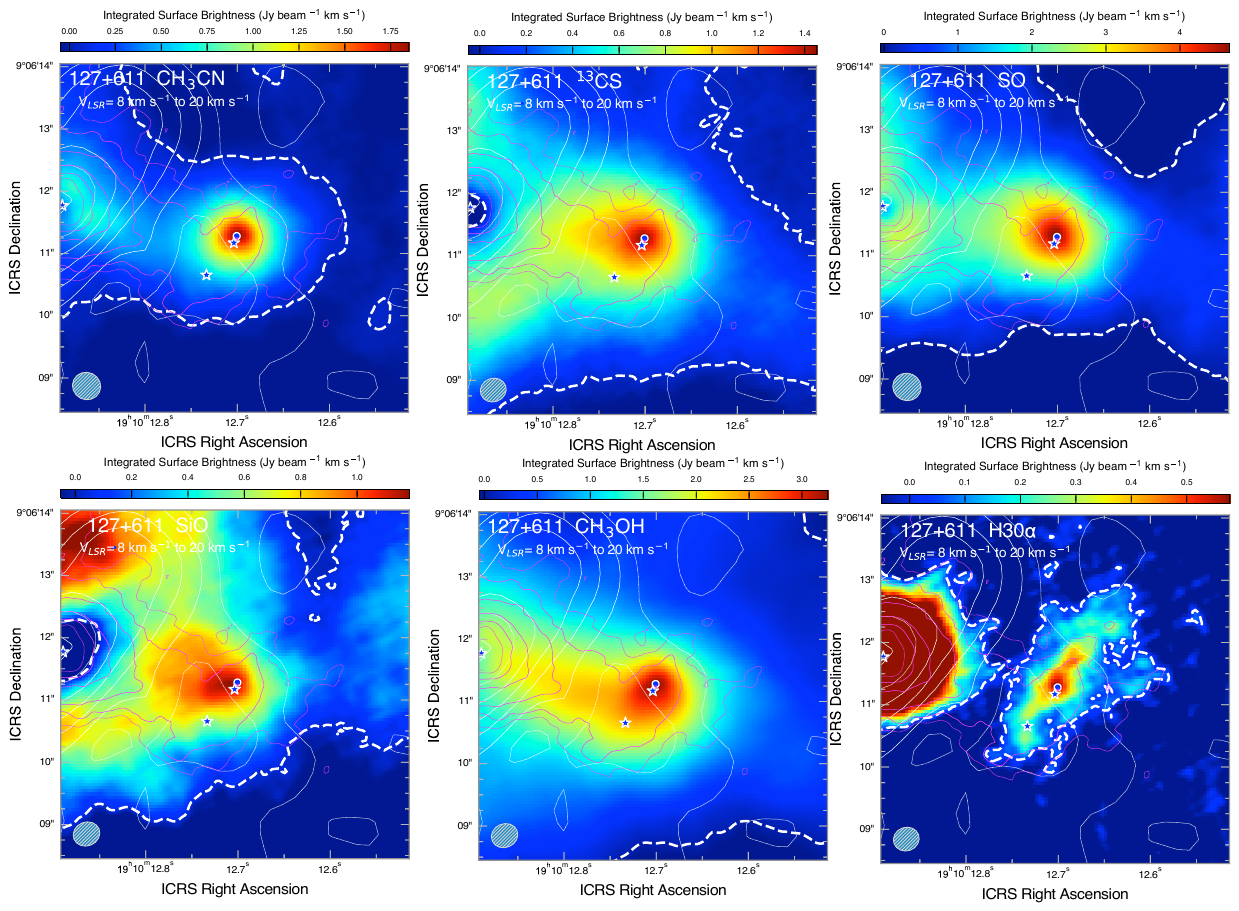}
\caption{Same as Fig~\ref{Fig109+517}, but for HMC127+611, integrated from $V_{\rm LSR} = 8$ to $20$~km\,s$^{-1}$ for molecular lines.
{Alt text: Six local moment~0 maps centered on HMC127+611 for CH$_3$CN, $^{13}$CS, SO, SiO, CH$_3$OH, and H30$\alpha$. Molecular lines are integrated over $V_{\rm LSR}$ = 8 to 20 km~s$^{-1}$, and H30$\alpha$ over $V_{\rm LSR}$ = -20 to +60 km~s$^{-1}$. Symbols: blue circle marks the CH$_3$CN peak; blue star marks the 1.3~mm continuum peak; magenta contours show 1.3~mm continuum (5-80\%); white contours show 3.6~cm continuum; red/blue dots mark OH/H$_2$O masers; yellow square/asterisk mark CH$_3$OH Class I/II masers.}
}
\label{Fig127+611}
\end{figure}

Figure~\ref{Fig127+611} presents maps of HMC127+611, located 2$\farcs$5 west of UCHII~A.
The 1.3~mm continuum map exhibits an isolated peak corresponding to C127+611 or N44, which is spatially coincident with the CH$_3$CN peak within 0\farcs11.

The emission line at the H30$\alpha$ frequency, whose spatial distribution is consistent with the 1.3~mm continuum, exhibits a skewed profile (see the bottom-right panel of Figure~\ref{Fig127+611}), including narrow spikes and possibly a broad pedestal with an FWHM of $\sim$30~km\,s$^{-1}$. 
The 3.6~cm continuum emission is detected in this area, with a local peak of $\sim$24~mJy\,beam$^{-1}$ located $\sim$0\farcs3 west of the 1.3~mm peak.
At least part of the nominal H30$\alpha$ line, namely the broad component, may originate from recombination emission, as supported by the detection of the H29$\alpha$ line toward this object.
Since COM emission lines were also detected from this HMC \citep{Madrid2013}, the observed spike features may represent a blend of recombination and COM lines.

All maps show compact, well-defined features with peaks coinciding with each other, except for those of DCN and C$^{18}$O.
The C$^{18}$O map suggests that this HMC is located at the western edge of a molecular cloud associated with UCHII~A.
Other molecular line maps also show weak emission features bridging the HMC and UCHII~A, where molecular emission tends to be depleted.

%:%%% Figure A09: HMC127+611
% Figure A09

\subsection{HMC131+618}\label{HMC131+618}

\begin{figure}[htbp]
\includegraphics[bb= 70 60 500 500, scale=0.4]{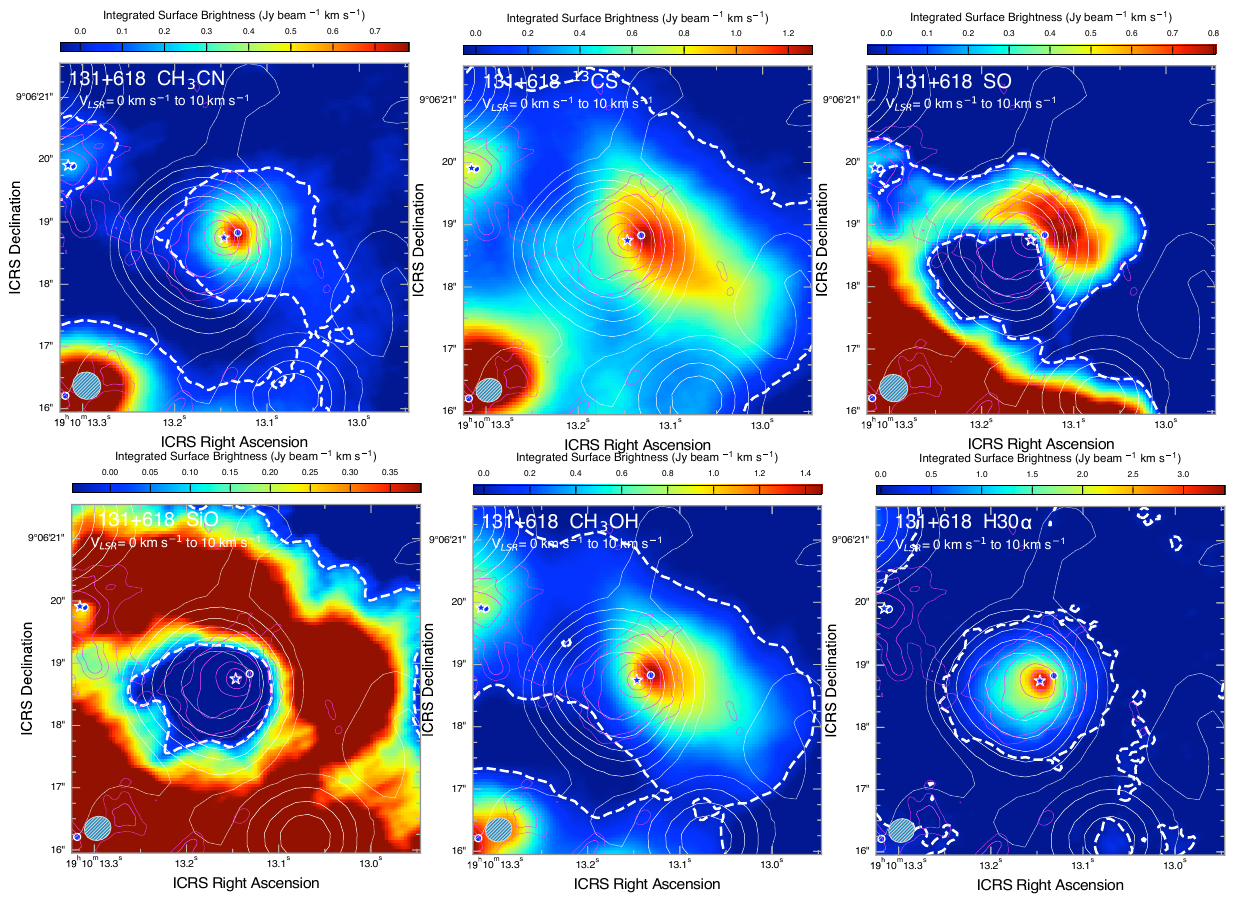}
\caption{Same as Fig~\ref{Fig109+517}, but for HMC131+618, integrated from $V_{\rm LSR}$ = 0 to 10~km\,s$^{-1}$ for molecular lines.
{Alt text: Six local moment~0 maps centered on HMC131+618 for CH$_3$CN, $^{13}$CS, SO, SiO, CH$_3$OH, and H30$\alpha$. Molecular lines are integrated over $V_{\rm LSR}$ = 0 to 10 km~s$^{-1}$, and H30$\alpha$ over $V_{\rm LSR}$ = -20 to +60 km~s$^{-1}$. Symbols: blue circle marks the CH$_3$CN peak; blue star marks the 1.3~mm continuum peak; magenta contours show 1.3~mm continuum (5-80\%); white contours show 3.6~cm continuum; red/blue dots mark OH/H$_2$O masers; yellow square/asterisk mark CH$_3$OH Class I/II masers.}
}
\label{Fig131+618}
\end{figure}

Figure~\ref{Fig131+618} shows maps of HMC131+618.
The CH$_3$CN emission peak is located 0\farcs22 west of the 1.3~mm continuum peak.
The peak positions of C131+618, N7, and UCHII C all coincide within 0\farcs1.
The strong and broad H30$\alpha$ line emission exhibits the same distribution as the 1.3~mm continuum.
Together with the flat SED of C131+618 shown in Figure~\ref{Fig: SEDs}, this suggests that the 1.3~mm emission is dominated by thermal free-free radiation.
The low-level contours of C131+618 are elongated toward the southeast, along the tail of the cometary-shaped UCHII C \citep{DePree2020}, where \citet{Nony2024} identified another continuum core, N72.

%:%%% Figure A10: HMC131+618
% Figure A10

The CH$_{3}$CN emission is compact, sharing its peak position with several other molecular emissions such as $^{13}$CS, CH$_{3}$OH, HNCO, HC$_{3}$N, and OCS.
Many of these molecular line maps exhibit extended features elongated in the northeast-southwest direction, along the periphery of the continuum source C131+618.
This elongation is oriented perpendicular to the large proper motion of UCHII C, corresponding to a velocity of $\sim$76~km\,s$^{-1}$ toward the northwest from the cometary head \citep{Rodriguez2020}.
The outflow tracers SiO and SO exhibit cavity-like structures toward the continuum source, with arc-shaped features located just outside the cometary head of UCHII C.
These characteristics may suggest an interaction between UCHII C131+618 and its adjacent HMC131+618.

\subsection{HMC131+613 and HMC131+614 (MCN-f)}
\label{MCN-f}

\begin{figure}[htbp]
\includegraphics[bb= 0 0 500 560, scale=0.29]{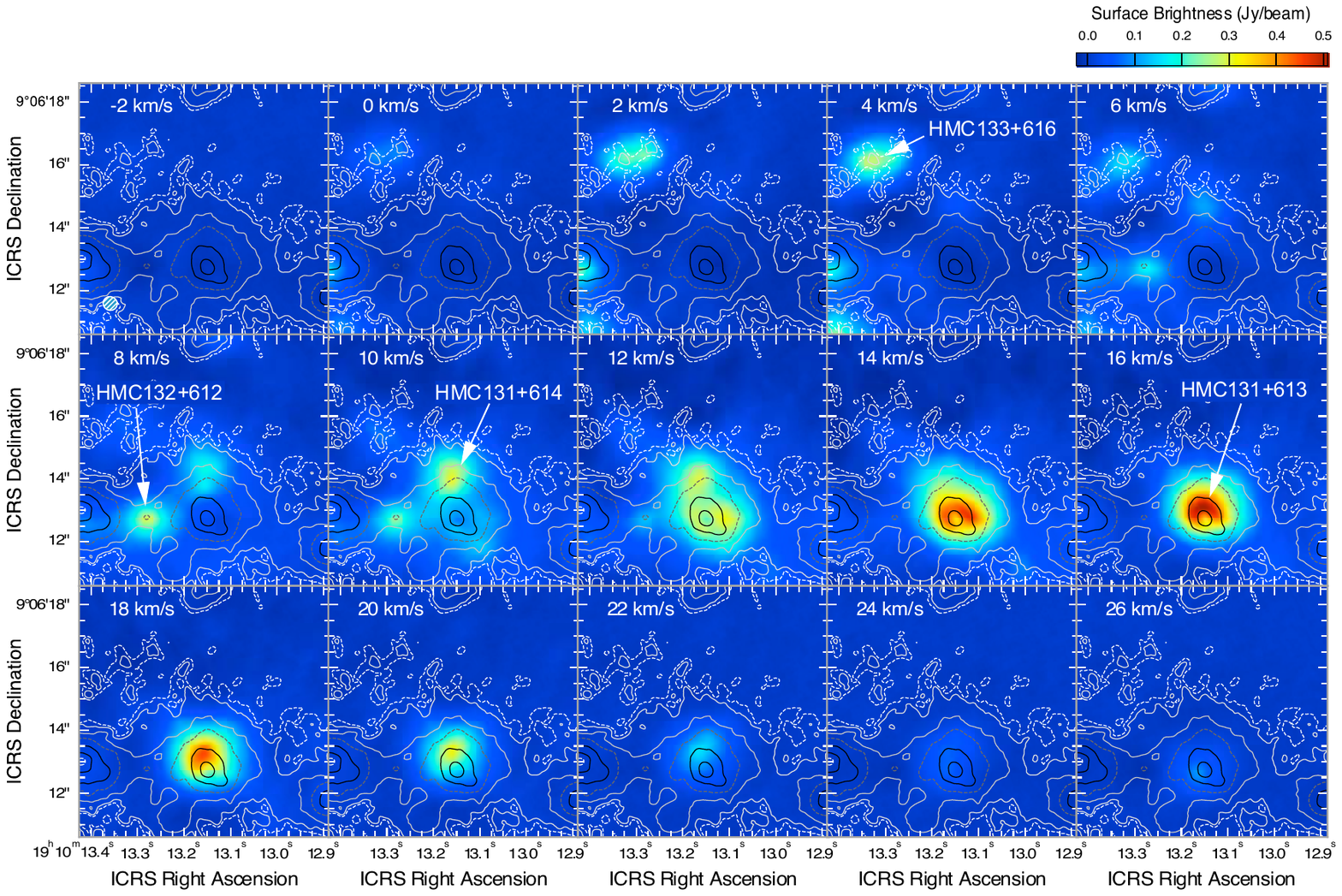}
\caption{Velocity channel maps of the CH$_3$CN line toward HMC131+613 and HMC131+614.
The contours represent the 1.3~mm continuum emission at 1, 2, 5, 10, 20, 40, and 80\% of the peak intensity.
HMC132+612 and HMC133+616 are also seen at $V_{\rm LSR}$ = 6--12~km\,s$^{-1}$ and 2--6~km\,s$^{-1}$, respectively.
{Alt text: Velocity-channel map sequence of CH$_3$CN emission toward HMC131+613 and HMC131+614. Each panel shows CH$_3$CN brightness in a single $V_{\rm LSR}$ channel, with 1.3~mm continuum contours overlaid at 1, 2, 5, 10, 20, 40, and 80\% of the peak. Emission from HMC132+612 appears in channels $V_{\rm LSR}$ = 6-12 km~s$^{-1}$ and from HMC133+616 in $V_{\rm LSR}$ = 2-6 km~s$^{-1}$.}}
\label{Fig131+613-4}
\end{figure}

MCN-f, a previously identified HMC in CH$_{3}$CN emission by \citet{Wilner2001}, encompasses two HMCs: HMC131+613 and HMC131+614.
These sources differ in both position and radial velocity, as shown in the channel maps (Figure~\ref{Fig131+613-4}): HMC131+613 has a radial velocity of $V_{\rm LSR}$ = 12--30~km\,s$^{-1}$, whereas HMC131+614 exhibits $V_{\rm LSR}$ = 4--12~km\,s$^{-1}$.

%:%%% Figure A11: HMC131+614 channel maps
% Figure A11

\begin{figure}[htbp]
\includegraphics[bb= 60 70 500 490, scale=0.40]{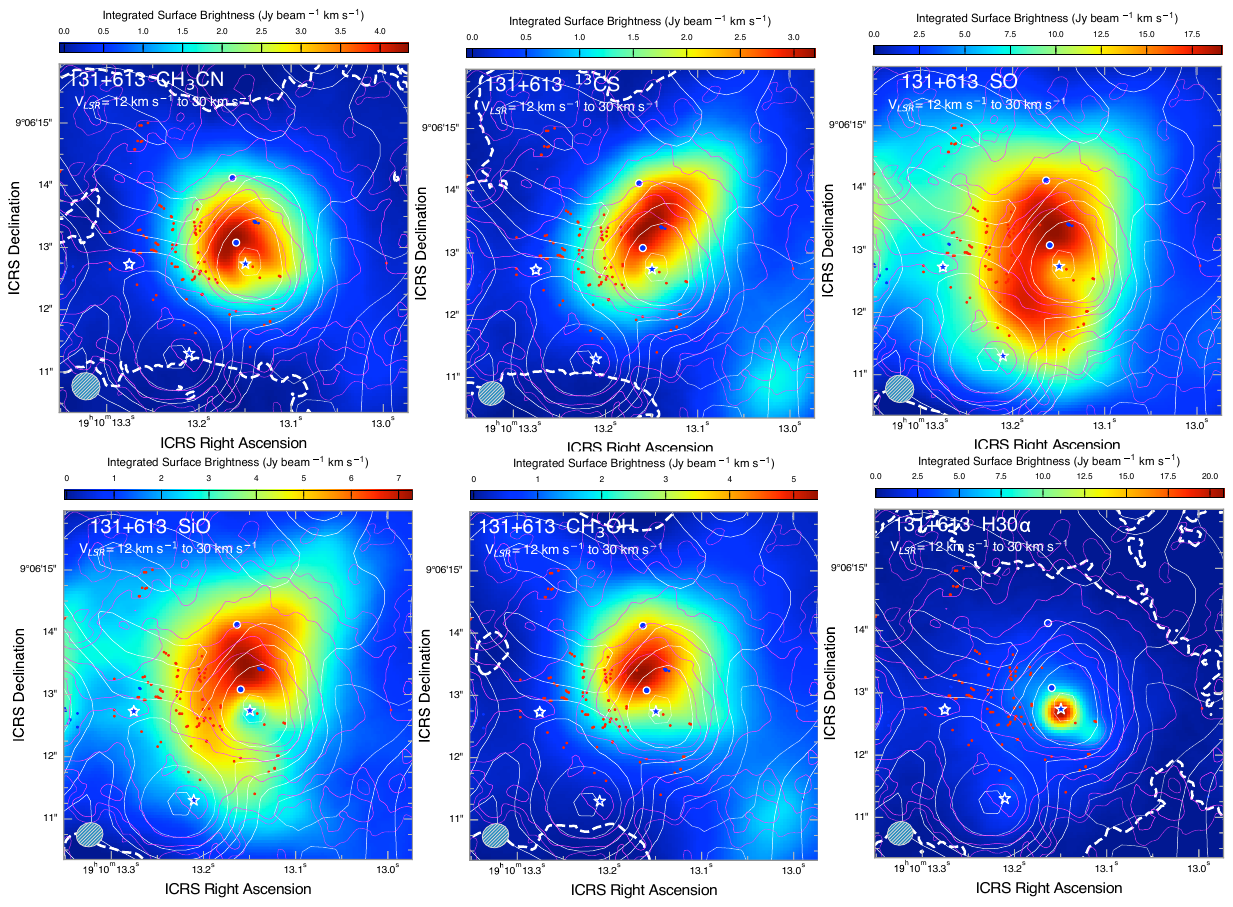}
\caption{Same as Fig~\ref{Fig109+517}, but for HMC131+613, integrated from $V_{\rm LSR}$ = 12 to 30~km\,s$^{-1}$ for molecular lines.
{Alt text: Six local moment~0 maps centered on HMC131+613 for CH$_3$CN, $^{13}$CS, SO, SiO, CH$_3$OH, and H30$\alpha$. Molecular lines are integrated over $V_{\rm LSR}$ = 12 to 30 km~s$^{-1}$, and H30$\alpha$ over $V_{\rm LSR}$ = -20 to +60 km~s$^{-1}$. Symbols: blue circle marks the CH$_3$CN peak; blue star marks the 1.3~mm continuum peak; magenta contours show 1.3~mm continuum (5-80\%); white contours show 3.6~cm continuum; red/blue dots mark OH/H$_2$O masers; yellow square/asterisk mark CH$_3$OH Class I/II masers.}
}
\label{Fig131+613}
\end{figure}

\label{HMC131+613}
Figure~\ref{Fig131+613} presents maps of HMC131+613.
The overall distribution of CH$_{3}$CN emission resembles that of the continuum source C131+612 (or N2), except near the continuum peak where a cavity in the molecular emission is evident.
The peak of the H30$\alpha$ emission coincides with the C131+612 peak, located $\sim$0\farcs45 south of the CH$_{3}$CN peak.
At the CH$_{3}$CN peak (HMC131+613), the H30$\alpha$ line profile shows a tapered shape with a broad wing, suggesting a mixture of recombination lines from a background HII region and from an HCHII region associated with C131+612.
A careful comparison of the CH$_{3}$CN and H30$\alpha$ maps indicates that, while the H30$\alpha$ emission is relatively compact and concentrated near the 3.6~cm peak (UCHII B2), the 1.3~mm continuum more closely follows the CH$_{3}$CN distribution around its peak position, implying that part of the 1.3~mm emission may originate from thermal dust.

%:%%% Figure A12: HMC131+613
% Figure A12

The cavity at the C131+612 peak is visible in all molecular line maps (see also Figure~\ref{Fig:Other_mol_lines_map01}).
The channel maps (Figure~\ref{Fig131+613-4}) at $V_{\rm LSR}$ = 12--16~km\,s$^{-1}$ suggest that the overall structure of the HMC has a ring- or shell-like morphology surrounding the 1.3~mm continuum peak.
Although this could imply that HMC131+613 is merely a remnant of the torus/disk and envelope that once surrounded C131+612, the strong asymmetry of the molecular distributions relative to the continuum peak may instead indicate that HMC131+613 harbors another massive star-forming site.
The molecular emission distributions of CH$_{3}$OH, HNCO, OCS, and SO$_{2}$ are approximately circularly symmetric with respect to the HMC131+613 peak position, further supporting the interpretation that this is an independent HMC distinct from C131+612, with its southeastern edge ionized by the HCHII region.

% CH$_3$CN, $^{13}$CS, SO, SiO, CH$_3$OH, and H30$\alpha$ lines.
The CH$_3$CN peak of HMC131+613 lies near the center of four 226 GHz continuum cores---N83, N32, N56, and N2---all within 0\farcs5 of the molecular peak.
The SiO and SO luminosities, the outflow signarture, are also strong, but their peaks are located between HMC131+613 and HMC131+614, with their low level emission surrounding the entire complex including C131+612 (UCHII B).
It is thus difficult to determine which of them are responsible for the outflow.

HMC131+613 is associated with OH maser spots, which are located at $\sim$0\farcs5 northwest of the HMC, near the peaks of the SO and SiO emissions.
The H$_2$O maser spots are found at the western edge of the HMC, well tracing the western part of the SiO and SO ring feature surrounding C131+612, whch is reasonable since both those masers and molecular lines are outflow tracers.

\begin{figure}[htbp]
\includegraphics[bb= 50 60 500 500, scale=0.4]{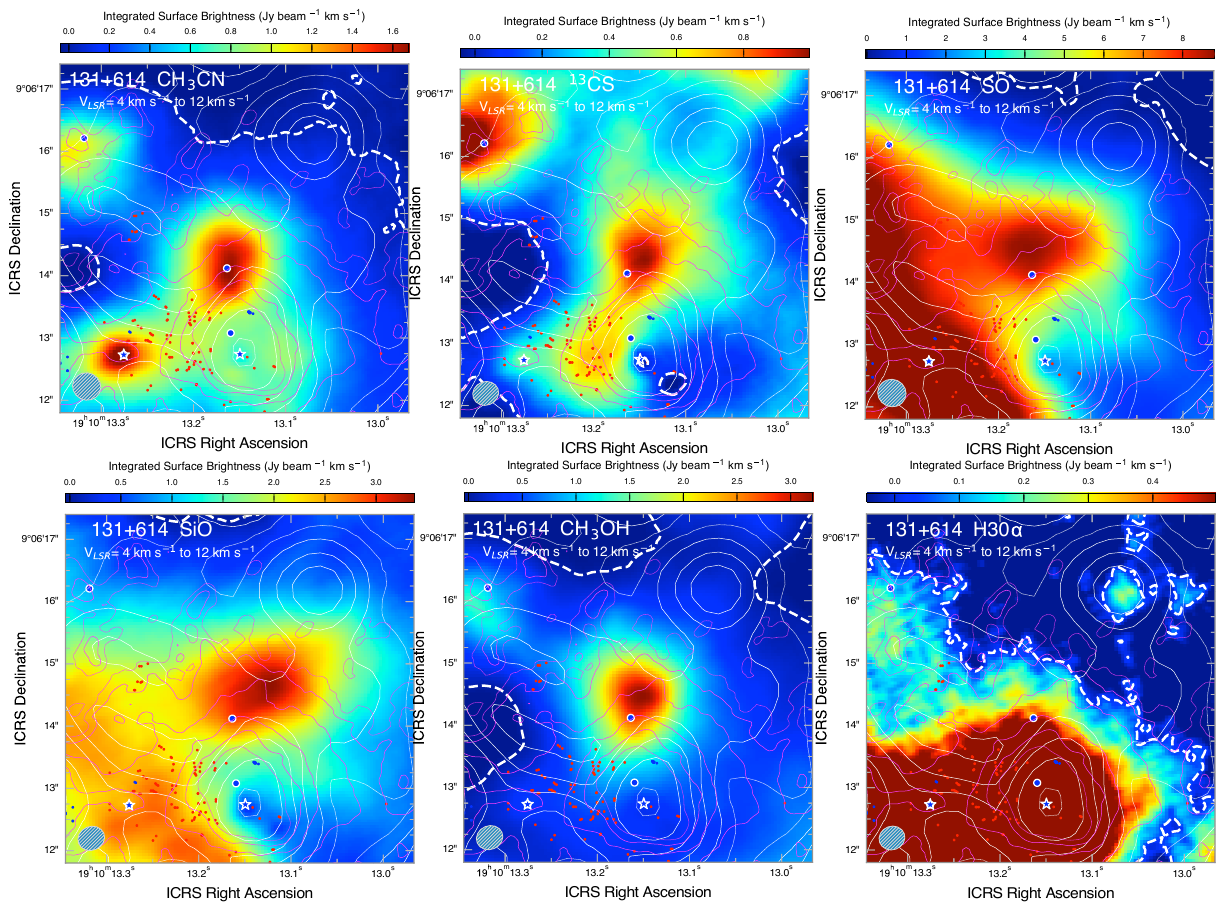}
\caption{Same as Fig~\ref{Fig109+517}, but for HMC131+614, integrated from $V_{\rm LSR}$ = 4 to 12~km\,s$^{-1}$ for molecular lines.
{Alt text: Six local moment~0 maps centered on HMC131+614 for CH$_3$CN, $^{13}$CS, SO, SiO, CH$_3$OH, and H30$\alpha$. Molecular lines are integrated over $V_{\rm LSR}$ = 4 to 12 km~s$^{-1}$, and H30$\alpha$ over $V_{\rm LSR}$ = -20 to +60 km~s$^{-1}$. Symbols: blue circle marks the CH$_3$CN peak; blue star marks the 1.3~mm continuum peak; magenta contours show 1.3~mm continuum (5-80\%); white contours show 3.6~cm continuum; red/blue dots mark OH/H$_2$O masers; yellow square/asterisk mark CH$_3$OH Class I/II masers.}
}
\label{Fig131+614}
\end{figure}

\label{HMC131+614}
Figure~\ref{Fig131+614} presents maps of HMC131+614, located 1\farcs5 north of HMC131+613.
This object is a CH$_{3}$CN source corresponding to the northern part of MCN-f \citep{Wilner2001}.
No peak of H30$\alpha$ emission is detected at its position. 
The H30$\alpha$ line is relatively strong but has a narrow width (8.6~km\,s$^{-1}$), suggesting an origin in the background HII region.
The 1.3~mm core N28 lies 0\farcs5 southwest of the CH$_{3}$CN peak of HMC131+614, at the edge of its molecular distribution.
The 1.3~mm continuum contours centered on C131+612 appear slightly elongated toward HMC131+614, possibly indicating that part of the 1.3~mm emission at this HMC originates from thermal dust.

Although this HMC is well defined in the CH$_{3}$CN emission, it is less clearly identified in other molecular tracers.
The $^{13}$CS, CH$_{3}$OH, and H$_{2}$CO maps show isolated emission features toward the HMC, but their peaks are shifted by $\sim$0\farcs5 to the north or northwest of the HMC131+614 peak.
Other molecular maps exhibit only weak peaks near the HMC, located at the northern edge of the stronger and more extended emission around UCHII B.
The H$_{2}$O and OH maser spots found to the south are more naturally associated with southern sources such as HMC131+613, UCHII B, or HMC131+612.

%:%%% Figure A13: HMC131+614  %%%
% Figure A13

\subsection{HMC132+612 (MCN-e)}\label{HMC132+612}

\begin{figure}[htbp]
\includegraphics[bb= 60 60 500 500, scale=0.4]{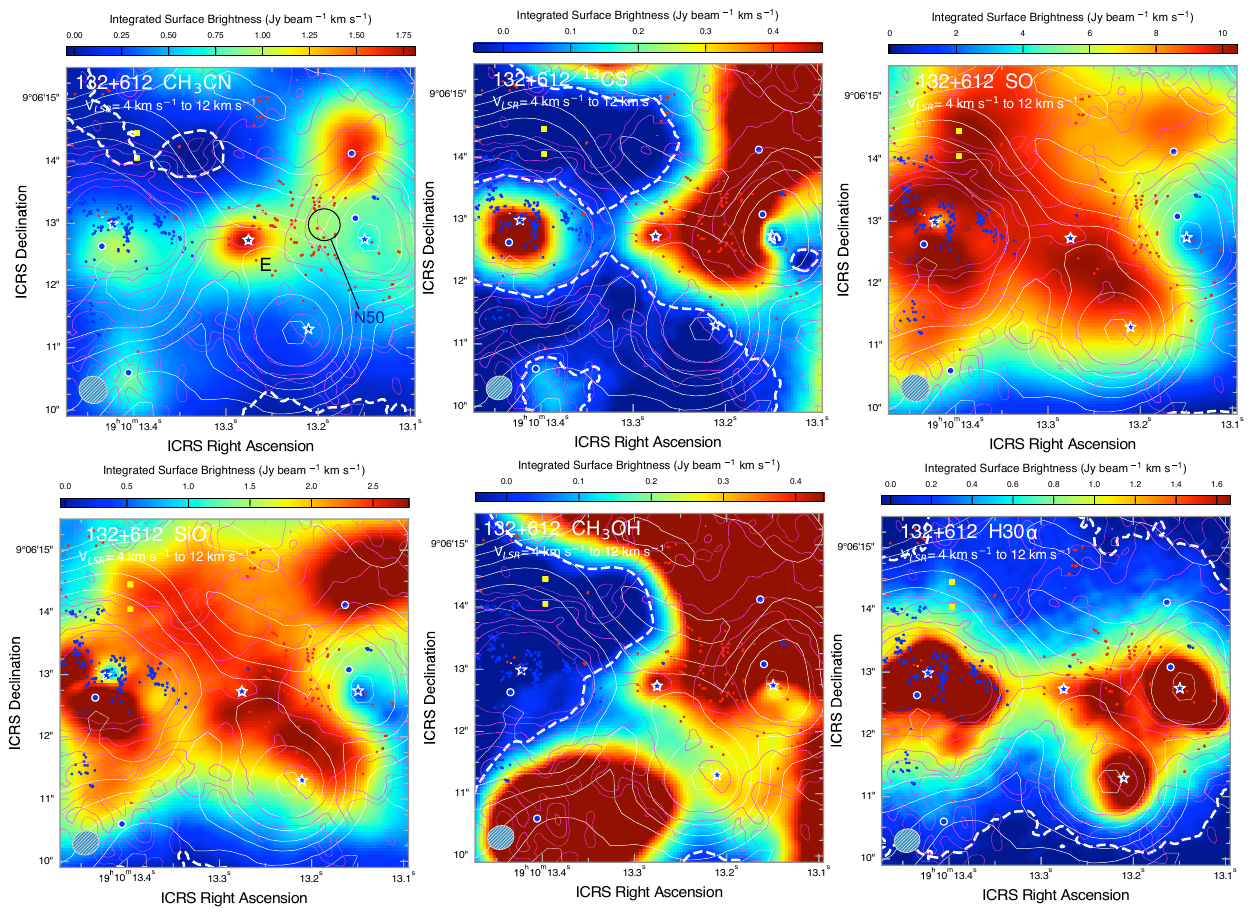}
\caption{Same as Fig~\ref{Fig109+517}, but for HMC132+612 (MCN-e), integrated from $V_{\rm LSR}$ = 4 to 12~km\,s$^{-1}$ for molecular lines.
{Alt text: Six local moment~0 maps centered on HMC132+612 (MCN-e) for CH$_3$CN, $^{13}$CS, SO, SiO, CH$_3$OH, and H30$\alpha$. Molecular lines are integrated over $V_{\rm LSR}$ = 4 to 12 km~s$^{-1}$, and H30$\alpha$ over $V_{\rm LSR}$ = -20 to +60 km~s$^{-1}$. Symbols: blue circle marks the CH$_3$CN peak; blue star marks the 1.3~mm continuum peak; magenta contours show 1.3~mm continuum (5-80\%); white contours show 3.6~cm continuum; red/blue dots mark OH/H$_2$O masers; yellow square/asterisk mark CH$_3$OH Class I/II masers.}
}
\label{Fig132+612}
\end{figure}

Figure~\ref{Fig132+612} shows maps of HMC132+612, which was previously identified as MCN-e \citep{Wilner2001}.
This source has a compact CH$_3$CN emission associated with a weak 1.3~mm continuum peak C132+612 (N20), located between the two bright UCHII regions B and G.

The H30$\alpha$ emission has been detected with a brightness temperature of $T_{\rm B}$ = 23~K and a velocity width of 14~km\,s$^{-1}$ (see the inset in the bottom-right panel of Figure~\ref{Fig132+612}). 
However, this emission may originate from the background HII region, as the H30$\alpha$ map shows no clear intensity enhancement at the source position, unlike the 1.3~mm continuum. 

The SED (see Figure~\ref{Fig: SEDs}) of C132+612  shows that the flux density increases with frequency as $F_\nu\propto\nu^{2.2}$ above 230~GHz, while no continuum emission is detected at lower frequencies. 
These observations suggest that the 226 GHz continuum emission mostly arises from hot dust particles in the HMC.

%:%%% Figure A14: HMC132+612  %%%
% Figure A14

% CH$_3$CN, $^{13}$CS, SO, SiO, CH$_3$OH, and H30$\alpha$ lines.
The molecular emissions generally peak at the position of HMC132+612, with the exception of SiO, SO, and C$^{18}$O.
The SiO and SO distributions are extended along the northeast-southwest direction, showing no distinct peak at the CH$_{3}$CN maximum.
In contrast, the C$^{18}$O map reveals a relatively well-defined peak located $\sim$0\farcs5 north of the HMC.
OH and H$_{2}$O maser spots are found northwest of the HMC peak, within the CH$_{3}$CN emitting region.

\subsection{HMC133+619}\label{HMC133+619}
\begin{figure}[htbp]
\includegraphics[bb= 80 70 500 510, scale=0.4]{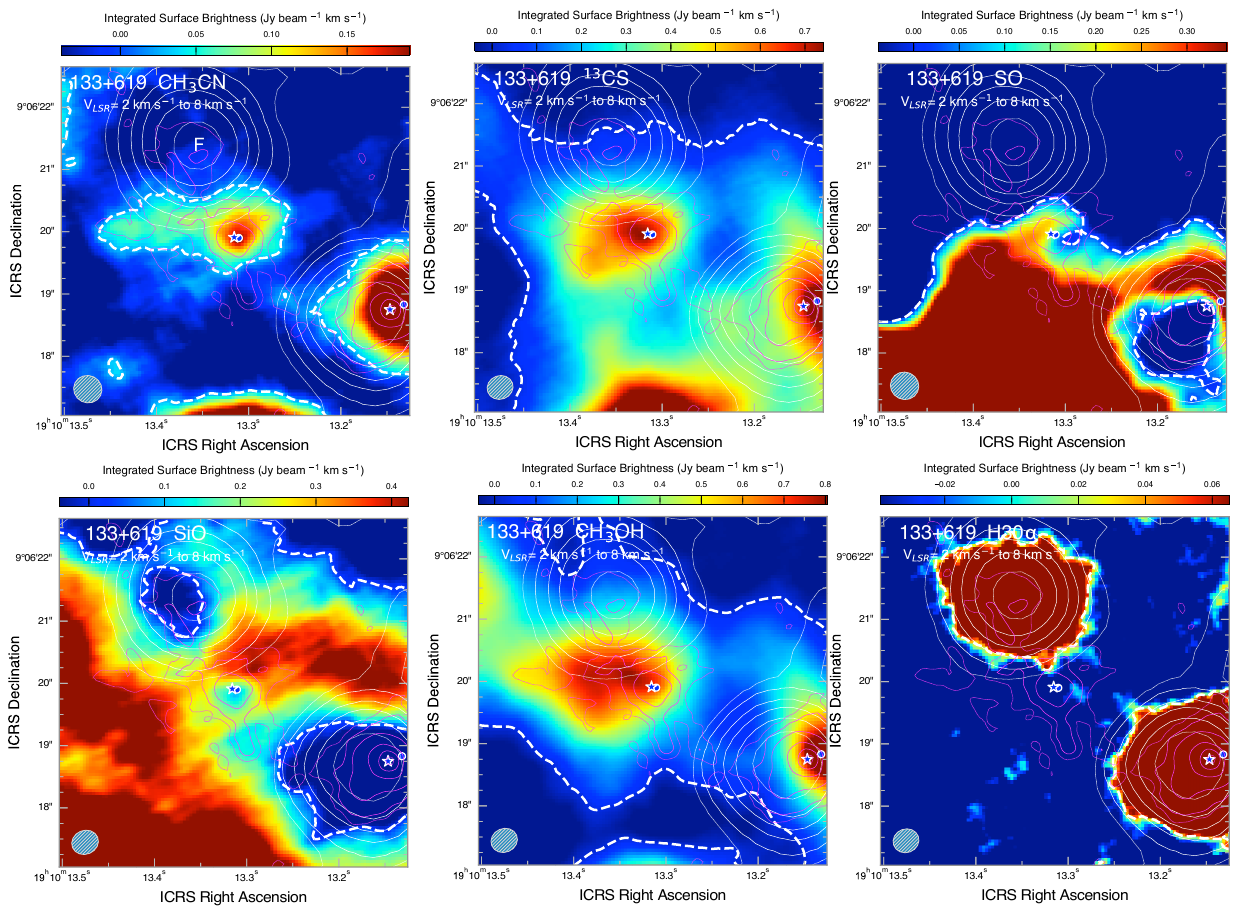}
\caption{Same as Fig~\ref{Fig109+517}, but for HMC133+619 integrated from $V_{\rm LSR}$ = 2 to 8~km\,s$^{-1}$ for molecular lines.
{Alt text: Six local moment~0 maps centered on HMC133+619 for CH$_3$CN, $^{13}$CS, SO, SiO, CH$_3$OH, and H30$\alpha$. Molecular lines are integrated over $V_{\rm LSR}$ = 2 to 8 km~s$^{-1}$, and H30$\alpha$ over $V_{\rm LSR}$ = -20 to +60 km~s$^{-1}$. Symbols: blue circle marks the CH$_3$CN peak; blue star marks the 1.3~mm continuum peak; magenta contours show 1.3~mm continuum (5-80\%); white contours show 3.6~cm continuum; red/blue dots mark OH/H$_2$O masers; yellow square/asterisk mark CH$_3$OH Class I/II masers.}
}
\label{Fig133+619}
\end{figure}

Figure~\ref{Fig133+619} presents maps of HMC133+619.
This object lies 1\farcs5 south of UCHII~F and shows no 3.6~cm continuum emission.
The absence of H30$\alpha$ emission further indicates that no HII region has yet developed in this source.
The continuum emission most likely originates from thermal dust, as suggested by its steeply rising SED toward shorter wavelengths (Figure~\ref{Fig: SEDs}).
It corresponds to N18 but was not identified as an HMC by \citet{Nony2024}.

The CH$_{3}$CN emission follows the 1.3~mm continuum distribution.
The $^{13}$CS, HNCO, HC$_{3}$N, and OCS maps show corresponding emission features, with peaks located near that of the CH$_{3}$CN.
In contrast, the SO and SiO lines exhibit no significant emission toward HMC133+619, implying that an outflow has not yet developed.
The extended gas traced by DCN and C$^{18}$O stretches eastward from the HMC.

%:%%% Figure A15: HMC133+619  %%%
% Figure A15

\subsection{HMC133+616 (MCN-d)}\label{HMC133+616}

\begin{figure}[htbp]
\includegraphics[bb= 80 80 500 500, scale=0.4]{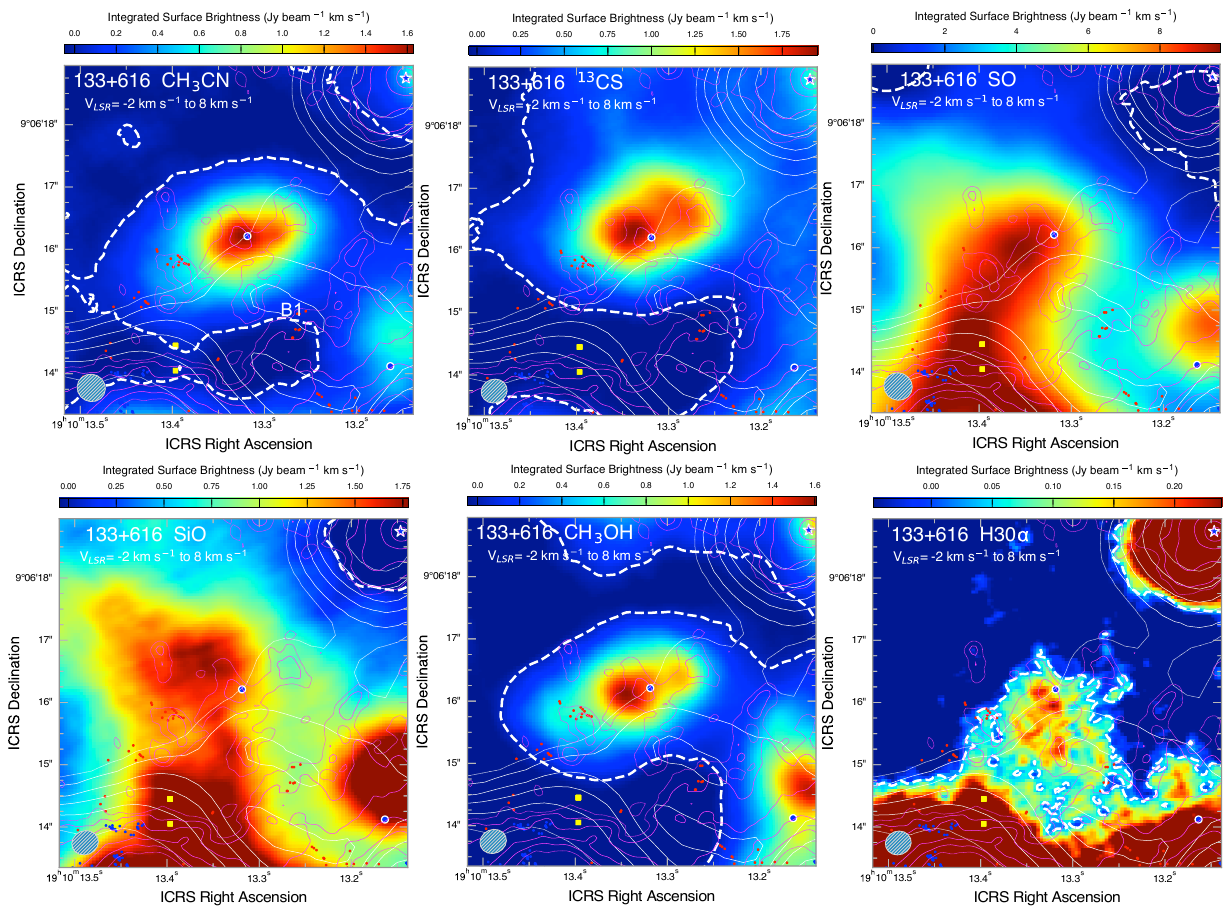}
\caption{Same as Fig~\ref{Fig109+517}, but for HMC133+616 (MCN-d) integrated from $V_{\rm LSR}$ = $-$2 to 8~km\,s$^{-1}$ for molecular lines.
{Alt text: Six local moment~0 maps centered on HMC133+616 (MCN-d) for CH$_3$CN, $^{13}$CS, SO, SiO, CH$_3$OH, and H30$\alpha$. Molecular lines are integrated over $V_{\rm LSR}$ = -2 to 8 km~s$^{-1}$, and H30$\alpha$ over $V_{\rm LSR}$ = -20 to +60 km~s$^{-1}$. Symbols: blue circle marks the CH$_3$CN peak; blue star marks the 1.3~mm continuum peak; magenta contours show 1.3~mm continuum (5-80\%); white contours show 3.6~cm continuum; red/blue dots mark OH/H$_2$O masers; yellow square/asterisk mark CH$_3$OH Class I/II masers.}
}
\label{Fig133+616}
\end{figure}

Figure~\ref{Fig133+616} shows maps of HMC133+616, which corresponds to MCN-d \citep{Wilner2001}. 
A weak 1.3~mm continuum peak, [C133+616], with a brightness of 15~mJy\,beam$^{-1}$ coincides with the CH$_{3}$CN peak within the beam size. 
The continuum brightness is comparable to that of other weak, noise-like features in the vicinity (see the right panel of Figure~\ref{Fig: 226G Comparison}).
\citet{Nony2024}, on the other hand, detected three continuum cores---N27, N54, and N43---toward this HMC. 
N54 coincides with our 1.3~mm continuum peak, while N27 matches another peak located 0\farcs76 west of N54. N43 lies 0\farcs46 away on the ridge of emission extending southeast from N54.

%:%%% Figure A16: HMC133+616  %%%
% Figure A16

No enhancement of the 3.6~cm continuum emission is seen toward HMC131+616.
A narrow emission line was detected at the H30$\alpha$ frequency, but it may not be the recombination line because the H29$\alpha$ line was not detected from this source.
These characteristics imply that HMC133+616 is not associated with a prominent H/UCHII region.
The weak 1.3~mm continuum may arise from thermal dust particles.

% CH$_3$CN, $^{13}$CS, SO, SiO, CH$_3$OH, and H30$\alpha$ lines.
Compact emission features elongated in the east-west direction are detected toward [C133+616] (N54) in the CH$_{3}$CN, $^{13}$CS, CH$_{3}$OH, HNCO, HC$_{3}$N, OCS, and H$_{2}$CO lines.
While the association of N54 with HMC133+616 is evident, the relationships of the other two continuum cores, N27 and N43, with the extended CH$_{3}$CN emission remain uncertain.
Both cores are located at intermediate levels of the CH$_{3}$CN brightness distribution---N27 on the western slope and N43 on the eastern slope.
In this paper, we regard this CH$_{3}$CN structure as a single HMC.
The SO and SiO emissions are spatially extended and appear to connect toward UCHII~G.
OH maser spots are located at $\sim1''$ east of the HMC, but their association with it is uncertain.

\subsection{HMC134+610 (MCN-c)}\label{HMC134+610}

\begin{figure}[htbp]
\includegraphics[bb= 70 70 500 500, scale=0.4]{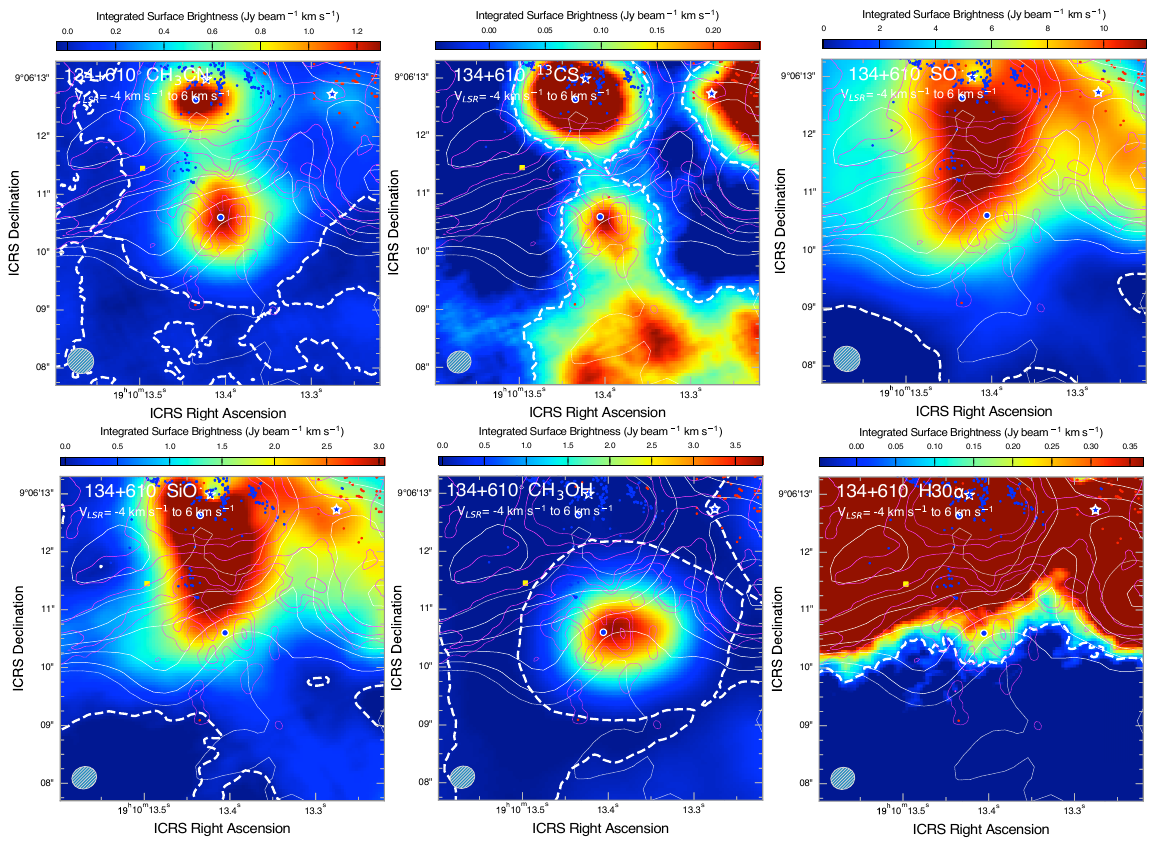}
\caption{Same as Fig~\ref{Fig109+517}, but for HMC134+610 (MCN-c) integrated from $V_{\rm LSR}$ = $-$4 to 6~km\,s$^{-1}$ for molecular lines.
{Alt text: Six local moment~0 maps centered on HMC134+610 (MCN-c) for CH$_3$CN, $^{13}$CS, SO, SiO, CH$_3$OH, and H30$\alpha$. Molecular lines are integrated over $V_{\rm LSR}$ = -4 to 6 km~s$^{-1}$, and H30$\alpha$ over $V_{\rm LSR}$ = -20 to +60 km~s$^{-1}$. Symbols: blue circle marks the CH$_3$CN peak; blue star marks the 1.3~mm continuum peak; magenta contours show 1.3~mm continuum (5-80\%); white contours show 3.6~cm continuum; red/blue dots mark OH/H$_2$O masers; yellow square/asterisk mark CH$_3$OH Class I/II masers.}
}
\label{Fig134+610}
\end{figure}

Figure~\ref{Fig134+610} shows maps of HMC134+610, corresponding to MCN-c, located approximately 2$''$ south of UCHII~G.
No enhancement is seen toward this HMC in the 1.3~mm continuum map.
Features at levels of $\lesssim$40~mJy\,beam$^{-1}$, indistinguishable from imaging artifacts, are distributed around the HMC peak but avoid the peak itself.
The 3.6~cm continuum map also shows no enhancement toward this object.
No 1.3~mm continuum core associated with this source was detected by \citet{Nony2024}.

The H30$\alpha$ emission may be present toward this object, but no enhancement is seen at or around the HMC peak, suggesting that the emission originates from a background HII region and that HMC134+610 has not yet developed a UCHII or HCHII region.
Emission lines of complex organic molecules (COMs) have been detected from this object \citep{Madrid2013}.
The narrow spike observed at the H30$\alpha$ frequency may therefore correspond to a COMs line.

%:%%% Figure A17: HMC134+610  %%%
% Figure A17

% CH$_3$CN, $^{13}$CS, SO, SiO, CH$_3$OH, and H30$\alpha$ lines.
The CH$_3$CN, $^{13}$CS, CH$_3$OH, HNCO, OCS, and H$_2$CO emissions are compact, with their peaks coinciding within the beam size.
The SO and SiO emissions are elongated northward toward UCHII~G but show no enhancement at the CH$_3$CN peak, possibly tracing outflows originating from one of the UCHII~G sources or HMC134+612.
H$_2$O maser spots located about $1''$ northeast of the HMC may be associated either with this HMC or with HMC134+612.
An OH maser spot located 0\farcs4 north of the HMC is also possibly associated with it.
A Class~I CH$_3$OH maser spot lies $\sim$1\farcs5 northeast of the HMC, in the region between this source and HMC134+612.
It is located within the outflow gas of HMC134+612 and is therefore likely associated with that HMC rather than with HMC134+610.

\subsection{HMC134+612 (MCN-b)}\label{HMC134+612}

\begin{figure}[htbp]
\includegraphics[bb= 80 80 500 510, scale=0.4]{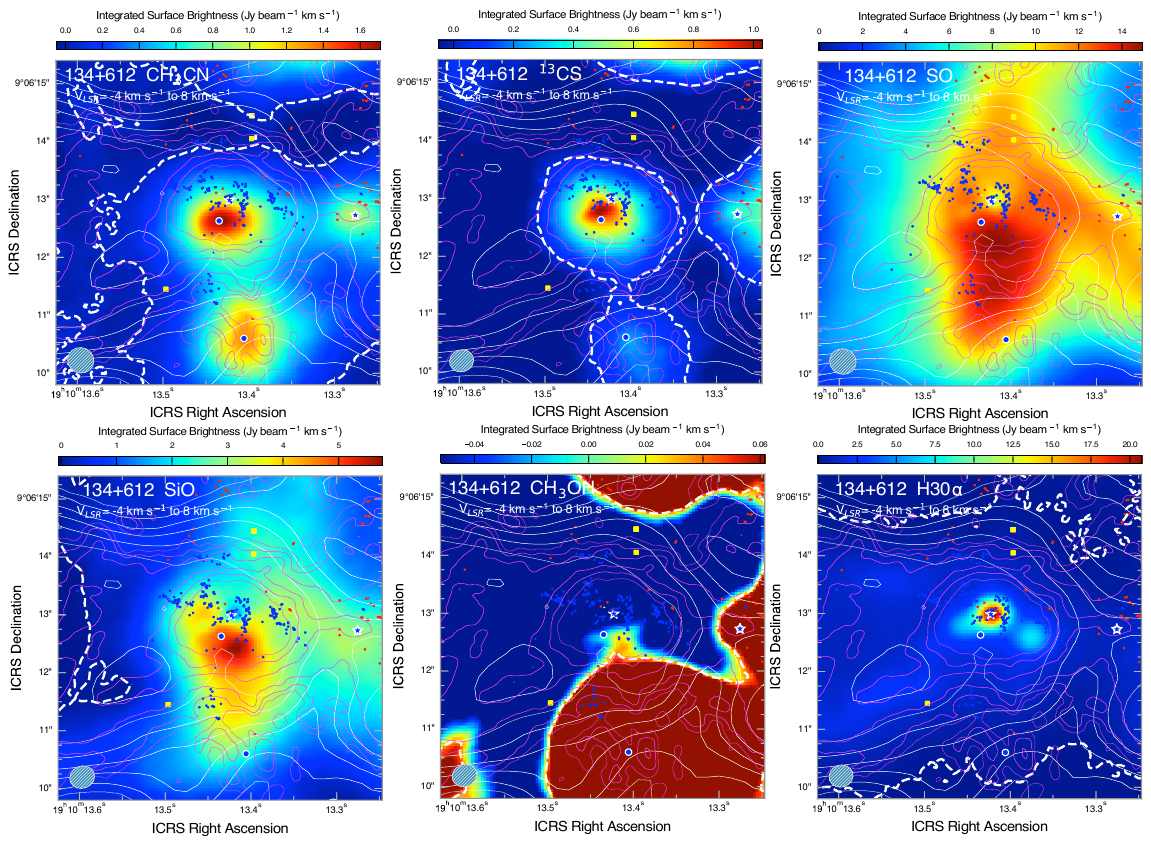}
\caption{Same as Fig~\ref{Fig109+517}, but for HMC134+612 (MCN-b) integrated from $V_{\rm LSR}$ = $-$4 to 8~km\,s$^{-1}$ for molecular lines.
{Alt text: Six local moment~0 maps centered on HMC134+612 (MCN-b) for CH$_3$CN, $^{13}$CS, SO, SiO, CH$_3$OH, and H30$\alpha$. Molecular lines are integrated over $V_{\rm LSR}$ = -4 to 8 km~s$^{-1}$, and H30$\alpha$ over $V_{\rm LSR}$ = -20 to +60 km~s$^{-1}$. Symbols: blue circle marks the CH$_3$CN peak; blue star marks the 1.3~mm continuum peak; magenta contours show 1.3~mm continuum (5-80\%); white contours show 3.6~cm continuum; red/blue dots mark OH/H$_2$O masers; yellow square/asterisk mark CH$_3$OH Class I/II masers.}}
\label{Fig134+612}
\end{figure}

Figure~\ref{Fig134+612} presents maps of HMC134+612, previously identified as MCN-b \citep{Wilner2001}.
The CH$_3$CN emission is distributed along the southern edge of UCHII~G2, which consists of G2a (N1), G2b, and G2c (N10).
The 1.3~mm continuum source C134+612 encompasses G2a and G2b; these two UCHII components are not resolved by our beam.
The CH$_3$CN peak of HMC134+612 is offset by $\sim$0\farcs4 to the south of C134+612, while UCHII~G2c is located $\sim$0\farcs4 northeast of the CH$_3$CN peak.
The HMC coincides with the mid-infrared source G:IRCS1, whose luminosity is estimated to be $\sim$3$\times$10$^{5}$~L$_\odot$ \citep{Smith2009}.

%:%%% Figure A18: HMC134+612  %%%
% Figure A18

The H30$\alpha$ emission line, both strong ($T_{\rm B} \sim 70$~K) and broad ($\Delta V = 59$~km,s$^{-1}$), closely follows the 1.3~mm continuum distribution of C134+612.
The low-level  H30$\alpha$ emission is elongated eastward, corresponding to G2c.
An independent peak located 1$''$ to the southwest is associated with G1, a shell-like UCHII region \citep{DePree2020}.
It remains uncertain whether HMC134+612 is associated with the H30$\alpha$ emission.
The absence of enhanced H30$\alpha$ intensity toward the peak of HMC134+612 may suggest that an HCHII/UCHII region has not yet developed at the center of the HMC.

The $^{13}$CS, HNCO, HC$_{3}$N, OCS, and SO$_{2}$ emissions are also compact, with the $^{13}$CS and HC$_{3}$N peaks shifted northward by $\sim$0\farcs2 from the CH$_3$CN peak.
In contrast, the SO and SiO emissions are more extended, with their peaks located south of the CH$_3$CN peak.
Both SO and SiO maps show reduced intensity toward UCHII~G2, forming a distinct emission cavity that suggests HMC134+612 is physically adjacent to the UCHII region.
CH$_3$OH emission is weak or absent toward the HMC.
%H$_2$O maser spots are scattered around UCHII~G1 and G2, and their characteristics will be discussed in detail in \S\ref\label{Individual HMCs} .
Three Class~I CH$_3$OH maser spots are located $\sim$1\farcs5 to the northwest and southeast of the HMC, in regions corresponding to the outflow gas traced by SiO and SO.

\subsection{HMC136+649}\label{HMC136+649}

\begin{figure}[htbp]
\includegraphics[bb= 60 50 500 490, scale=0.4]{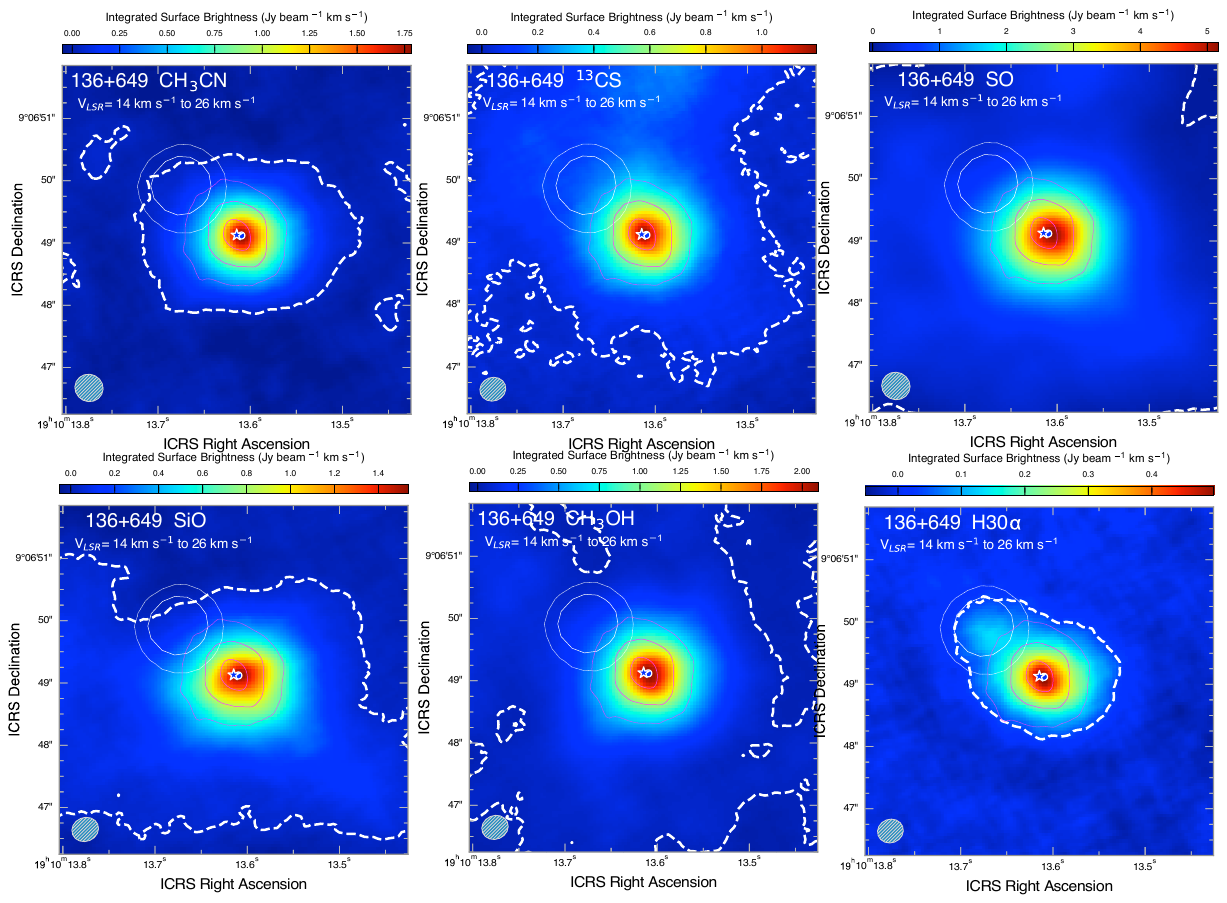}
\caption{Same as Fig~\ref{Fig109+517}, but for HMC136+649 integrated from $V_{\rm LSR}$ = 14 to 26~km\,s$^{-1}$ for molecular lines.
{Alt text: Six local moment~0 maps centered on HMC136+649 for CH$_3$CN, $^{13}$CS, SO, SiO, CH$_3$OH, and H30$\alpha$. Molecular lines are integrated over $V_{\rm LSR}$ = 14 to 26 km~s$^{-1}$, and H30$\alpha$ over $V_{\rm LSR}$ = -20 to +60 km~s$^{-1}$. Symbols: blue circle marks the CH$_3$CN peak; blue star marks the 1.3~mm continuum peak; magenta contours show 1.3~mm continuum (5-80\%); white contours show 3.6~cm continuum; red/blue dots mark OH/H$_2$O masers; yellow square/asterisk mark CH$_3$OH Class I/II masers.}
}
\label{Fig136+649}
\end{figure}

Figure~\ref{Fig136+649} shows maps of HMC136+649.
This is an isolated HMC coinciding with the 1.3~mm continuum source C136+649 and N6, located 1\farcs5 southwest of the 3.6~cm UCHII~GG.
The continuum emission from C136+649 was not detected at frequencies less than 1.3~mm (see Figure~\ref{Fig: SEDs}).
It has a rapidly rising SED (spectral index \ga4) toward higher frequency, suggesting the thermal dust origin of the 1.3~mm continuum.
All the molecular line emissions, as well as the 1.3~mm continuum emission, exhibit simple and compact spatial distributions, with their emission peaks closely coinciding.

%:%%% Figure A19: HMC136+649  %%%
% Figure A19

A weak emission line is detected at the H30$\alpha$ frequency.
Its distribution coincides with that of the 1.3~mm continuum, which originates from thermal dust.
The H50$\beta$ line is not detected toward this HMC.
The emission line toward HMC136+649 at the H30$\alpha$ frequency may therefore not be associated with ionized gas, but rather with COMs.

\subsection{HMC141+624 (MCN-a)}\label{HMC141+624}

\begin{figure}[htbp]
\includegraphics[bb= 80 70 500 490, scale=0.4]{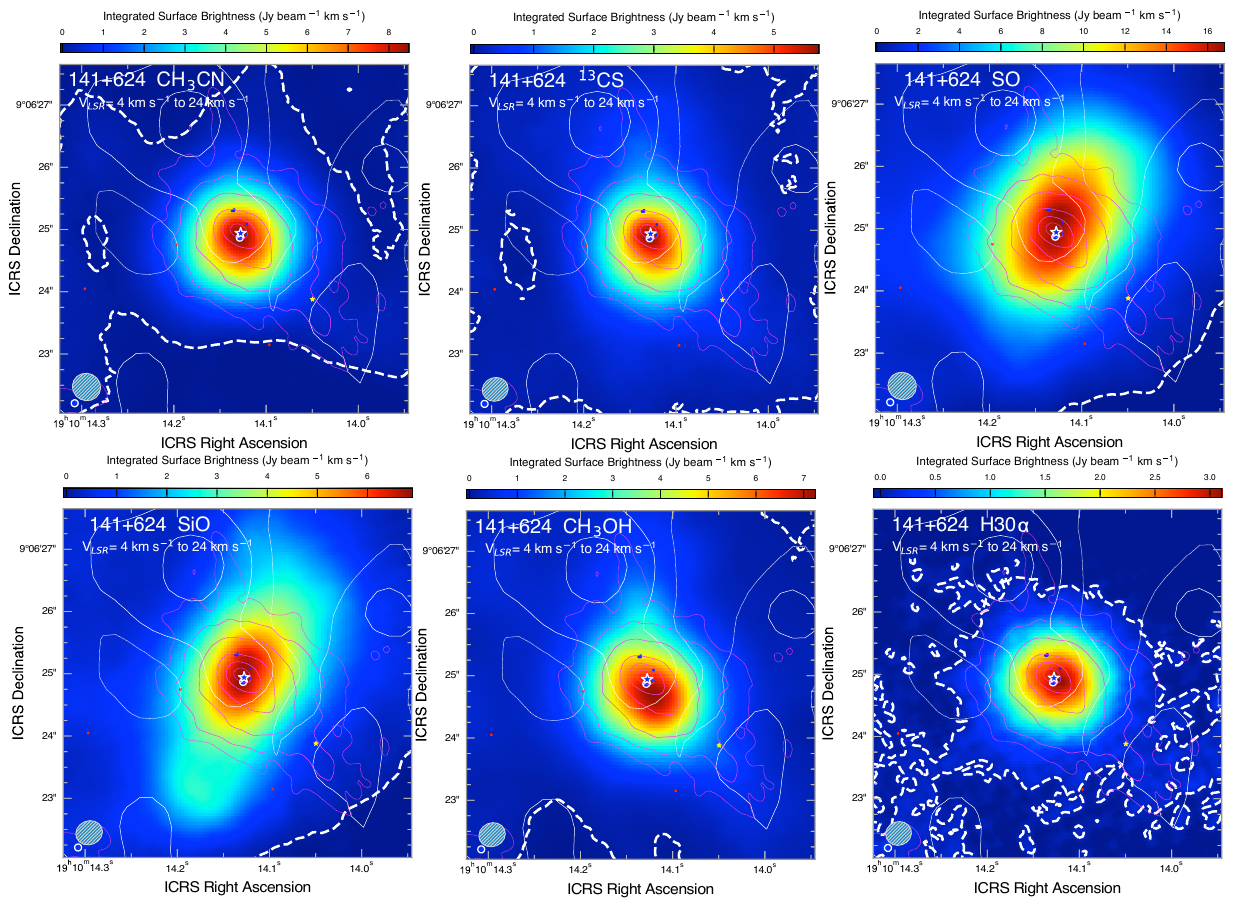}
\caption{Same as Fig~\ref{Fig109+517}, but for HMC141+624 (MCN-a) integrated from $V_{\rm LSR}$ = 4 to 24~km\,s$^{-1}$ for molecular lines.
{Alt text: Six local moment~0 maps centered on HMC141+624 (MCN-a) for CH$_3$CN, $^{13}$CS, SO, SiO, CH$_3$OH, and H30$\alpha$. Molecular lines are integrated over $V_{\rm LSR}$ = 4 to 24 km~s$^{-1}$, and H30$\alpha$ over $V_{\rm LSR}$ = -20 to +60 km~s$^{-1}$. Symbols: blue circle marks the CH$_3$CN peak; blue star marks the 1.3~mm continuum peak; magenta contours show 1.3~mm continuum (5-80\%); white contours show 3.6~cm continuum; red/blue dots mark OH/H$_2$O masers; yellow square/asterisk mark CH$_3$OH Class I/II masers.}
}
\label{Fig141+624}
\end{figure}

Figure~\ref{Fig141+624} shows maps of HMC141+624, which was identified as MCN-a by \citet{Wilner2001}.
This HMC corresponds to the 226 GHz continuum source C141+624 and to N8 identified by \citet{Nony2024}, who also listed three weaker 1.3~mm continuum cores (N75, N104, and N133) around this source.
These weaker cores have 226 GHz flux densities more than ten times lower than that of N8 and are not apparent as distinct peaks in our 226 GHz continuum map.

%:%%% Figure A20: HMC141+624  %%%
% Figure A20

% CH$_3$CN, $^{13}$CS, SO, SiO, CH$_3$OH, and H30$\alpha$ lines.
This source was extensively studied in a previous paper \citep{Miyawaki2022b}, which revealed a gravitationally unstable, rotating torus of molecular gas, within which a central accreting disk may reside.
All molecular lines show compact and isolated distributions, with their peaks coinciding with each other.
The SO and SiO emissions are elongated along the northwest-southeast direction, likely influenced by bipolar outflows originating from the central region \citep{Miyawaki2022b}.

The line detected at the nominal H30$\alpha$ frequency exhibits a spatial distribution similar to those of the molecular lines.
Although this HMC corresponds to the weak 3.6~cm UCHII region J1, the spectral feature is not the recombination line H30$\alpha$, as discussed in \citet{Miyawaki2022b}.
Several H$_2$O maser spots are associated with the HMC, some lying within the beam size of the peak position.
In contrast, the OH maser spots are located more than 1$''$ away, and their association with the HMC remains uncertain.

\subsection{HMC143+622}\label{HMC143+622}

\begin{figure}[htbp]
\includegraphics[bb= 80 65 500 500, scale=0.4]{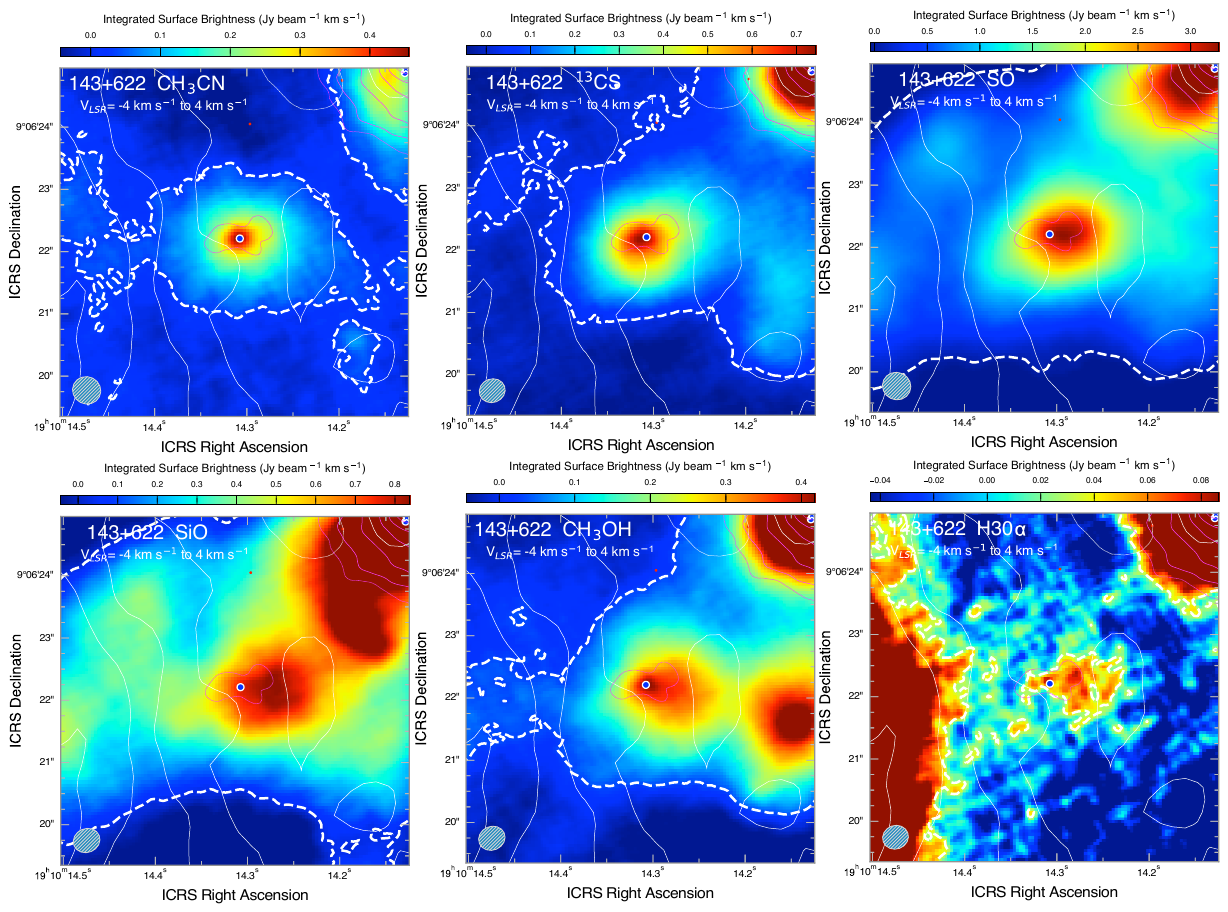}
\caption{Same as Fig~\ref{Fig109+517}, but for HMC143+622 integrated from $V_{\rm LSR}$ = $-$4 to 4~km\,s$^{-1}$ for molecular lines.
{Alt text: Six local moment~0 maps centered on HMC143+622 for CH$_3$CN, $^{13}$CS, SO, SiO, CH$_3$OH, and H30$\alpha$. Molecular lines are integrated over $V_{\rm LSR}$ = -4 to 4 km~s$^{-1}$, and H30$\alpha$ over $V_{\rm LSR}$ = -20 to +60 km~s$^{-1}$. Symbols: blue circle marks the CH$_3$CN peak; blue star marks the 1.3~mm continuum peak; magenta contours show 1.3~mm continuum (5-80\%); white contours show 3.6~cm continuum; red/blue dots mark OH/H$_2$O masers; yellow square/asterisk mark CH$_3$OH Class I/II masers.}
}
\label{Fig143+622}
\end{figure}

Figure~\ref{Fig143+622} shows maps of HMC143+622. 
This source corresponds to N31 in \citet{Nony2024}, although it was not identified as an HMC in their study. 
The 1.3~mm continuum emission, shown in magenta contours in Figure~\ref{Fig143+622}, has a marginal peak coincident with the HMC peak. 
This feature, [C143+622], has a peak brightness of 12~mJy\,beam$^{-1}$ (with the surrounding noise level of 4.3~mJy\,beam$^{-1}$) and is elongated in the east-west direction. 
The 1.3~mm continuum contour shows a protrusion toward the northwest. 
This location corresponds to the position of another 1.3~mm continuum core, N24, located 0\farcs4 west of the CH$_{3}$CN peak.

%:%%% Figure A21: HMC143+622  %%%
% Figure A21

A weak and relatively broad emission line is detected at the nominal H30$\alpha$ frequency.
Its spatial distribution follows that of the 226 GHz emission, but not the 3.6~cm continuum, suggesting that it may not be a hydrogen recombination line.

The emission features of CH$_3$CN, $^{13}$CS, CH$_3$OH, HNCO, HC$_3$N, OCS, and H$_2$CO are compact, with their peaks coinciding with the weak 1.3~mm continuum emission.
In contrast, the emission features of the outflow tracers SO and SiO and the extended gas tracers DCN and C$^{18}$O are shifted westward by $\sim$1$''$.
An OH maser spot is located about 2$''$ north of the HMC, but its association with the source remains uncertain.

\subsection{HMC146+625}\label{HMC146+625}

\begin{figure}[htbp]
\includegraphics[bb= 60 80 500 520, scale=0.4]{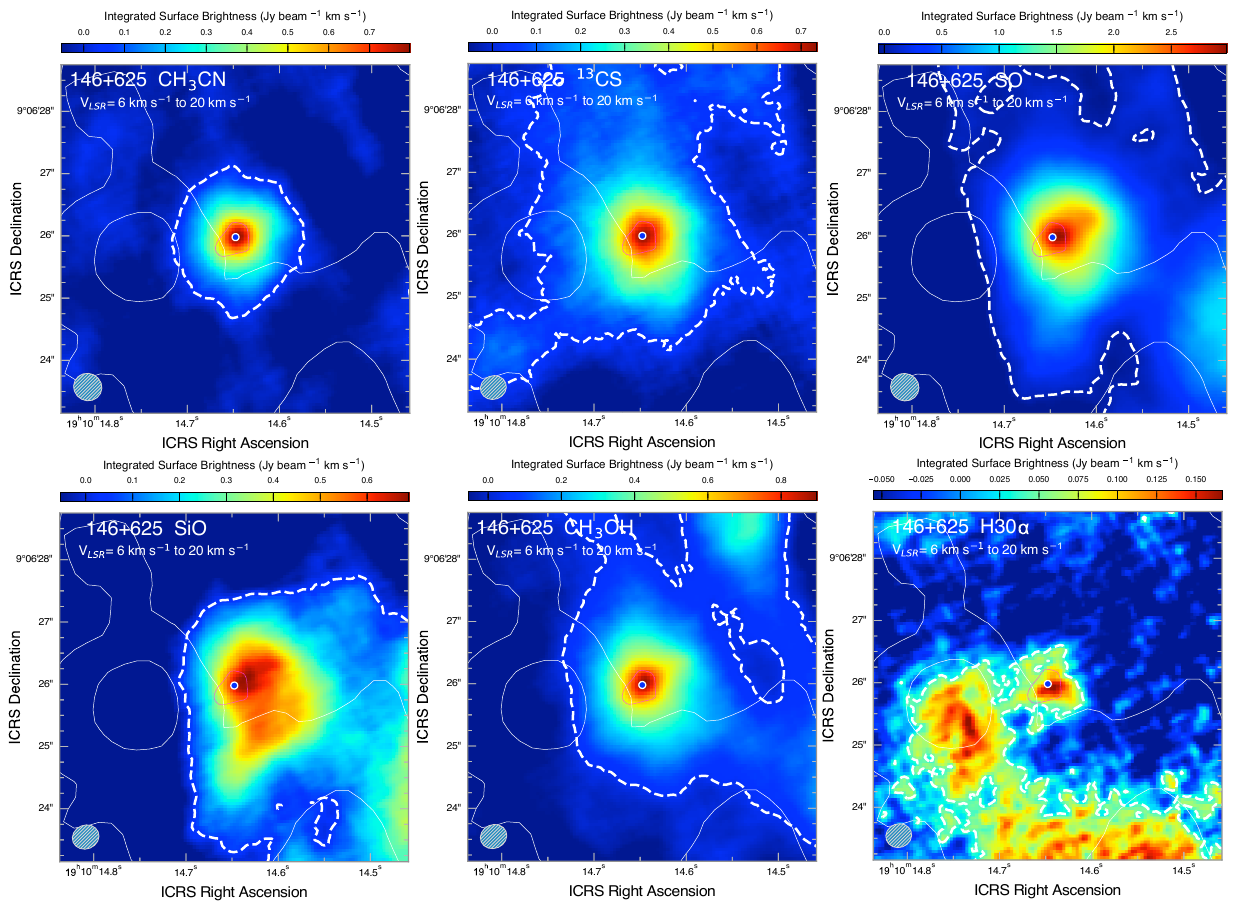}
\caption{
Same as Fig~\ref{Fig109+517}, but for HMC146+625 integrated from $V_{\rm LSR}$ = 6 to 20~km\,s$^{-1}$ for molecular lines.
{Alt text: Six local moment~0 maps centered on HMC146+625 for CH$_3$CN, $^{13}$CS, SO, SiO, CH$_3$OH, and H30$\alpha$. Molecular lines are integrated over $V_{\rm LSR}$ = 6 to 10 km~s$^{-1}$, and H30$\alpha$ over $V_{\rm LSR}$ = -20 to +60 km~s$^{-1}$. Symbols: blue circle marks the CH$_3$CN peak; blue star marks the 1.3~mm continuum peak; magenta contours show 1.3~mm continuum (5-80\%); white contours show 3.6~cm continuum; red/blue dots mark OH/H$_2$O masers; yellow square/asterisk mark CH$_3$OH Class I/II masers.}
}
\label{Fig146+625}
\end{figure}

Figure~\ref{Fig146+625} shows maps of HMC146+625, located 3$''$ west of the 3.6~cm UCHII region M.
Its peak position coincides with the 226 GHz continuum source [C146+625], shown in the magenta contours in Figure\ref{Fig146+625}, or with N19 within 0\farcs1. 
Its fitted peak brightness is 12~mJy\,beam$^{-1}$. 
A weak emission line is detected at the nominal H30$\alpha$ frequency. 
It may not be a hydrogen recombination line, as its line width is narrow ($\Delta V_{\rm FWHM}=8.2$~km\,s$^{-1}$).

All molecular lines exhibit simple and compact spatial distributions,
with their peak positions coinciding with each other except for SO and SiO.
The SO and SiO emissions show peaks shifted by 0\farcs1-0\farcs4 to the west or northwest of the CH$_3$CN peak, with weaker emission extending toward the northwest and south.
No maser emission has been detected toward this HMC.

%:%%% Figure A22: HMC146+625  %%%
% Figure A22

\citet{Nony2024} identified two additional 1.3~mm continuum cores, N63 and N42, within the CH$_{3}$CN distribution of the HMC. 
N63 is located 0\farcs7 northwest of HMC146+625. 
The SO and SiO emissions are also extended in this direction, but their physical relationship with N63 remains unknown. 
N42 is located 0\farcs5 southeast of the HMC. No enhancement in molecular line or continuum emission is seen toward this core.

\subsection{HMC153+615}\label{HMC153+615}

\begin{figure}[htbp]
\includegraphics[bb= 80 80 500 510, scale=0.408]{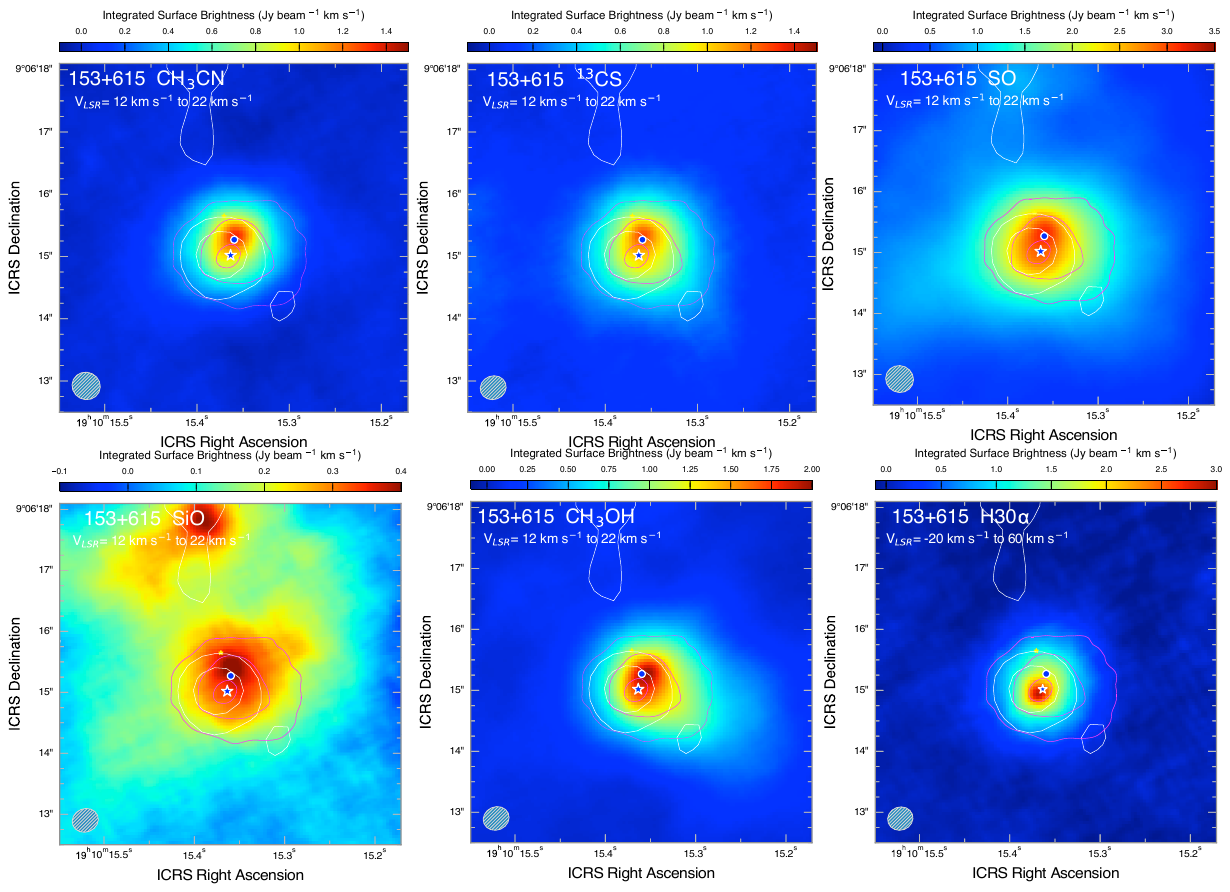}
\caption{
Same as Fig~\ref{Fig109+517}, but for HMC153+615 integrated from $V_{\rm LSR}$ = 12 to 22~km\,s$^{-1}$ for molecular lines.
{Alt text: Six local moment~0 maps centered on HMC153+615 for CH$_3$CN, $^{13}$CS, SO, SiO, CH$_3$OH, and H30$\alpha$. Molecular lines are integrated over $V_{\rm LSR}$ = 12 to 22 km~s$^{-1}$, and H30$\alpha$ over $V_{\rm LSR}$ = -20 to +60 km~s$^{-1}$. Symbols: blue circle marks the CH$_3$CN peak; blue star marks the 1.3~mm continuum peak; magenta contours show 1.3~mm continuum (5-80\%); white contours show 3.6~cm continuum; red/blue dots mark OH/H$_2$O masers; yellow square/asterisk mark CH$_3$OH Class I/II masers.}
}
\label{Fig153+615}
\end{figure}

Figure~\ref{Fig153+615} shows maps of HMC153+615, which coincides with N25 within the beam size.
The 1.3~mm continuum source C153+614, corresponding to N11 or UCHII~N, is located $\sim0\farcs3$ south of the HMC peak.
The distribution of the nominal H30$\alpha$ emission is consistent with those of the 8~GHz (UCHII~N) and N11 continuum emissions, suggesting that the line primarily originates from ionized gas.
The excess 1.3~mm continuum emission observed toward HMC153+615, located 0\farcs3 north of C153+614, may be partially associated with the HMC and originate from hot dust, since the flux density of C153+614 integrated over its solid angle exhibits a rapidly rising SED between 93~GHz and 1.3~mm.

All molecular lines except SiO, DCN and C$^{18}$O exhibit simple and compact emission features, with their peaks coinciding with the CH$_{3}$CN peak.
The maps of SiO, DCN and C$^{18}$O share the same peak position as the other molecular lines, but have additional features at the north or east of the main feature.
Class~II CH$_3$OH maser spots are located about 0$\farcs$8 north of the continuum peak.

%:%%% Figure A23: HMC153+615  %%%
% Figure A23

\subsection{HMC163+606}\label{HMC163+606 }

\begin{figure}[htbp]
\includegraphics[bb= 60 50 500 500, scale=0.4]{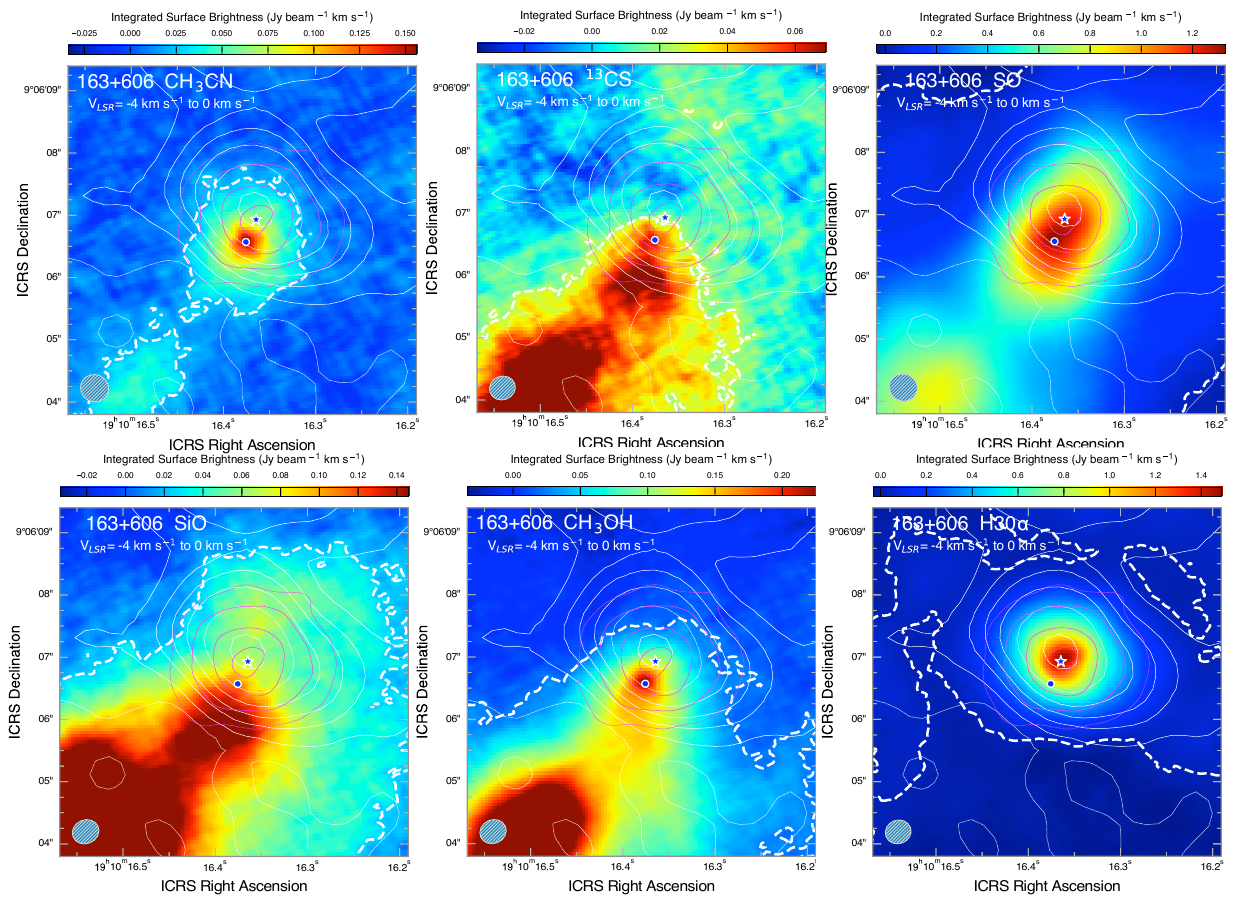}
\caption{
Same as Fig~\ref{Fig109+517}, but for HMC163+606 integrated from $V_{\rm LSR}$ = $-$4 to 0~km\,s$^{-1}$ for molecular lines.
{Alt text: Six local moment~0 maps centered on HMC163+606 for CH$_3$CN, $^{13}$CS, SO, SiO, CH$_3$OH, and H30$\alpha$. Molecular lines are integrated over $V_{\rm LSR}$ = -4 to 0 km~s$^{-1}$, and H30$\alpha$ over $V_{\rm LSR}$ = -20 to +60 km~s$^{-1}$. Symbols: blue circle marks the CH$_3$CN peak; blue star marks the 1.3~mm continuum peak; magenta contours show 1.3~mm continuum (5-80\%); white contours show 3.6~cm continuum; red/blue dots mark OH/H$_2$O masers; yellow square/asterisk mark CH$_3$OH Class I/II masers.}
}
\label{Fig163+606}
\end{figure}

Figure~\ref{Fig163+606} presents maps of HMC163+606.
The CH$_3$CN emission exhibits a compact feature located 0$\farcs$4 south of the 1.3~mm continuum source C163+606, which corresponds to N12 and may also include N5, situated 0$\farcs$24 northwest of C163+606.
The 3.6~cm continuum (UCHII~O) coincides with C163+606, although the 3.6~cm peak is slightly shifted ($\sim$0$\farcs$2) northward, close to N5.
The H30$\alpha$ emission, with a broad linewidth of $\Delta V_{\rm FWHM}=30$~km\,s$^{-1}$, shows a good spatial correspondence with the 3.6~cm and 1.3~mm continuum distributions, suggesting that the 1.3~mm emission of C163+606 primarily originates from ionized gas.
The HMC appears to be unaffected by C163+606, and the two may represent independent sources.
It remains, however, uncertain whether HMC163+606 is associated with any 1.3~mm continuum emission, either of free-free or dust origin.

The $^{13}$CS, SO, CH$_3$OH, HC$_3$N, OCS, and H$_2$CO emissions peak at the same position as CH$_3$CN, whereas the SiO, HNCO, DCN, and C$^{18}$O emissions show no corresponding peaks.
Additional peaks and/or extended emission features are present in several molecular lines to the southeast of this HMC.

%:%%% Figure A24: HMC163+606  %%%
% Figure A24

%:Appendix 4
\section{Tables and Maps of Additional Molecular Lines}
\label{Additional tables and maps}
%\section{The HMC parameters of various molecules}
Tables~A1-A12 present the HMC parameters, analogous to those in Table~\ref{Table: HMC candidates}, for molecular lines other than CH$_{3}$CN and H30$\alpha$.
The positional offsets are measured relative to the CH$_{3}$CN peak positions.
In Table~A5, the FWHM line widths are derived from the emission detected at the H30$\alpha$ frequency.
Figures~\ref{Fig:Other_mol_lines_map01} and \ref{Fig:Other_mol_lines_map02}  show the maps of HNCO, HC$_3$N, OCS, H$_2$CO, SO$_2$, DCN, and C$^{18}$O for the 18 HMCs.
\clearpage
%*********************

%:%%% Table 13CS %%
\begin{table}[htbp]
\caption{$^{13}$CS HMC properties}
\label{Table:13CS_HMC_properties}
\begin{center}
\scalebox{0.5}[0.5]
{
\begin{tabular}{cccccccl}

\hline\hline
Name & \multicolumn{2}{c}{Position} & Brightness &  FWHM Size & PA & Line & Remarks \\
 & $\Delta$RA & $\Delta$DEC & fitted peak & major  $\times$ minor  && Luminosity &  \\
 & (arcsec) & (arcsec) &($\mathrm{Jy\,beam^{-1}\,km\,s^{-1}}$) & ($''\,\times\,''$) & (degree) & (L$_\odot$) &  \\
 \hline

HMC109+517 & -0.08 & -0.03 & 0.423 & 1.25  $\times$ 1.01 & 49.6 & 1.49$\times$10$^{-4}$ \\
HMC110+520 & -0.02 & -0.14 & 0.176 & 1.56  $\times$ 1.42 & 12.6 & 1.09$\times$10$^{-4}$ \\
HMC120+614 & 0.11 & -0.02 & 0.175 & 1.06  $\times$ 0.89 & 71.1 & 4.60$\times$10$^{-5}$ \\
HMC127+611 & -0.01 & 0.02 & 1.338 & 1.58  $\times$ 1.42 & 81.5 & 8.38$\times$10$^{-4}$ \\
HMC131+618 & -0.06 & 0.03 & 1.213 & 1.94  $\times$ 1.32 & 38.8 & 8.67$\times$10$^{-4}$ \\
HMC131+613 & -0.05 & 0.32 & 3.159 & 2.73  $\times$ 1.44 & 142.9 & 3.46$\times$10$^{-3}$ \\
HMC131+614 & -0.15 & -0.32 & 0.960 & 1.83  $\times$ 1.38 & 166.6 & 6.76$\times$10$^{-4}$ \\
HMC132+612 & -0.10 & 0.04 & 0.487 & 1.41  $\times$ 1.01 & 87.1 & 1.93$\times$10$^{-4}$ \\
HMC133+619 & 0.21 & 0.00 & 0.717 & 1.76  $\times$ 1.26 & 99.1 & 4.44$\times$10$^{-4}$ \\
HMC133+616 & 0.24 & 0.15 & 1.927 & 1.72  $\times$ 1.26 & 108.4 & 1.17$\times$10$^{-3}$ \\
HMC134+610 & -0.04 & -0.06 & 0.224 & 1.11  $\times$ 0.93 & 28.9 & 6.45$\times$10$^{-5}$ \\
HMC134+612 & 0.08 & 0.17 & 0.922 & 1.15  $\times$ 1.06 & 59.3 & 3.13$\times$10$^{-4}$ \\
HMC136+649 & 0.01 & 0.03 & 1.085 & 1.15  $\times$ 1.06 & 59.3 & 3.69$\times$10$^{-4}$ \\
HMC141+624 & -0.03 & 0.00 & 5.511 & 1.38  $\times$ 1.24 & 28.5 & 2.63$\times$10$^{-3}$ \\
HMC143+622 & 0.07 & -0.08 & 0.681 & 1.26  $\times$ 1.07 & 108.6 & 2.56$\times$10$^{-4}$ \\
HMC146+625 & -0.02 & -0.01 & 0.678 & 1.26  $\times$ 1.13 & 0.8 & 2.69$\times$10$^{-4}$ \\
HMC153+615 & -0.02 & -0.04 & 1.179 & 1.29  $\times$ 1.18 & 164.6 & 5.01$\times$10$^{-4}$ \\
HMC163+606 & 0.30 & -0.85 & 0.075 & 2.20  $\times$ 1.18 & 161.7 & 5.40$\times$10$^{-5}$ \\
\hline\noalign{\vspace{1pt}}
Average& & & &  1.55  $\times$ 1.19    & &  7.21$\times$10$^{-4}$  \\
\hline

\end{tabular}
}
\end{center}

\end{table}

%:%%% Table SO %%
\begin{table}[htbp]
\caption{SO HMC properties}
\label{Table:SO_HMC_properties}
\begin{center}
\scalebox{0.5}[0.5]
{
\begin{tabular}{cccccccl}

\hline\hline
Name & \multicolumn{2}{c}{Position}  & Brightness & FWHM Size & PA & Line & Remarks \\
 & $\Delta$RA & $\Delta$DEC & fitted peak & major  $\times$ minor  && Luminosity &  \\
 & (arcsec) & (arcsec) & (Jy beam$^{-1}$ km\, s$^{-1}$ )& ($''\,\times\,''$) & (degree) & (L$_\odot$) &  \\
 \hline

HMC109+517 & -0.03 & 0.07 & 5.4574 & 1.46 $\times$ 1.31 & 56.4 & 2.91$\times$10$^{-3}$ \\
HMC110+520 & -0.06 & -0.25 & 1.2018 & 1.41 $\times$ 1.22 & 176.8 & 5.77$\times$10$^{-4}$ \\
HMC120+614 & -0.45 & -0.21 & 0.455 & 1.80 $\times$ 1.55 & 70.6 & 3.54$\times$10$^{-4}$ \\
HMC127+611 & 0.08 & 0.05 & 4.3574 & 1.47 $\times$ 1.29 & 61.1 & 2.31$\times$10$^{-3}$ \\
HMC131+618 & -0.26 & 0.22 & 0.863 & 2.46 $\times$ 0.79 & 36.8 & 4.68$\times$10$^{-4}$ \\
HMC131+613 & 0.15 & 0.17 & 19.008 & 2.62 $\times$ 2.33 & 92.7 & 3.24$\times$10$^{-2}$ \\
HMC131+614 & 0.22 & -0.12 & 8.8271 & 2.96 $\times$ 2.13 & 96.4 & 1.55$\times$10$^{-2}$ \\
HMC132+612 & -0.63 & -0.51 & 10.4449 & 6.88 $\times$ 3.20 & 69.9 & 6.42$\times$10$^{-2}$ \\
HMC133+619 & --- & --- & --- &--- & --- & $<$4.12$\times$10$^{-4}$ \\
HMC133+616 & 0.60 & -0.30 & 9.6109 & 3.55 $\times$ 2.33 & 134.1 & 2.22$\times$10$^{-2}$ \\
HMC134+610 & 0.20 & 1.00 & 12.5327 & 3.86 $\times$ 2.04 & 159.6 & 2.75$\times$10$^{-2}$ \\
HMC134+612 & -0.22 & -0.44 & 14.9199 & 3.15 $\times$ 2.60 & 72.7 & 3.41$\times$10$^{-2}$ \\
HMC136+649 & -0.05 & -0.02 & 4.9369 & 1.30 $\times$ 1.15 & 86.9 & 2.06$\times$10$^{-3}$ \\
HMC141+624 & -0.04 & 0.22 & 16.6894 & 2.77$\times$ 1.67 & 157.8 & 2.15$\times$10$^{-2}$ \\
HMC143+622 & -0.26 & 0.00 & 3.2069 & 1.80 $\times$ 1.34 & 97.8 & 2.16$\times$10$^{-3}$ \\
HMC146+625 & -0.16 & 0.06 & 2.7586 & 1.34 $\times$ 1.25 & 133.6 & 1.29$\times$10$^{-3}$ \\
HMC153+615 & 0.04 & -0.09 & 3.0656 & 1.67 $\times$ 1.48 & 159.8 & 2.11$\times$10$^{-3}$ \\
HMC163+606 & -0.07 & 0.10 & 1.2938 & 2.14 $\times$ 1.46 & 143.5 & 1.13$\times$10$^{-3}$ \\

\hline\noalign{\vspace{1pt}}
Average& & & &  2.51  $\times$ 1.71    &&  1.37$\times$10$^{-2}$  \\
\hline

\end{tabular}
}
\end{center}

\end{table}

%:%%% Table SiO %%
\begin{table}[htbp]
\caption{SiO HMC properties}
\label{Table:SiO_HMC_properties}
\begin{center}
\scalebox{0.5}[0.5]
{
\begin{tabular}{cccccccl}

\hline\hline
Name & \multicolumn{2}{c}{Position} & Brightness & FWHM Size & PA & Line & Remarks \\
 & $\Delta$RA & $\Delta$DEC & fitted peak & major  $\times$ minor  && Luminosity &  \\
 & (arcsec) & (arcsec) & (Jy beam$^{-1}$ km\, s$^{-1}$ )& ($''\,\times\,''$) & (degree) & (L$_\odot$) &  \\
 \hline
HMC109+517 & 0.02 & 0.06 & 0.7955 & 1.57 $\times$ 1.48 & 82.6 & 5.16$\times$10$^{-4}$ \\
HMC110+520 & -0.03 & -0.62 & 0.2181 & 1.78 $\times$ 1.38 & 139.6 & 1.49$\times$10$^{-4}$ \\
HMC120+614 & -0.79 & -0.29 & 0.1572 & 2.50 $\times$ 1.26 & 138.9 & 1.38$\times$10$^{-4}$ \\
HMC127+611 & 0.31 & 0.00 & 1.1464 & 1.62 $\times$ 1.27 & 82.1 & 6.58$\times$10$^{-4}$ \\
HMC131+618 &--- & --- & --- &--- & --- & $<$ 4.19 $\times$10$^{-4}$ & Cavity\\
HMC131+613 & 0.03 & 0.34 & 7.1817 & 2.31 $\times$ 2.17 & 116.1 & 1.00$\times$10$^{-2}$ \\
HMC131+614 & -0.35 & -0.04 & 3.3352 & 2.88 $\times$ 1.90 & 112.0 & 5.09$\times$10$^{-3}$ \\
HMC132+612 & -0.52 & -0.23 & 2.7400 & 4.19 $\times$ 3.27 & 24.2 & 1.05$\times$10$^{-2}$ \\
HMC133+619 & -0.40 & 0.55 & 0.4046 & 2.82$\times$ 1.59 & 69.8 & 5.06$\times$10$^{-4}$ \\
HMC133+616 & 0.66 & 0.42 & 1.7593 & 3.39$\times$ 2.25 & 88.8 & 3.74$\times$10$^{-3}$ \\
HMC134+610 & 0.14 & 1.05 & 3.6180 & 2.82 $\times$ 1.59 & 160.5 & 4.53$\times$10$^{-3}$ \\
HMC134+612 & -0.12 & -0.24 & 5.4842 & 1.73 $\times$ 1.41 & 79.7 & 3.73$\times$10$^{-3}$ \\
HMC136+649 & -0.04 & -0.03 & 1.4201 & 1.30 $\times$ 1.04 & 98.0 & 5.36$\times$10$^{-4}$ \\
HMC141+624 & -0.01 & 0.16 & 6.8172 & 2.17 $\times$ 1.35 & 159.5 & 5.57$\times$10$^{-3}$ \\
HMC143+622 & -0.55 & -0.04 & 0.8431 & 2.55 $\times$ 1.87 & 108.8 & 1.12$\times$10$^{-3}$ \\
HMC146+625 & -0.30 & 0.11 & 0.6744 & 1.58 $\times$ 1.37 & 156.0 & 4.07$\times$10$^{-4}$ \\
HMC153+615 & -0.05 & 0.11 & 0.3927 & 1.75 $\times$ 1.65 & 24.6 & 3.16$\times$10$^{-4}$ \\
HMC163+606 & 0.31 & -0.83 & 0.1717 & 2.05 $\times$ 1.85 & 156.1 & 1.82$\times$10$^{-4}$ \\
\hline
\noalign{\vspace{1pt}}
Average& & & &  2.29  $\times$ 1.69    &&  2.81$\times$10$^{-3}$  \\
\hline

\end{tabular}
}
\end{center}

\end{table}

%:%%% Table CH3OH %%
\begin{table}[htbp]
\caption{CH$_3$OH HMC properties}
\label{Table:CH3OH_HMC_properties}
\begin{center}
\scalebox{0.5}[0.5]
{
\begin{tabular}{cccccccl}

\hline\hline
Name & \multicolumn{2}{c}{Position}  & Brightness & FWHM Size & PA & Line & Remarks \\
 & $\Delta$RA & $\Delta$DEC & fitted peak & major  $\times$ minor  && Luminosity &  \\
 & (arcsec) & (arcsec) & (Jy beam$^{-1}$ km\, s$^{-1}$ )& ($''\,\times\,''$) & (degree) & (L$_\odot$) &  \\
 \hline
HMC109+517 & 0.03 & -0.02 & 0.7955 & 1.17 $\times$ 0.95 & 58.5 & 3.55$\times$10$^{-4}$ \\
HMC110+520 & 0.02 & -0.06 & 0.2181 & 1.09 $\times$ 0.99 & 140.2 & 1.46$\times$10$^{-4}$ \\
HMC120+614 & -0.07 & -0.07 & 0.1572 & 1.17 $\times$ 0.97 & 67.3 & 1.30$\times$10$^{-4}$ \\
HMC127+611 & 0.08 & -0.14 & 1.1464 & 1.61 $\times$ 1.42 & 107.7 & 1.99$\times$10$^{-3}$ \\
HMC131+618 & -0.13 & 0.04 & 5.0736 & 1.29 $\times$ 1.07 & 38.3 & 5.23$\times$10$^{-4}$ \\
HMC131+613 & 0.11 & 0.25 & 7.1817 & 1.88 $\times$ 1.55 & 115.6 & 4.31$\times$10$^{-3}$ \\
HMC131+614 & 0.01 & -0.21 & 3.3352 & 1.42 $\times$ 1.34 & 3.4 & 1.69$\times$10$^{-3}$ \\
HMC132+612 & -0.47 & 0.15 & 2.7400 & 1.51 $\times$ 1.01 & 104.7 & 1.90$\times$10$^{-4}$ \\
HMC133+619 & 0.75 & 0.18 & 0.4046 & 3.18 $\times$ 1.53 & 91.1 & 1.07$\times$10$^{-3}$ \\
HMC133+616 & 0.28 & -0.05 & 1.7593 & 1.35 $\times$ 1.07 & 96.8 & 6.31$\times$10$^{-4}$ \\
HMC134+610 & -0.24 & 0.03 & 3.6180 & 1.65 $\times$ 1.29 & 100.5 & 2.27$\times$10$^{-3}$ \\
HMC134+612 & --- & ---&  --- &  --- &  ---&  --- \\
HMC136+649 & -0.02 & 0.00 & 1.4201 & 1.65 $\times$ 1.39 & 74.7 & 1.30$\times$10$^{-3}$ \\
HMC141+624 & -0.07 & -0.13 & 6.8172 & 1.65 $\times$ 1.39 & 46.3 & 4.63$\times$10$^{-3}$ \\
HMC143+622 & -0.23 & -0.04 & 0.8431 & 1.47 $\times$ 1.21 & 98.4 & 1.96$\times$10$^{-4}$ \\
HMC146+625 & -0.02 & 0.04 & 0.6744 & 1.16 $\times$ 1.07 & 147.1 & 2.76$\times$10$^{-4}$ \\
HMC153+615 & -0.09 & -0.11 & 0.3927 & 1.26 $\times$ 1.13 & 162.4 & 7.54$\times$10$^{-4}$ \\
HMC163+606 & 0.07 & -0.11 & 0.1717 & 1.30 $\times$ 1.04 & 144.1 & 7.21$\times$10$^{-5}$ \\
\hline
\noalign{\vspace{1pt}}
Average& & & &  1.52  $\times$ 1.18    &&  1.16$\times$10$^{-3}$  \\
\hline

\end{tabular}
}
\end{center}

\end{table}

\clearpage

%:%%% Table H30alpha %%

\refstepcounter{table}
\label{Table:H30alpha_HMC_properties}

\begin{center}
{\bfseries Table \thetable. H30$\alpha$ HMC properties}

\vspace{1ex}

\scalebox{0.4}[0.4]
{
\begin{tabular}{ccccccccl}

\hline\hline
Name & \multicolumn{2}{c}{Position}
 & Brightness & FWHM Size & PA & Line
 & $\Delta V$(H30$\alpha$) & \hfill Remarks \\
 & $\Delta$RA & $\Delta$DEC
 & fitted peak & major $\times$ minor
 & & Luminosity & & \\
 & (arcsec) & (arcsec)
 & (Jy beam$^{-1}$ km\,s$^{-1}$)
 & ($''\,\times\,''$)
 & (degree) & (L$_\odot$)
 & (km\,s$^{-1}$) & \\
\hline
HMC109+517 & -0.01 & 0.11 & 9.2607 & 0.82$\times$ 0.78 & 98.0 & 1.19$\times$10$^{-3}$ & 39 &\\
HMC110+520 & -0.10 & -0.07 & 4.063 & 0.92 $\times$ 0.79 & 119.2 & 7.40$\times$10$^{-4}$ & 29 &\\
HMC120+614 & 0.22 & 0.24 & 0.7795 & 1.15 $\times$ 0.96 & 98.1 & 1.79$\times$10$^{-4}$ & 37 & Overlapping narrow lines\\
HMC127+611 & -0.04 & 0.15 & 1.968 & 1.30 $\times$ 0.64 & 136.5 & 3.93$\times$10$^{-4}$ & 29 &\\
HMC131+618 & 0.26 & -0.10 & 12.2732 & 0.86 $\times$ 0.85 & 100.7 & 2.10$\times$10$^{-3}$ & 34 &\\
HMC131+613 & -0.08 & -0.47 & 52.2121 & 0.75 $\times$ 0.68 & 53.8 & 6.42$\times$10$^{-3}$ & 37 &\\
HMC131+614 &--- & --- &  --- &  --- &  --- &  $<$1.58$\times$10$^{-3}$ & 8.6 &\\
HMC132+612 &--- & --- &  --- &  --- &  --- & $<$8.04$\times$10$^{-4}$ & 14 &\\
HMC133+619 & --- & --- &  --- &  --- &  --- & $<$5.54$\times$10$^{-5}$ & --- & Absorption\\
HMC133+616 & -0.16 & -0.50 & 0.7475 & 1.68 $\times$ 1.17 & 11.0 & 4.89$\times$10$^{-4}$ & 6.0 &\\
HMC134+610 & --- & --- &  --- &  --- &  --- &  $<$4.87$\times$10$^{-4}$ & 31 &\\
HMC134+612 & -0.12 & 0.36 & 96.036 & 0.62 $\times$ 0.50 & 85.9 & 7.52$\times$10$^{-3}$ & 59 &\\
HMC136+649 & -0.04 & -0.04 & 1.5784 & 1.13 $\times$ 0.91 & 85.9 & 4.23$\times$10$^{-4}$ & 13 &\\
HMC141+624 & -0.02 & 0.07 & 8.754 & 1.38 $\times$ 1.26 & 74.6 & 4.17$\times$10$^{-3}$ & 11 &\\
HMC143+622 & -0.24 & 0.12 & 0.7325 & 1.85 $\times$ 0.66 & 102.0 & 1.63$\times$10$^{-4}$ & --- &Unable to fit\\
HMC146+625 & -0.03 & -0.06 & 0.597 & 0.97 $\times$ 0.70 & 54.0 & 9.67$\times$10$^{-5}$ & 8.2 &\\
HMC153+615 & 0.10 & -0.26 & 2.8711 & 0.99 $\times$ 0.90 & 141.5 & 5.96$\times$10$^{-4}$ & 24 &\\
HMC163+606 & -0.19 & 0.38 & 8.1952 & 1.12 $\times$ 1.07 & 68.5 & 2.63$\times$10$^{-3}$ & 34 &\\

\hline\noalign{\vspace{1pt}}
Average& & & &  1.06  $\times$ 0.89    &&  1.99$\times$10$^{-3}$  \\
\hline
\end{tabular}
}
\end{center}

%:%%% Table HNCO %%
\clearpage
\refstepcounter{table}
\label{Table:HNCO_HMC_properties}
\begin{center}
{\bfseries Table \thetable. HNCO HMC properties}
\vspace{1ex}
\scalebox{0.5}[0.5]
{
\begin{tabular}{cccccccl}
\hline\hline
Name & \multicolumn{2}{c}{Position}  & Brightness & FWHM Size & PA & Line & Remarks \\
 & $\Delta$RA & $\Delta$DEC & fitted peak & major  $\times$ minor  && Luminosity &  \\
 & (arcsec) & (arcsec) & (Jy beam$^{-1}$ km\, s$^{-1}$ )& ($''\,\times\,''$) & (degree) & (L$_\odot$) &  \\
 \hline
HMC109+517 & -0.03 & 0.00 &0.9357& 1.02 $\times$ 0.95 & 67.9 & 2.53$\times$10$^{-4}$ & \\
HMC110+520 & 0.11 & -0.46 &0.0815& 1.78 $\times$ 1.02 & 153.5 & 4.13$\times$10$^{-5}$ & \\
HMC120+614 & -0.09 & -0.03 & 0.1337&1.13 $\times$ 0.79 & 66.5 & 3.33$\times$10$^{-5}$ & \\
HMC127+611 & 0.01 & 0.00 &1.0543& 1.10 $\times$ 0.97 & 104.7 & 3.14$\times$10$^{-4}$ & \\
HMC131+618 & -0.05 & 0.00 &0.5667& 0.96 $\times$ 0.82 & 48.6 & 1.24$\times$10$^{-4}$ & \\
HMC131+613 & 0.31 & -0.15 &7.6675& 2.13 $\times$ 1.76 & 166.3 & 8.02$\times$10$^{-3}$ & \\
HMC131+614 & --- & --- &  --- &  --- &  --- &  $<$7.91$\times$10$^{-4}$  & \\
HMC132+612 & --- & --- &  --- &  --- &  --- &  $<$1.95$\times$10$^{-3}$  &\\
HMC133+619 & -0.03 & 0.11 &0.1225& 1.57 $\times$ 0.89 & 94.7 & 4.78$\times$10$^{-5}$ & \\
HMC133+616& -0.17 & -0.19  &0.9154& 1.60 $\times$ 1.24 & 115.2 & 5.07$\times$10$^{-4}$ & \\
HMC134+610 & 0.01 & 0.22 &1.2486& 1.56 $\times$ 1.36 & 5.7 & 7.39$\times$10$^{-4}$ & \\
HMC134+612 & -0.12 & -0.03&1.8648 & 1.93 $\times$ 1.15 & 91.9 & 1.15$\times$10$^{-3}$ & \\
HMC136+649 & 0.00 & -0.01 & 1.1637 & 0.97 $\times$ 0.92 & 87 & 2.90$\times$10$^{-4}$ & \\
HMC141+624 & 0.02 & 0.10 &7.761& 1.38 $\times$ 1.35 & 110.7 & 4.03$\times$10$^{-3}$ & \\
HMC143+622 & -0.20 & 0.04 &0.2728& 1.43 $\times$ 1.05 & 93.1 & 1.14$\times$10$^{-4}$ & \\
HMC146+625 & -0.04 & 0.02&0.4004 & 0.94 $\times$ 0.90 & 109.8 & 9.45$\times$10$^{-5}$ & \\
HMC153+615 & -0.06 & 0.08 & 0.6043 & 1.11 $\times$ 1.03 & 135.8 & 1.93$\times$10$^{-4}$ & \\
HMC163+606 & -0.04 & -0.37 &0.0328& 2.25 $\times$ 1.34 & 173.2 & 2.76$\times$10$^{-5}$ & \\
\hline
\noalign{\vspace{1pt}}
Average& & & &  1.43  $\times$ 1.10    &&  9.99$\times$10$^{-4}$  \\
\hline

\end{tabular}
}
\end{center}
%\clearpage
%%% Table10%%%

%:%%% Table HC3N %%
\clearpage
\refstepcounter{table}
\label{Table:HC3N_HMC_properties}
\begin{center}
{\bfseries Table \thetable. HC$_3$N HMC properties}
\vspace{1ex}
\scalebox{0.5}[0.5]
{
\begin{tabular}{cccccccl}

\hline\hline
Name & \multicolumn{2}{c}{Position}  & Brightness & FWHM Size & PA & Line & Remarks \\
 & $\Delta$RA & $\Delta$DEC & fitted peak & major  $\times$ minor  && Luminosity &  \\
 & (arcsec) & (arcsec) & (Jy beam$^{-1}$ km\, s$^{-1}$ )& ($''\,\times\,''$) & (degree) & (L$_\odot$) &  \\
 \hline
HMC109+517 & 0.00 & 0.05 & 0.4759 & 1.18 $\times$ 1.1 & 34.2 & 1.72$\times$10$^{-4}$ \\
HMC110+520 & 0.01 & -0.11 & 0.3410 & 1.22 $\times$ 1.14 & 132.8 & 1.32$\times$10$^{-4}$ \\
HMC120+614 & -0.09 & -0.02 & 0.188 & 1.09$\times$ 0.90 & 78.4 & 5.15$\times$10$^{-5}$ \\
HMC127+611 & 0.11 & -0.08 & 1.5897 & 1.49 $\times$ 1.20 & 88.9 & 7.93$\times$10$^{-4}$ \\
HMC131+618 & -0.02 & -0.06 & 0.2887 & 0.51$\times$ 0.49 & 47.5 & 2.01$\times$10$^{-5}$ \\
HMC131+613 & 0.34 & -0.32 & 9.1916 & 1.85 $\times$ 1.45 & 168.9 & 6.88$\times$10$^{-3}$ \\
HMC131+614 & --- & --- & --- & --- & --- & $<$6.57$\times$10$^{-4}$ & \\
HMC132+612 & --- & --- & --- & --- & --- & $<$1.41$\times$10$^{-3}$  & \\
HMC133+619 & --- & --- & --- & --- & --- & $<$2.94$\times$10$^{-5}$  &  \\
HMC133+616 & -0.06 & -0.01 & 1.7758 & 1.44 $\times$ 0.99 & 94.4 & 7.06$\times$10$^{-4}$ \\
HMC134+610 & -0.02 & -0.42 & 1.3086 & 1.81 $\times$ 1.34 & 14.4 & 8.86$\times$10$^{-4}$ \\
HMC134+612 & 0.12 & 0.13 & 3.4218 & 1.14 $\times$ 0.95 & 87.1 & 1.03$\times$10$^{-3}$ \\
HMC136+649 & 0.00 & -0.01 & 1.6135 & 0.96 $\times$ 0.90 & 80.5 & 3.89$\times$10$^{-4}$ \\
HMC141+624 & 0.00 & 0.10 & 10.6704 & 1.45 $\times$ 1.19 & 155.5 & 5.14$\times$10$^{-3}$ \\
HMC143+622 & 0.11 & -0.01 & 0.4164 & 1.14 $\times$ 0.92 & 106.8 & 1.22$\times$10$^{-4}$ \\
HMC146+625 & 0.00 & -0.03 & 1.3117 & 1.06 $\times$ 0.96 & 134.0 & 3.72$\times$10$^{-4}$ \\
HMC153+615 & 0.03 & 0.02 & 1.131 & 1.20 $\times$ 1.04 & 141.1 & 3.94$\times$10$^{-4}$ \\
HMC163+606 & 0.04 & -0.11 & 0.1445 & 1.26 $\times$ 1.04 & 128.9 & 5.28$\times$10$^{-5}$ \\
\hline
\noalign{\vspace{1pt}}
Average& & & &  1.25  $\times$ 1.04    &&  1.14$\times$10$^{-3}$  \\
\hline

\end{tabular}
}
\end{center}

%%% Table11 %%%

%:%%% Table OCS %%
\clearpage
\refstepcounter{table}
\label{Table:OCS_HMC_properties}
\begin{center}
{\bfseries Table \thetable. OCS HMC properties}
\vspace{1ex}
\scalebox{0.5}[0.5]
{
\begin{tabular}{cccccccl}

\hline\hline
Name & \multicolumn{2}{c}{Position}  & Brightness & FWHM Size & PA & Line & Remarks \\
 & $\Delta$RA & $\Delta$DEC & fitted peak & major  $\times$ minor  && Luminosity &  \\
 & (arcsec) & (arcsec) & (Jy beam$^{-1}$ km\, s$^{-1}$ )& ($''\,\times\,''$) & (degree) & (L$_\odot$) &  \\
 \hline
HMC109+517 & -0.01 & 0.01 &0.7482& 1.10 $\times$ 0.99 & 65.1 & 2.27$\times$10$^{-4}$ \\
HMC110+520 & 0.01 & -0.01 &0.3139& 0.95 $\times$ 0.94 & 155.8 & 7.82$\times$10$^{-5}$ \\
HMC120+614 & 0.05 & -0.05&0.2396 & 1.09 $\times$ 0.83 & 69.9 & 6.05$\times$10$^{-5}$ \\
HMC127+611 & -0.03 & -0.01&2.0409 & 1.06 $\times$ 1.01 & 91.8 & 6.10$\times$10$^{-4}$ \\
HMC131+618 & -0.10 & 0.00&1.1165 & 1.29 $\times$ 1.12 & 35.0 & 4.50$\times$10$^{-4}$ \\
HMC131+613 & 0.18 & -0.10 &7.8425& 2.01 $\times$ 1.54 & 145.2 & 6.77$\times$10$^{-3}$ \\
HMC131+614 & ---& --- &  --- &  --- &  --- & $<$1.06$\times$10$^{-3}$  \\
HMC132+612 & --- & ---&  --- &  --- &  --- &  $<$1.15$\times$10$^{-3}$ \\
HMC133+619 &0.23 &0.707& 0.02 & 1.46 $\times$ 1.12 & 81.1 & 3.23$\times$10$^{-4}$ \\
HMC133+616 & 0.04 & -0.02 & 2.2118& 1.79 $\times$ 1.07 & 103.8 & 1.18$\times$10$^{-3}$ \\
HMC134+610& 0.01 & 0.05 & 1.4349 &1.34 $\times$ 1.18 & 159.5 & 6.33$\times$10$^{-4}$ \\
HMC134+612 & 0.10 & 0.08 &2.6502& 1.44$\times$ 1.03 & 82.9 & 1.10$\times$10$^{-3}$ \\
HMC136+649 & 0.00 & 0.01 &1.7125& 0.97 $\times$ 0.91 & 78.1 & 4.22$\times$10$^{-4}$ \\
HMC141+624 & 0.00 & 0.00 &10.4692& 1.53 $\times$ 1.37 & 164.4 & 6.12$\times$10$^{-3}$ \\
HMC143+622 & 0.01 & 0.00  &0.5716& 1.13 $\times$ 0.92 & 102 & 1.66$\times$10$^{-4}$ \\
HMC146+625 & -0.04 & 0.03 &0.8249& 0.84 $\times$ 0.81 & 145.3 & 1.57$\times$10$^{-4}$ \\
HMC153+615 & 0.01 & -0.01&1.4326 & 1.11 $\times$ 0.97 & 147.3 & 4.30$\times$10$^{-4}$ \\
HMC163+606& 0.01 & -0.03  &0.0969& 0.95 $\times$ 0.88 & 168 & 2.26$\times$10$^{-5}$ \\

\hline
\noalign{\vspace{1pt}}
Average& & & &  1.25  $\times$ 1.04    &&  1.17$\times$10$^{-3}$  \\
\hline

\end{tabular}
}
\end{center}

%%% Table12%%%

%:%%% Table SO2 %%
\clearpage
\refstepcounter{table}
\label{Table:SO2_HMC_properties}
\begin{center}
{\bfseries Table \thetable. SO$_2$ HMC properties}
\vspace{1ex}
\scalebox{0.5}[0.5]
{
\begin{tabular}{cccccccl}

\hline\hline
Name & \multicolumn{2}{c}{Position}  & Brightness & FWHM Size & PA & Line & Remarks \\
 & $\Delta$RA & $\Delta$DEC & fitted peak & major  $\times$ minor  && Luminosity &  \\
 & (arcsec) & (arcsec) & (Jy beam$^{-1}$ km\, s$^{-1}$ )& ($''\,\times\,''$) & (degree) & (L$_\odot$) &  \\
 \hline
HMC109+517 & -0.04 & 0.01 &1.4434& 0.95 $\times$  0.91 & 77.9 & 3.48$\times$10$^{-4}$ \\
HMC110+520 & 0.00 & -0.13 &0.0929& 0.94 $\times$ 0.82 & 94.5 & 2.00$\times$10$^{-5}$ \\
HMC120+614 & 0.14 & 0.04 &0.1751& 1.56 $\times$ 1.14 & 78.5 & 8.69$\times$10$^{-5}$ \\
HMC127+611 & -0.02 & 0.04 &0.8222& 0.54 $\times$ 0.43 & 97.9 & 5.33$\times$10$^{-5}$ \\
HMC131+618 & 0.08 & -0.03 &0.5309& 0.36 $\times$ 0.3 & 39.4 & 1.60$\times$10$^{-5}$ \\
HMC131+613 & 0.15 & -0.07 &13.5391& 2.82 $\times$ 1.77 & 92.5 & 1.89$\times$10$^{-2}$ \\
HMC131+614 & 0.39 & -1.44 &6.0649& 2.58 $\times$ 2.08 & 177.4 & 9.08$\times$10$^{-3}$ \\
HMC132+612& --- &  --- &  --- &  --- &  ---  &  $<$6.64$\times$10$^{-4}$\\
HMC133+619 & --- &  --- &  --- &  --- &  ---  &   $<$3.66$\times$10$^{-6}$\\
HMC133+616& --- &  --- &  --- &  --- &  ---  &   $<$1.63$\times$10$^{-3}$\\
HMC134+610 & --- &  --- &  --- &  --- &  ---  & $<$6.54$\times$10$^{-4}$\\
HMC134+612 & 0.03 & 0.09 &13.6715& 2.07 $\times$ 1.42 & 82.2 & 1.12$\times$10$^{-2}$ \\
HMC136+649 & 0.00 & -0.02 &1.7463& 1.44 $\times$ 1.33 & 153.3 & 8.19$\times$10$^{-4}$ \\
HMC141+624 & 0.01 & 0.13 &9.9521& 0.72 $\times$ 0.66 & 93.2 & 1.32$\times$10$^{-3}$ \\
HMC143+622 & -0.23 & 0.06 &0.6084& 1.27 $\times$ 0.84 & 93.2 & 1.68$\times$10$^{-4}$ \\
HMC146+625 & -0.06 & 0.04 &0.8693& 0.72 $\times$ 0.66 & 110.6 & 9.04$\times$10$^{-5}$ \\
HMC153+615 & 0.03 & 0.00 &0.8693& 1.18 $\times$ 0.96 & 150.2 & 2.43$\times$10$^{-4}$ \\
HMC163+606 & -0.12 & 0.25 &0.0401& 1.00 $\times$ 0.86 & 104.8 & 7.94$\times$10$^{-6}$ \\

\hline
\vspace{1pt}
Average& & & &  1.20  $\times$ 0.93    &&  2.56$\times$10$^{-3}$  \\
\hline

\end{tabular}
}
\end{center}

\clearpage
%%% Table13%%%

%:%%% Table DCN %%
%Peak Temp or integ にしたほうがよいようだ
\clearpage
\refstepcounter{table}
\label{Table:DCN_HMC_properties}
\begin{center}
{\bfseries Table \thetable. DCN HMC properties}
\vspace{1ex}
\scalebox{0.5}[0.5]
{
\begin{tabular}{cccccccl}

\hline\hline
Name & \multicolumn{2}{c}{Position}  & Brightness & FWHM Size & PA & Line & Remarks \\
 & $\Delta$RA & $\Delta$DEC & fitted peak & major  $\times$ minor  && Luminosity &  \\
 & (arcsec) & (arcsec) & (Jy beam$^{-1}$ km\, s$^{-1}$ )& ($''\,\times\,''$) & (degree) & (L$_\odot$) &  \\
 \hline
HMC109+517 & 0.01 & 0.01 & 0.2379 & 1.83 $\times$ 1.21 & 41.1 & 1.47$\times$10$^{-4}$ \\
HMC110+520 & -0.07 & -0.58 & 0.1472 & 1.63 $\times$ 1.34 & 5.5 & 8.97$\times$10$^{-5}$ \\
HMC120+614 & 0.05 & -0.16 & 0.1702 & 1.31 $\times$ 1.13 & 115.1 & 7.03$\times$10$^{-5}$ \\
HMC127+611 & -0.02 & -0.25 & 0.7045 & 4.61 $\times$ 1.88 & 78.9 & 1.70$\times$10$^{-3}$ \\
HMC131+618 & -0.90 & -0.38 & 0.3182 & 4.72 $\times$ 1.62 & 48.5 & 6.79$\times$10$^{-4}$ \\
HMC131+613 & 0.47 & 0.05 & 1.2537 & 2.76 $\times$ 1.29 & 155.2 & 1.25$\times$10$^{-3}$ \\
HMC131+614& --- &  --- &  --- &  --- &  ---  &   $<$1.49$\times$10$^{-3}$\\
HMC132+612& --- &  --- &  --- &  --- &  ---  &  $<$3.65$\times$10$^{-3}$\\
HMC133+619& --- &  --- &  --- &  --- &  ---  &  $<$8.20$\times$10$^{-3}$\\
HMC133+616 & 0.12 & -0.15 & 0.1478 & 1.52 $\times$ 1.23 & 156.8 & 7.71$\times$10$^{-5}$ \\
HMC134+610 & -0.05 & -0.07 & 0.2153 & 1.67 $\times$ 1.19 & 1.30 & 1.19$\times$10$^{-4}$ \\
HMC134+612 & -0.10 & -0.23 & 0.1520 & 1.30 $\times$ 0.77 & 65.8 & 4.25$\times$10$^{-5}$ \\
HMC136+649 & 0.00 & 0.11 & 0.8311 & 1.37 $\times$ 1.24 & 98.9 & 3.94$\times$10$^{-4}$ \\
HMC141+624 & -0.05 & 0.05 & 3.1158 & 1.32 $\times$ 1.27 & 53.6 & 1.46$\times$10$^{-3}$ \\
HMC143+622 & -0.58 & -0.29 & 0.1505 & 2.61$\times$ 2.12 & 89.2 & 2.32$\times$10$^{-4}$ &\\
HMC146+625 & 0.00 & 0.05 & 0.4051 & 1.32 $\times$ 1.15 & 17.6 & 1.72$\times$10$^{-4}$ \\
HMC153+615 & -0.16 & -0.06 & 0.3397 & 1.52 $\times$ 1.35 & 17.6 & 1.94$\times$10$^{-4}$ \\
HMC163+606 & 0.29 & -1.02 & 0.1259 & 2.16 $\times$ 1.67 & 158.9 & 1.27$\times$10$^{-4}$ \\
\hline
\vspace{1pt}
Average& & & & 2.15  $\times$ 1.37    && 4.61$\times$10$^{-4}$  \\
\hline

\end{tabular}
}
\end{center}

%%% Table14%%%

%:%%% Table H2CO %%
\clearpage
\refstepcounter{table}
\label{Table:H2CO_HMC_properties}
\begin{center}
{\bfseries Table \thetable. H$_2$CO HMC properties}
\vspace{1ex}
\scalebox{0.5}[0.5]
{
\begin{tabular}{cccccccl}

\hline\hline
Name & \multicolumn{2}{c}{Position}  & Brightness & FWHM Size & PA & Line & Remarks \\
 & $\Delta$RA & $\Delta$DEC & fitted peak & major  $\times$ minor  && Luminosity &  \\
 & (arcsec) & (arcsec) & (Jy beam$^{-1}$ km\, s$^{-1}$ )& ($''\,\times\,''$) & (degree) & (L$_\odot$) &  \\
 \hline
HMC109+517 & 0.03 & 0.02 & 0.7736 & 1.39 $\times$ 1.15 & 58.9 & 3.45$\times$10$^{-4}$ \\
HMC110+520 & 0.02 & -0.13 & 0.4129 & 1.40$\times$ 1.17 & 156.2 & 1.89$\times$10$^{-4}$ \\
HMC120+614 & -0.01 & -0.06 & 0.2703 & 1.20 $\times$ 0.94 & 65.1 & 8.51$\times$10$^{-5}$ \\
HMC127+611 & 0.07 & 0.01 & 1.4522 & 1.50 $\times$ 1.34 & 70.8 & 8.14$\times$10$^{-4}$ \\
HMC131+618 & -0.84 & -0.36 & 0.4368 & 2.85 $\times$ 1.15 & 32.3 & 3.99$\times$10$^{-4}$ \\
HMC131+613 & 0.21 & 0.41 & 3.3649 & 3.47 $\times$ 1.78 & 115.3 & 5.80$\times$10$^{-3}$ \\
HMC131+614 & 0.05 & -0.01 & 1.603 & 1.88 $\times$ 1.62 & 94.3 & 1.36$\times$10$^{-3}$ \\
HMC132+612 & -0.27 & -0.01 & 1.0684 & 1.98 $\times$ 1.53 & 125.1 & 9.03$\times$10$^{-4}$ \\
HMC133+619 & 0.20 & 0.57 & 0.1064 & 2.81 $\times$ 0.61 & 122.1 & 5.09$\times$10$^{-5}$ \\
HMC133+616 & 0.30 & -0.02 & 1.7862 & 2.04 $\times$ 1.45 & 93.9 & 1.47$\times$10$^{-3}$ \\
HMC134+610 & -0.03 & -0.01 & 2.8853 & 1.85$\times$ 1.52 & 149.3 & 2.26$\times$10$^{-3}$ \\
HMC134+612 & --- &  --- &  --- &  --- &  ---  &  $<$1.75$\times$10$^{-4}$ \\
HMC136+649 & -0.02 & -0.01 & 1.2566 & 1.18 $\times$ 1.09 & 85.1 & 4.51$\times$10$^{-4}$ \\
HMC141+624 & 0.01 & 0.08 & 5.9914 & 1.61 $\times$ 1.47 & 165.8 & 3.96$\times$10$^{-3}$ \\
HMC143+622 & -0.10 & -0.03 & 0.3374 & 1.58 $\times$ 1.23 & 101.0 & 1.83$\times$10$^{-4}$ \\
HMC146+625 & -0.05 & 0.02 & 0.8464 & 1.16 $\times$ 1.07 & 148.2 & 2.93$\times$10$^{-4}$ \\
HMC153+615 & -0.04 & -0.07 & 0.9012 & 1.40 $\times$ 1.23 & 166.0 & 4.33$\times$10$^{-4}$ \\
HMC163+606 & 0.08 & -0.13 & 0.3383 & 1.45 $\times$ 1.31 & 138.3 & 1.79$\times$10$^{-4}$ \\
 \hline
\vspace{1pt}
Average& & & & 1.81  $\times$ 1.27    && 1.13$\times$10$^{-3}$  \\
\hline

\end{tabular}
}
\end{center}

%%% Table14%%%

%:%%% Table C18O %%
\clearpage
\refstepcounter{table}
\label{Table:C18O_HMC_properties}
\begin{center}
{\bfseries Table \thetable. C$^{18}$O HMC properties}
\vspace{1ex}
\scalebox{0.5}[0.5]
{
\begin{tabular}{cccccccl}

\hline\hline
Name & \multicolumn{2}{c}{Position}  & Brightness & FWHM Size & PA & Line & Remarks \\
 & $\Delta$RA & $\Delta$DEC & fitted peak & major  $\times$ minor  && Luminosity &  \\
 & (arcsec) & (arcsec) & (Jy beam$^{-1}$ km\, s$^{-1}$ )& ($''\,\times\,''$) & (degree) & (L$_\odot$) &  \\
 \hline
 HMC109+517 & -0.08 & -0.16 & 1.4004 & 4.28 $\times$ 1.76 & 56.8 & 2.94$\times$10$^{-3}$ \\
HMC110+520 & -0.01 & -0.43 & 0.9666 & 2.16 $\times$ 1.99 & 19.6 & 1.16$\times$10$^{-3}$ \\
HMC120+614& --- &  --- &  --- &  --- &  ---  & $<$3.78$\times$10$^{-4}$ \\
HMC127+611 & 1.44 & 0.18 & 2.7795 & 7.95 $\times$ 3.07 & 86.2 & 1.89$\times$10$^{-2}$ \\
HMC131+618 & -0.49 & -0.29 & 2.4333 & 3.10 $\times$ 2.18 & 38.5 & 4.59$\times$10$^{-3}$ \\
HMC131+613 & -0.03 & 0.29 & 8.0529 & 3.40 $\times$ 2.12 & 102.0 & 1.62$\times$10$^{-2}$ \\
HMC131+614 & 0.00 & 0.15 & 2.5685 & 3.41 $\times$ 2.72 & 157.7 & 6.65$\times$10$^{-3}$ \\
HMC132+612 & -0.16 & 0.36 & 4.1671 & 2.71$\times$ 1.88 & 72.5 & 5.92$\times$10$^{-3}$ \\
HMC133+619 & 0.83 & -0.27 & 3.0434 & 4.04 $\times$ 2.94 & 85.2 & 1.01$\times$10$^{-2}$ \\
HMC133+616 & 0.99 & -0.31 & 3.1699 & 3.66 $\times$ 1.94 & 117.8 & 6.28$\times$10$^{-3}$ \\
HMC134+610& --- &  --- &  --- &  --- &  ---  &   $<$2.43$\times$10$^{-3}$ \\
HMC134+612 & 0.81 & 0.13 & 2.2734 & 2.34 $\times$ 1.31 & 72.2 & 1.94$\times$10$^{-3}$ \\
HMC136+649 & 0.00 & -0.13 & 2.5321 & 2.45 $\times$ 1.99 & 7.3 & 3.44$\times$10$^{-3}$ \\
HMC141+624 & -0.07 & -0.01 & 5.4463 & 2.42 $\times$ 2.11 & 76.9 & 7.76$\times$10$^{-3}$ \\
HMC143+622 & -0.54 & -0.05 & 2.1872 & 2.71 $\times$ 2.12 & 2.1 & 3.51$\times$10$^{-3}$ \\
HMC146+625 & -0.05 & 0.08 & 2.0733 & 2.49 $\times$ 2.09 & 2.9 & 3.01$\times$10$^{-3}$ \\
HMC153+615 & 0.14 & -0.17 & 2.4867 & 3.20 $\times$ 2.86 & 37.9 & 6.35$\times$10$^{-3}$ \\
HMC163+606 & 0.43 & -0.11 & 1.632 & 4.02 $\times$ 2.10 & 122.2 & 3.84$\times$10$^{-3}$ \\
\hline
\vspace{1pt}
Average& & & & 3.53  $\times$ 2.20    && 1.13$\times$10$^{-3}$  \\
\hline

\end{tabular}
}
\end{center}

\clearpage

%%% Table15%%%

%****************
%\clearpage

%:Appnedix Maps

%\section{HNCO, HC$_3$N, OCS, H$_2$CO, SO$_2$, DCN, and C$^{18}$O maps}

%:%%% Figure A25: Other_mol_lines_map01
\clearpage
\onecolumn

\begingroup\makeatletter
\def\@captype{figure}\makeatother
\begin{center}
\includegraphics[width=\textwidth,  height=0.88\textheight,  keepaspectratio]{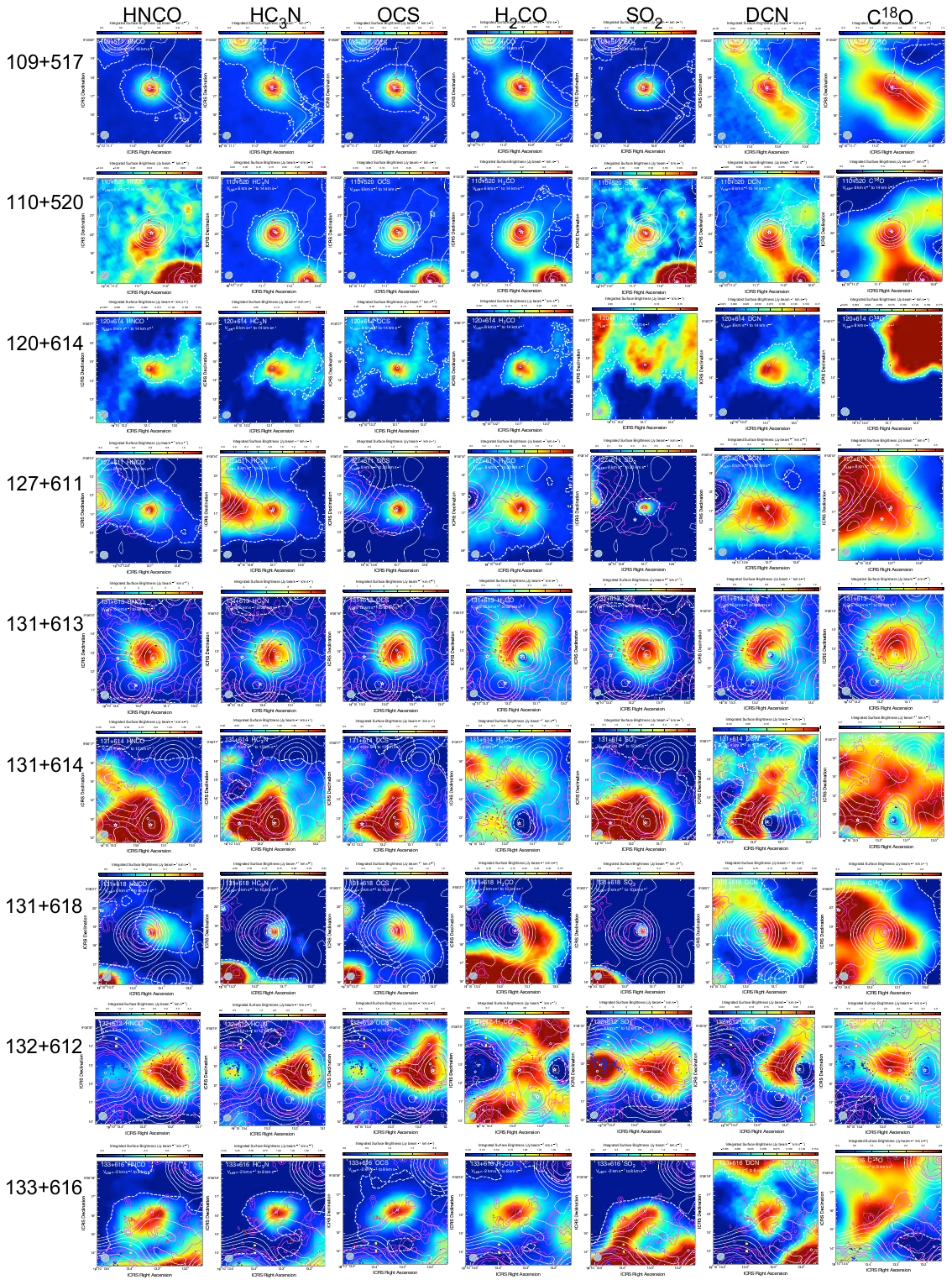}

\caption{HNCO, HC$_3$N, OCS, H$_2$CO, SO$_2$, DCN, and C$^{18}$O maps of HMC 109+517, 110+520, 120+614, 127+611, 131+613, 131+614, 131+618, 132+612, and 133+616,.
{Alt text: Multi-panel set of local integrated-intensity maps for several HMCs (109+517, 110+520, 120+614, 127+611, 131+613, 131+614, 131+618, 132+612, and 133+616), shown for the lines HNCO, HC$_3$N, OCS, H$_2$CO, SO$_2$, DCN, and C$^{18}$O. Panels allow visual comparison of the spatial distribution of these tracers across the listed HMCs.}
}\label{Fig:Other_mol_lines_map01}
\end{center}

\endgroup
\clearpage
% Figure A25: Other_mol_lines_map01

%:%%% Figure A26: Other_mol_lines_map02
\onecolumn
\begingroup
\makeatletter\def\@captype{figure}\makeatother
\begin{center}
\includegraphics[width=\textwidth,height=0.85\textheight,keepaspectratio]{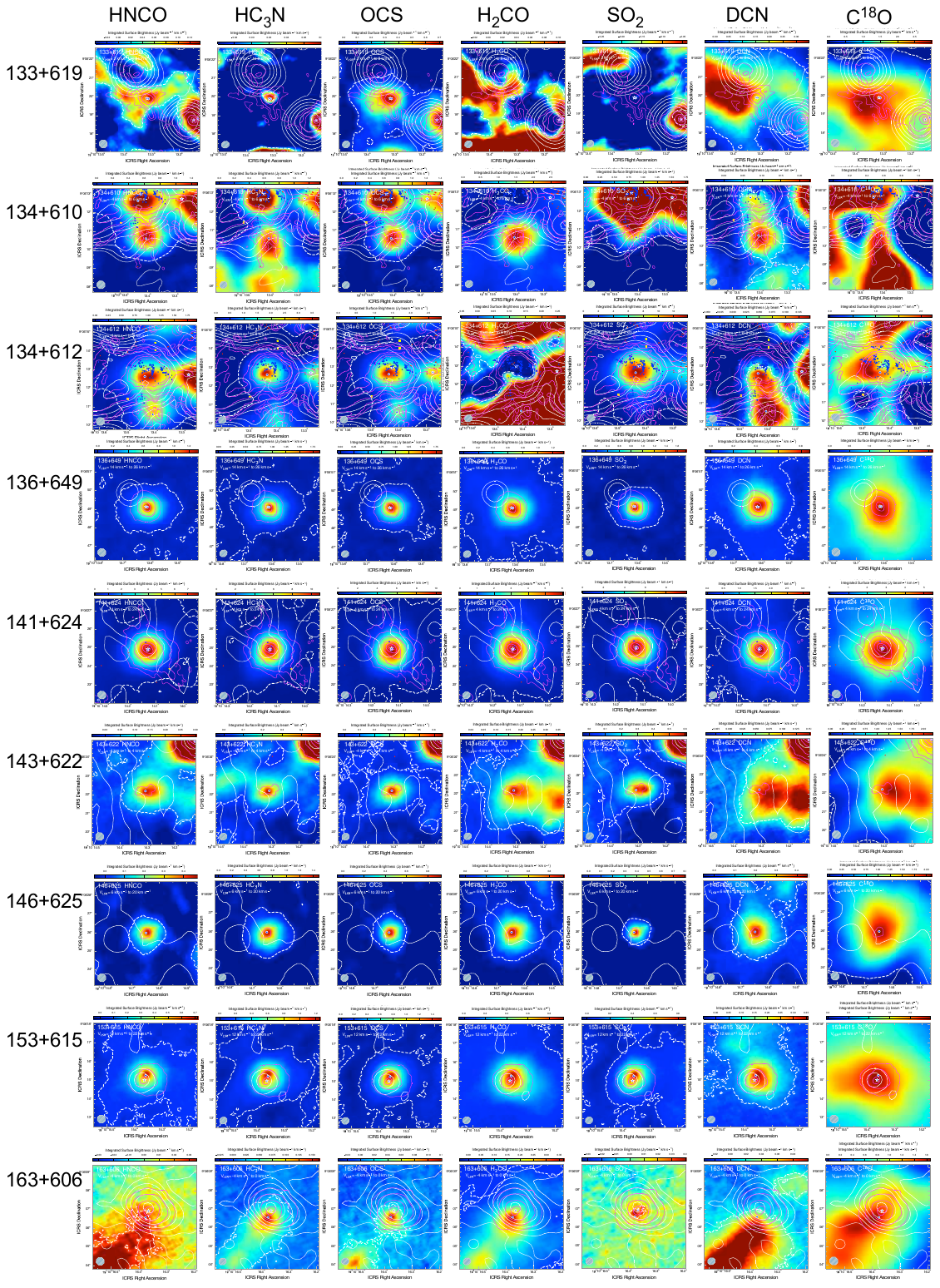}
\caption{HNCO, HC$_3$N, OCS, H$_2$CO, SO$_2$, DCN, and C$^{18}$O maps of HMC  133+619, 134+610, 134+612, 136+649, 141+624, 143+622, 146+625, 153+615, and 163+606.
{Alt text: Multi-panel set of local integrated-intensity maps for several HMCs (133+619, 134+610, 134+612, 136+649, 141+624, 143+622, 146+625, 153+615, and 163+606), shown for the lines HNCO, HC$_3$N, OCS, H$_2$CO, SO$_2$, DCN, and C$^{18}$O. Panels allow visual comparison of the spatial distribution of these tracers across the listed HMCs.}
}
\label{Fig:Other_mol_lines_map02}
\end{center}
\endgroup
\clearpage
\twocolumn

\end{appendix}

%:REFERENCES


\begin{thebibliography}{}\label{REFERENCES}

\bibitem[ALMA Partnership(2023)]{ALMA2023}
ALMA Partnership 2023, ALMA Technical Handbook, Cycle 10,
https://almascience.eso.org/documents-and-tools/cycle10/alma-technical-handbook

 \bibitem[Araya et al.(2005)]{Araya2005}
Araya, E.,  Hofner, P.,  Kurtz, S.,  Bronfman, L., \& DeDeo, S. 2005, \apjs, 157, 279

\bibitem[Asanok et al.(2023)]{Asanok2023}
Asanok, K. et al. 2023, \apj, 943, 79

 \bibitem[Asanok et al.(2024)]{Asanok2024}
Asanok, K. et al. 2024, IAUS, 380, 202

\bibitem[Battersby et al.(2017)]{Battersby2017}
Battersby, C., Bally, J., \& Svoboda, B. 2017, \apj, 835, 263

\bibitem[Barrett et al.(1971)]{Barrett1971}
Barrett, A. H., Schwartz, P. R. \& Waters, J. W. 1971, \apj, 168, 101

\bibitem[Bayandina et al.(2025)]{Bayandina2025}
Bayandina, O. S., Moscadelli, L., Cesaroni, R., Beltr\'{a}n, M. T., Sanna, A.,\& Goddi, C. 2025, \aap, 694, 92B

\bibitem[Beuther et al.(2002)]{Beuther2002}
Beuther, H., Walsh, A., Schilke, P., Sridharan, T. K., Menten, K. M., \& Wyrowski, F. 2002 \aap, 390, 289

\bibitem[Beuther et al.(2007)]{Beuther2007}
Beuther, H., Churchwell, E., McKee, C., \& Tan, J. 2007, in Protostars and Planets V, ed. B. Reipurth, D. Jewitt, \& K. Keil (Tucson: University of Arizona Press), 165 

\bibitem[Beuther et al.(2018)]{Beuther2018}
Beuther, H.et al. 2018 \aap, 617, 100B

\bibitem[Beuther et al.(2019)]{Beuther2019}
Beuther, H.et al. 2019 \aap, 628, A90

\bibitem[Beuther, (2025)]{Beuther2025} %yet
Beuther, H.,  Kuiper, R., \& Tafalla, M. 2025 ARA\&A, 63, 1B

\bibitem[Bonnell et al.(1997)]{Bonnell1997}
Bonnell I. A., Bate M. R., Clarke C. J., \& Pringle J. E., 1997, MNRAS, 285, 201

\bibitem[Bonnell et al.(2001)]{Bonnell2001}
Bonnell, I. A.,  Bate, M. R., Clarke, C. J., \& Pringle, J. E. 2001, MNRAS, 323, 785

\bibitem[Bonnell, Vine, \&  Bate(2004)]{Bonnell2004}
Bonnell, I. A., Vine, S. G. \& Bate, M. R. 2004, MNRAS, 349, 735

\bibitem[Blaszkiewicz \& Kus(2004)]{Blaszkiewicz2004}
Blaszkiewicz, L. \& Kus, A. J. 2004, \aap, 413, 233

\bibitem[Breen et al.(2015)]{Breen2015}
Breen, S. L. et al. 2015, \mnras, 450, 4109

\bibitem[Brouillet et al.(2022)]{Brouillet2022}
Brouillet, N., et al. 2022, A\&A, 665, A140

\bibitem[Burns et al.(2023)]{Burns2023}
Burns, R. A. et al. 2023,  Nature Astronomy, 7, 557

\bibitem[The CASA Team et al.(2022)]{CASA2022}
THE CASA TEAM, Bean, B., Bhatnagar, S., Castro, S., Meyer, J. D., Emonts, B., Garcia, E., Garwood, R., et al.  2022, \pasp, 134, 1041

\bibitem[Caswell et al. (1995)]{Caswell1995}
Caswell, J. L.Vaile, R. A.Ellingsen, S. P. \& Norris, R. P. 1995, \mnras, 272, 96

\bibitem[Comrie et al.(2021)]{Comrie2021}
Comrie, A., Wang, K.-S., Ford, P., et al. 2021, CARTA:TheCubeAnalysis and Rendering Tool for Astronomy, v2.0.0, Zenodo, doi:10.5281/zenodo. 4905459

\bibitem[Contreras et al. (2013)]{Contreras2013}
Contreras, Y., et al., 2013, A\&A, 549, A45

\bibitem[Cragg et al.(2005)]{Cragg2005}
Cragg, D. M., Sobolev, A. M., \& Godfrey, P. D. 2005, \mnras, 360, 533

\bibitem[Csengeri et al.(2017)]{Csengeri2017}
Csengeri, T. 2017, \aap, 600, 10C

\bibitem[Deshpande et al.(2013)]{Deshpande2013}
Deshpande,  A. A., Goss, W. M., \& Mendoza-Torres, J. E.2013, \apj, 775, 36

\bibitem[de la Fuente et al.(2018)]{Fuente2018}
de la Fuente, E., Trinidad, M. A., Porras, A., Rodr\'{i}guez-Rico, C., Araya, E. D., Kurtz, S., Hofner, P., \& Nigoche-Netro, A. 2018, Revista Mexicana de Astronom\'ia y Astrof\'isica, 54, 129

\bibitem[De~Pree et al.(1997)]{DePree1997}
De Pree, C. G., Mehringer, D. M., \& Goss, W. M. 1997, \apj, 482, 307

\bibitem[De~Pree et al.(2000)]{DePree2000}
 De Pree, C. G., Wilner, D. J., Goss, W. M., Welch, W. J., \& McGrath, E.  2000, \apj, 540, 308

\bibitem[De~Pree et al.(2004)]{DePree2004}
De Pree, C. G., Wilner, D. J., Mercer, A. J., Davis, L. E., Goss, W. M., \& Kurtz, S.  2004, \apj, 600, 286

\bibitem[De~Pree et al.(2020)]{DePree2020}
De Pree, C. G., et al.  2020, AJ,160, 234

\bibitem[Edris et al.(2007)]{Edris2007}
Edris, K. A., Fuller, G. A., \& Cohen, R. J. 2007, \aap, 465, 865

\bibitem[Forster \& Caswell (1989)]{Forster1989}
 Forster, J. R. \& Caswell, J. L.  1989, \aap, 213, 339

\bibitem[Fontani et al.(2010)]{Fontani2010}
Fontani, F., Cesaroni, R., \& Furuya, R. S., 2010, \aap, 517, 56

\bibitem[Francis et al. (2021)]{Francis2021}
Francis, L. , Johnstone, D., Herczeg,G. , Hunter, T. R.,\& Harsono, D. 2020, \aj, 160, 270

\bibitem[Fuente et al.(2014)]{Fuente2014}
Fuente, A., et al. 2014, \aap, 568, A65

\bibitem[Furuya et al.(2011)]{Furuya2011}
Furuya, R.S., Cesaroni, R., \& Shinnaga, H. 2011, \aap, 525, 72

\bibitem[Gan et al.(2013)]{Gan2013}
Gan, Cong-Gui Chen, Xi Shen, Zhi-Qiang Xu, Ye Ju, Bing-Gang 2013, \apj, 763, 2

\bibitem[Garay \& Lizano(1999)]{Garay1999}
Garay, G., \& Lizano, S. 1999, \pasp, 111, 1049

\bibitem[Gieser et al.(2019)]{Gieser2019}
Gerner, T., Beuther, H., Semenov, D., et al. 2014, A\&A, 563, A97

\bibitem[Goldsmith \& Langer(1999)]{Goldsmith1999}
Goldsmith, P. F., \& Langer, W. D. 1999, \apj, 517, 209

\bibitem[Galv\'{a}n-Madrid et al.(2013)]{Madrid2013}
Galv\'{a}n-Madrid, R., Liu, H. B., Zhang, Z.-Y.,  Pineda, J. E., Peng T.-C., Zhang, Q., Keto,  E. R.,
Ho, P. T. P., Rodr\'{i}guez, L., Zapata,F.,  Peters, T., \& De~Pree, C. G.  2013, \apj, 779, 121

\bibitem[Giani et al.(2023)]{Giani2023}
Giani, L., Ceccarelli, C., Mancini, L. et al. 2023, \mnras, 526, 4535

\bibitem[Green \& McClure-Griffiths (2011)]{Green2011}
Green, J. A. \& McClure-Griffiths, N. M. 2011, \mnras, 417, 2500

\bibitem[Gwinn et al. (1992)]{Gwinn1992}
Gwinn, C. R., Moran, J. M., \& Reid, M. J. 1992, \apj, 393, 149

\bibitem[Hanson(1997)]{Hanson1997}		
Hanson M. M. Howarth I. D., \& Conti P. S. 1997, \apj, 489, 698

\bibitem[Haschick et al. (1990)]{Haschick1990}
Haschick, Aubrey D. Menten, Karl M. Baan, \& Willem A. 1990, \apj, 354, 556

\bibitem[He et al.(2021)]{He2021}
He, Z. Z., Li, G-. X., \& Zhang, C. 2021, \raa, 21, 207

\bibitem[Herbst \& van Dishoeck(2009)]{Herbst2009}		
Herbst, E. \& van Dishoeck, E. F., 2009, \aap, 47, 427


\bibitem[Hern\'andez-Hern\'andez et al.(2014)]{Hernandez2014}
Hern\'andez-Hern\'andez, V., Zapata, L., Kurtz, S. \& Garay, G., 2014, \apj, 786, 38

\bibitem[Hirota et al.(2017)]{Hirota2017}
Hirota, T., Machida, M. N., Matsushita, Y., Motogi, K.,
Matsumoto, N., Kim, M. K., Burns, R. A., \& Honma, M. 2017, Nature Astronomy, 1, 0146

\bibitem[Hildebrand(1983)]{Hildebrand1983}
Hildebrand, R. H. 1983, QJRAS, 24, 267

\bibitem[Hollis(1982)]{Hollis1982}
Hollis, J.~M.,  1982, \apj, 260, 159

\bibitem[Hosokawa et al.(2010)]{Hosokawa2010}
Hosokawa, T., Yorke, H.~W., \& Omukai, K. 2010, \apj, 721, 478

\bibitem[Hu et al.(2016)]{Hu2016}
Hu, B., Menten, K. M. Wu, Y. Bartkiewicz, A. Rygl, K. Reid, M. J. Urquhart, J. S. \& Zheng, X.  2016, \apj, 833, 18

\bibitem[Ilee et al.(2018)]{Ilee2018}	
Ilee, J. D., Cyganowski, C. J., Brogan, C. L., Hunter, T. R., Forgan, D. H., Haworth, T. J., Clarke, C. J.,  \& Harries, T. J. 2018,  \apj, 869, 24

\bibitem[Jim\'{e}nez-Serra et al.(2012)]{Serra2012}		
Jim\'{e}nez-Serra, I., Zhang, Q., Viti, S., Martin-Pintado, J., \& de Wit, W.-J. 2012, \apj., 753, 34

\bibitem[Kalenskii et al.(2001)]{Kalenskii2001}		
Kalenskii, S. V. Slysh, V. I. Val'tts, I. E. Winnberg, A. \& Johansson, L. E. 2001, Astronomy Reports, 45, 26

\bibitem[Kauffman et al.(1998)]{Kauffman1998}		
Kauffman, M. J., Hollenbach, D. J., \& Tielens, A. G. G. M. 1998, \apj., 497, 276

\bibitem[Kavak et al.(2021)]{Kavak2021}	
Kavak, \"{U}. et al. 2024, \aap, 645, A29

\bibitem[Keene, Hildebrand \& Whitcomb(1982)]{Keene1982} 
Keene, J., Hildebrand, R. H., \& Whitcomb, S. E. 1982, \apj, 252, L11

\bibitem[Keto(2002)]{Keto2002}
Keto, E. 2002, \apj, 580, 980

\bibitem[Keto(2003)]{Keto2003}
Keto, E. 2003, \apj, 599, 1196

\bibitem[Kurtz et al.(2000)]{Kurtz2000}
Kurtz, S., Cesaroni, R., Churchwell, E., Hofner, P., \& Walmsley, C. M. 2000, in Protostars and Planets IV, eds Mannings, V., Boss, A.P., Russell, S. S. (Tucson: University of Arizona Press), 299

\bibitem[Le~Gal et al.(2019)]{LeGal2019}
Le~Gal, R., \"{O}berg, K. I., Loomis, R. A., Pegues, J., \& Bergner, J. B. 2019, \apj, 876, 72

\bibitem[Li et al.(2020)]{Li2020}
Li, S., et al., 2020, \apj 903, 119

\bibitem[Li et al.(2023)]{Li2023}
Li, S., et al., 2023, \apj, 949, 109

\bibitem[Lin et al.(2022)]{Lin2022}
Lin, Y.., Wyrowski, F.,  Liu, H. B., Izquierdo, A. F., Csengeri, T., Leurini,  S. \& Menten, K. M.  2023, \aap, 658, A128

\bibitem[Liechti \& Wilson(1996)]{Liechti1996}
 Liechti, S. \& Wilson, T. L. 1996, \aap, 314, 615

\bibitem[Loren \& Mundy(1984)]{Loren1984}
Loren, R.~B., \& Mundy L.~G., 1984, \apj, 286, 232

\bibitem[McKee \& Tan(2002)]{McKee2002}
McKee, C. F., \& Tan, J. C., 2002, Nature, 416, 59

\bibitem[McKee \& Tan(2003)]{McKee2003}	
McKee, C. F., \& Tan, J. C., 2003, \apj, 585, 850

\bibitem[Malyshev \& Sobolev (2003)]{Malyshev2003}
Malyshev, A. V. \& Sobolev, A. M.  2003, A\&AT, 22, 1 

\bibitem[McGrath et al.(2004)]{McGrath2004}	
McGrath, E. J., Goss, W. M. \&De~Pree, C. G. 2004, \apjs, 155, 577

\bibitem[Mendoza-Torres,  Ju\'{a}rez-Gama, \& Rodr\'{i}guez-Esnard (2023)]{Mendoza2023}
Mendoza-Torres, J. E., Ju\'{a}rez-Gama, M., \& Rodr\'{i}guez-Esnard, I. T.  2023, \aap, 669, 100 

\bibitem[Meng et al.(2026)]{Meng2026} 
Meng, D., Liu, T., Esimbek, J., et al.\ 2026, \apj, 997, 340

\bibitem[Menten (1991)]{Menten1991}
Menten, K. M. 1991, \apj, 380, L75

\bibitem[Men'shchikov (2021)]{Menshchikov2021}
Men'shchikov, A. 2021, A\&A, 649, A89

\bibitem[Minier et al. (2005)]{Minier2005}
Minier, V., Burton, M. G., Hill, T., Pestalozzi, M. R., Purcell, C., Garay, G., Walsh, A., \& Longmore, S., 2005, \aap, 429, 945

\bibitem[Minissale et al.(2022)]{Minissale2022}
Minissale, M., Aikawa, Y., Bergin, E., et al. 2022, ACS Earth and Space Chemistry, 6, 597

\bibitem[Miyawaki et al.(2009)]{Miyawaki2009}
Miyawaki, R., Hayashi, M., \& Hasegawa, T., 2009, PASJ, 61, 39

\bibitem[Miyawaki et al.(2021)]{Miyawaki2021}
Miyawaki, R.,Tsuboi, M., Uehara, K., \& Miyazaki, A., 2021, PASJ, 73, 943

\bibitem[Miyawaki, Hayashi, \& Hasegawa(2022a)]{Miyawaki2022a}
Miyawaki, R., Hayashi, M., \& Hasegawa, T., 2022a, PASJ, 74, 128

\bibitem[Miyawaki, Hayashi, \& Hasegawa(2022b)]{Miyawaki2022b}
Miyawaki, R., Hayashi, M., \& Hasegawa, T., 2022b, PASJ, 74, 705

\bibitem[Miyawaki, Hayashi, \& Hasegawa(2023)]{Miyawaki2023}
Miyawaki, R., Hayashi, M., \& Hasegawa, T., 2023, PASJ, 75, 225

\bibitem[Moscadelli et al. (2022)]{Moscadelli2022}
Moscadelli, L., Sanna, A., Beuther, H., Oliva, A., \& Kuiper, R. 2022, Nature Astronomy, 6, 1068

\bibitem[Motte et al.(2018)]{Motte2018}
Motte, F.,  Bontemps, S., \& Louvet, F. 2018, ARA \& A, 56, 41

\bibitem[Morii et al.(2021)]{Morii2021}
Morii, K. et al. 2021, \apj, 923, 147

\bibitem[Nguyen et al.(2022)]{Nguyen2022}
Nguyen, H. R.  et al. 2022, \aap, 666, 59

\bibitem[Nomura \& Millar(2004)]{Nomura2004}
Nomura, H. \& Millar, T. J., 2004, \aa, 414, 409

\bibitem[Nony et al.(2024)]{Nony2024}
Nony, T.  et al. 2004, \aap, 687, 84

\bibitem[Nummelin et al.(2000)]{Nummelin2000}
Nummelin, A., Bergman, P., Hjalmarson, \AA., Friberg, P., Irvine, W. M., Millar, T. J., Ohishi, M., \& Saito, S. 2000, ApJS, 128, 213

\bibitem[Osorio et al. (1999)]{Osorio1999}
Osorio, M., Lizano, S., \& D\'Alessio, P., \apj, 525, 808

\bibitem[Pandian et al.(2011)]{Pandian2011}
Pandian, J. D.Momjian, E.Xu, Y.Menten, K. M. \& Goldsmith, P. F. 2011, \apj, 730, 55

\bibitem[Palau et al.(2015)]{Palau2015}
Palau, A. et al.  2015, \mnras, 453, 3785

\bibitem[Pestalozzi et al.(2002)]{Pestalozzi2002}
Pestalozzi, M., Humphreys, E. M. L., \& Booth, R. S., 2002, \aap, 384, L15

\bibitem[Plambeck \& Menten(1990)]{Plambeck1990}
Plambeck R. L., \& Menten K. M., 1990 \apj, 364, 555

\bibitem[Pols et al.(2018)]{Pols2018}
 Pols, S., Schw\"{o}rer, A.,  Schilke, P., Schmiedeke, A., S\'{a}nchez-Monge, \'{A}., \& M\"{o}ller, Th. 2018, \aap, 614, A123

\bibitem[Purcell(2006)]{Purcell2006}
Purcell, C. R. 2006, Doctoral dissertation, The Unviersity of South Wales

\bibitem[Redaelli et al.(2022)]{Redaelli2022}
Redaelli, E., et al., 2022, \apj, 936, 169

\bibitem[Remijan et al.(2004)]{Remijan2004}
Remijan, A., Sutton, E. C.,Snyder,  L. E., Friedel, D. N.,  Liu, S.-Y., \& Pei, C.-C. 2004, \apj, 606, 917

\bibitem[Rolffs et al.(2011)]{Rolffs2011}
Rolffs, R., Schilke, P., Zhang, Q., \& Zapata, L. 2011, \aap, 536, 33

\bibitem[Rollig et al.(1999)]{Rollig1999}
R\"{o}llig, M. Kegel, W. H. Mauersberger, \& R. Doerr, C. 1999, \aap, 343, 939

\bibitem[Rodr\'{i}guez et al.(2020)]{Rodriguez2020}
 Rodr\'{i}guez, L., F., Galv\'{a}n-Madrid, R., Sanchez-Bermudez, J., \& De~Pree, C. G. 2020, \apj, 890, 165

\bibitem[S\'anchez-Monge et al.(2018)]{SanchezMonge2018}
S\'anchez-Monge, \'A., Schmiedeke, A., et al. 2018, \aap, 609, A101

\bibitem[Sanhueza et al.(2017)]{Sanhueza2017}
Sanhueza et al. 2017, \apj, 841, 97

\bibitem[Sanna et al.(2014)]{Sanna2014}
Sanna, A., Cesaroni, R., Moscadelli, L., Zhang, Q., Menten, K. M., Molinari, S., Caratti, A., o Garatti, A., \& De~Buizer, J. M. 2014, \aap, 565, 34

\bibitem[Sarma et al.(2002)]{Sarma2002}
Sarma, A. P.Troland, T. H.Crutcher, R. M. \& Roberts, D. A. 2004, \apj, 580, 928

\bibitem[Schmiedeke et al.(2016)]{Schmiedeke2016}
Schmiedeke, A., et al. 2016, \aap, 588, A143

\bibitem[Shimonishi et al.(2021)]{Shimonishi2021}
Shimonishi, T., Izumi, N., Furuya, K.,  \& Yasui, C. 2021, \apj, 922, 206

\bibitem[Silva et al.(2017)]{Silva2017}
Silva, A., , Zhang, Q., Sanhueza, P., Lu, X., Beltr\'{a}n, M. T., Fallscheer, C., Beuther, H., Sridharan, T. K., \& Cesaroni, R. 2017, \apj, 847, 87

\bibitem[Slysh et al. (1999)]{Slysh1999}
Slysh, V. I. Kalenskii, S. V. Val'TTS, I. E. Golubev, V. V.  \& Mead, K. 1999, \apjs, 123, 515

\bibitem[Smith et al. (2009)]{Smith2009}
Smith, N., Whitney, B. A., Conti P. S., De Pree, C. G., \& Jackson, J. M. 2009, \mnras, 399, 952

\bibitem[Sridharan et al.(2002)]{Sridharan2002}
Sridharan, T. K.,  Beuther, H., Schilke, P., Menten, K. M., \& Wyrowski, F. 2002, \apj, .566, 931

\bibitem[St\'{e}phan et al.(2018)]{Stephan2018}
St\'{e}phan, G., S., Schilke, P., Le Bourlot, J., Schmiedeke, A., Choudhury, R., Godard, B.,  \& S\'{a}nchez-Monge, \'{A}.  2018, \aap, 617, 60

\bibitem[Szymczak et al. (2000) ]{Szymczak2000}
Szymczak, M. Hrynek, G. \& Kus, A. J. 2000, \aaps, 143, 269

\bibitem[Tanaka et al.(2016)]{Tanaka2016}
Tanaka, K. E. I. , Tan, J. C., Zhang, Y. 2016, \apj, 818, 52

\bibitem[Torrelles et al.(2003)]{Torrelles2003}
Torrelles, J. M. et al. 2003, \apj, 598, 115
%Torrelles, J. M. Patel, N. A. ; Anglada, G. ; G\UTF{00F3}mez, J. F.  ; Ho, P. T. P.  ; Lara, L. ; Alberdi, A. ; Cant\UTF{00F3}, J. ; Curiel, S. ; Garay, G. ; Rodr\UTF{00ED}guez, L. F. 

\bibitem[Turner(1991)]{Turner1991}
Turner, B.~E. 1991,\apj, 76, 617

\bibitem[Valtts et al.(1995)]{Valtts1995}
Valtts, I. E. Dzyura, A. M. Kalenskii, S. V. Slysh, V. I. Bus, R. \& Vinnberg, A. 1995, Astronomicheskii Zhurnal, 72, 22. 

\bibitem[V\'{a}zquez-Semadeni et al.(2019)]{Vazquez2019}
V\'{a}zquez-Semadeni, E., Palau, A., Ballesteros-Paredes, J., G\'{o}mez, G. C., Zamora-Avil\'{e}s, M. 2019, MNRAS, 490, 3061

\bibitem[Walmsley(1995)]{Walmsley1995} 
Walmsley, C.~M.\ 1995, in ASP Conf. Ser. 80, The Physics of the Interstellar Medium and Intergalactic Medium, ed. A. Ferrara et al. (San Francisco, CA: ASP), 321

\bibitem[Walsh et al.(2003)]{Walsh2003}
Walsh, A. J., Macdonald, G. H., Alvey, N. D. S., Burton, M. G., \& Lee, J.-K., 2003, \aap, 410, 597

\bibitem[Wilner et al.(2001)]{Wilner2001}	
Wilner, D. J., De Pree, C. G., Welch, W. J., \& Goss, W. M. 2001, \apj, 550, L81

\bibitem[Yang et al.(2017)]{Yang2017}
Yang, Wenjin, Xu, Ye, Chen, Xi, Ellingsen, S. P., Lu, Dengrong, Ju, Binggang, \& Li, Yingjie 2017, \apjs, 231, 20

\bibitem[Yonekura et al.(2016)]{Yonekura2016}
Yonekura, Y. et al.  2016, PASJ, 68, 74
%Yonekura, YoshinoriSaito, YuSugiyama, KoichiroSoon, Kang LouMomose, MunetakeYokosawa, MasayoshiOgawa, HideoKimura, KimihiroAbe, YasuhiroNishimura, AtsushiHasegawa, YutakaFujisawa, KentaOhyama, TomoakiKono, YusukeMiyamoto, YusukeSawada-Satoh, SatokoKobayashi, HideyukiKawaguchi, NoriyukiHonma, MarekiShibata, Katsunori M.Sato, KatsuhisaUeno, YujiJike, TakaakiTamura, YoshiakiHirota, TomoyaMiyazaki, AtsushiNiinuma, KotaroSorai, KazuoTakaba, HiroshiHachisuka, KazuyaKondo, TetsuroSekido, MamoruMurata, YasuhiroNakai, NaomasaOmodaka, Toshihiro

\bibitem[Zapata et al.(2012)]{Zapata2012}
Zapata, L. A., Rodr\'iguez, L. F., Schmid-Burgk, J., Loinard, L., Menten, K. M., \& Curiel, S. 2012, \apj, 754, L17

\bibitem[Zhang et al.(2013)]{Zhang2013}
Zhang, B., Reid M. J., Menten, K. M., Zheng, X. W., Brunthaler, A., Dame, T. M., \& Xu, Y. 2013, \apj, 775, 79

\bibitem[Zhang et al.(2019)]{Zhang2019a}
Zhang, Y.,  et al. 2019, \apj, 873, 73 

\bibitem[Zhang et al.(2019)]{Zhang2019b}
Zhang, Y.,  et al. 2019, Nature Astronomy, 3, 517

\bibitem[Zhang et al.(2024)]{Zhang2024}
Zhang, W.,  et al. 2024, \aap, 688, A99

\end{thebibliography}
\end{document}